%% file: paper.tex
\documentclass[prd,aps,showpacs,floats,floatfix,superscriptaddress,twocolumn,nofootinbib]{revtex4}

\usepackage[dvips]{graphicx}
\usepackage{amsmath} 
\usepackage{longtable}
\usepackage{dcolumn,epsfig}

\def\laq{\raise 0.4ex\hbox{$<$}\kern -0.8em\lower 0.62ex\hbox{$\sim$}}
\def\gaq{\raise 0.4ex\hbox{$>$}\kern -0.7em\lower 0.62ex\hbox{$\sim$}}

\newcommand{\beq}{\begin{equation}}
\newcommand{\eeq}{\end{equation}}
\newcommand{\bea}{\begin{eqnarray}} 
\newcommand{\eea}{\end{eqnarray}}
\newcommand{\ba}{\begin{array}}
\newcommand{\ea}{\end{array}}

\newcommand{\lp}{\mbox{\scriptsize
    low-power}}
\newcommand{\comment}[1]{}

\newlength{\sizeonefig}
\newlength{\sizetwofig}
\newlength{\sizeonefigb}
\newlength{\sizetwofigb}
\begin{document}

\title{Improving the sensitivity to gravitational-wave sources by modifying the input-output 
optics of advanced interferometers}

\author{Alessandra Buonanno} 

\affiliation{Groupe de Gravitation et Cosmologie (GReCO), 
Institut d'Astrophysique de Paris (CNRS), 
98$^{\rm bis}$ Boulevard Arago, 
75014 Paris, France}

\author{Yanbei Chen} 

\affiliation{Theoretical Astrophysics, California 
Institute of Technology, Pasadena, CA 91125}

\begin{abstract}
To improve the sensitivity of laser-interferometer gravitational-wave
(GW) detectors,  
experimental techniques of generating squeezed vacuum in the GW
frequency band are 
being developed. The squeezed vacuum generated from non-linear optics
have constant squeeze angle and squeeze factor, while optimal use of
squeezing 
usually requires frequency dependent (FD) squeeze angle and/or homodyne 
detection phase. This frequency dependence can be realized by
filtering the input squeezed vacuum or the output light through detuned
Fabry-Perot cavities. In this paper, we study FD input-output schemes for 
signal-recycling interferometers, the baseline design of Advanced LIGO and 
the currently operational configuration of GEO\,600. 
Complementary to a recent proposal by Harms et al.\ to use FD input squeezing 
and ordinary homodyne detection, we explore a scheme which uses 
ordinary squeezed vacuum, but FD readout. Both schemes, which are sub-optimal 
among all possible input-output schemes, provide a global noise suppression 
by the power squeeze factor. At high frequencies, the two schemes
are equivalent, while at low frequencies the scheme studied in this paper gives better 
performance than the Harms et al.\ scheme, and is nearly fully
optimal. We then study the sensitivity improvement achievable by these schemes 
in Advanced LIGO era (with 30-m filter cavities and current estimates of filter-mirror
losses and thermal noise),  for  
neutron star binary inspirals, for low-mass X-ray binaries and known radio pulsars. 
Optical losses are shown to be a major obstacle for the actual implementation of 
these techniques in Advanced LIGO. 
On time scales of third-generation interferometers, like EURO/LIGO-III ($\sim 2012$),  
with kilometer-scale filter cavities and/or mirrors with lower losses, 
a signal-recycling interferometer with the FD readout 
scheme explored in this paper can 
have performances comparable to existing proposals.
\end{abstract}

\pacs{04.80.Nn, 03.65.ta, 42.50.Dv, 95.55.Ym}
\maketitle

\section{Introduction}
\label{sec1}

The first generation of kilometer-scale, ground-based 
laser-interferometer gravitational-wave 
(GW) detectors (interferometers for short), located
in the United States (LIGO~\cite{LIGO,S1inst}), Europe (VIRGO~\cite{VIRGO} 
and GEO\,600~\cite{GEO,S1inst}) and Japan (TAMA\,300~\cite{TAMA}), have begun 
their search for gravitational radiation and have
yielded first scientific results~\cite{S1continuous,S1inspiral,S1burst,S1stochastic}. The 
 development of interferometers of the second generation, such as Advanced
LIGO (to be operative around 2008~\cite{ALIGO}), and future generations (such as
EURO and LIGO-III), is underway. 
In this paper we explore the possibility of improving
the sensitivity of signal-recycling (SR) interferometers~\cite{SR,M95}, 
the baseline design of Advanced LIGO~\cite{ALIGO} and the current optical 
configuration of GEO\,600~\cite{GEO}, when squeezed vacuum is injected 
into the antisymmetric port (the ``input port'', as we
shall refer to  it in this paper\footnote{This is
  the same port from which the GW signal light exits, but here the squeezed
  vacuum propagates {\it into} the interferometer, instead of coming
  {\it out of}  it. }). 

In the early 1980s, building on works of Caves~\cite{Caves}, 
Unruh~\cite{Unruh} proposed the first design of a squeezed-input 
interferometer, which can beat the free-mass Standard Quantum Limit (SQL)~\cite{SQL}.
Other theoretical studies of input squeezing followed~\cite{Others}.
If generated from non-linear optics, squeezed vacuum will have
frequency independent squeeze angle and squeeze factor in the GW 
frequency band~\cite{SQZ}. The above theoretical works, as well as past experiments
employing squeezed vacuum to enhance interferometer performances~\cite{SQZexp}, all assume
frequency independent squeezing.  
In the 1990s, Vyatchanin, Matsko and Zubova~\cite{FDH} realized that
the sensitivity of GW interferometers can also be improved, beating the
SQL, by measuring an optimal output quadrature, which is usually
frequency dependent. Later, Kimble, Levin, Matsko, Thorne and
Vyatchanin (KLMTV)~\cite{KLMTV00} made a comprehensive, unified
theoretical study of improving the sensitivity of  conventional 
interferometers\footnote{By conventional interferometer we mean a
Michelson interferometer without SR cavity, or with
SR cavity on resonance or antiresonance with the laser frequency.} by
injecting squeezed vacuum into the input port and/or performing frequency
dependent (FD) homodyne detection at the output port. They showed that, for
conventional interferometers, in order to obtain a noise suppression 
proportional to the power squeeze factor at all frequencies (optimal 
input squeezed vacuum), either the squeezed vacuum must have 
a FD squeeze angle ({\it squeezed-input interferometer}), or FD 
homodyne detection has to be applied at the output port 
({\it squeezed-variational interferometer}). 
[Of course, combinations of those optical configurations can also be 
used, but they will be experimentally more challenging.] 
KLMTV proposed a practical way of implementing FD homodyne
detection, as well as converting squeezed vacuum with constant
squeeze angle into squeezed vacuum with FD squeeze angle, by 
filtering the output light or input squeezed vacuum 
through two detuned (with respect to the laser carrier frequency) 
FP cavities (KLMTV filters).  
KLMTV constructed the explicit filter parameters that provide the 
desired frequency dependence for squeezed-input and squeezed-variational interferometers, 
showing that the latter provides a better ideal performance than the
former, but is more susceptible to optical losses. Purdue and  Chen (PC) studied the KLMTV filters
further, and worked out the most general FD squeeze angle and homodyne phase 
that a sequence of filters can provide~\cite{PC02}.  

Experimental programs on generating squeezed vacuum {\it in the GW
frequency band} and injecting it into GW interferometers have
already started in several groups, for example, at the Australian National
University~\cite{ANU}, at the Massachusetts Institute of Technology, USA~\cite{CM}, and at the
Albert Einstein Institut in Hannover, Germany~\cite{AEI}. Their goal
in the next several years is to inject squeezed vacuum with $\sim$10\,dB  
squeeze factor  (as a net result after optical losses) into an 
interferometer. It is very likely that the squeezed vacuum they 
obtain has a constant squeeze angle in the GW frequency band.

This paper contains three relatively independent parts. 
In the first part, we  generalize the work by KLMTV on 
FD input-output optics to SR interferometers. 
Recently, Harms et al.~\cite{Harms03} applied the KLMTV 
squeezed-input scheme, combining FD input squeezed vacuum 
with ordinary homodyne detection to SR interferometers, 
achieving a global noise suppression equal to the power squeeze factor. 
Harms et al. also showed that their FD squeezed-input scheme is only sub-optimal; 
the fully optimal scheme, however, requires
complicated frequency dependence in both input squeeze angle and
homodyne phase, and {\it cannot} be achieved by KLMTV filters.   
Complementary to Harms et al.'s scheme, we explore here the scheme which combines 
ordinary input squeezed vacuum with FD homodyne detection (henceforth, 
the BC scheme). This scheme, which can be thought of as a generalization 
of the KLMTV squeezed-variational scheme, can also
provide a global noise suppression by the power squeeze factor. In 
addition, at high frequencies (above $\sim 200$Hz for typical Advanced
LIGO configurations), it is equivalent to the Harms et al.\ scheme, 
while at low frequencies (below $\sim 200$\,Hz) it is to a very good
approximation fully optimal, and thus provides a better sensitivity
than the Harms et al.\ scheme. 

In the second part of this paper we apply these FD input-output schemes to 
Advanced LIGO (2008), assuming that the generation and injection of squeezed vacuum 
might have already (or partially) become available at that time scale. 
The major obstacle in using FD input-output techniques 
in the facilities of Advanced LIGO is the constraint that 
the filter cavities cannot be longer than $\sim$30\,meters
--- the shorter the filter cavities, the larger the optical losses. 
In our analyses we assume that filter losses dominate 
over internal interferometer losses, and comment only briefly  
on the effects of the latter. To quantify the improvement 
in sensitivity due to FD techniques, we consider three classes of 
astrophysical sources: neutron-star (NS) binary inspirals,  
low-mass X-ray binaries (LMXBs) and known radio pulsars. 
In addition to the ideal quantum noise and filter optical losses, we
also take into account  current estimates of thermal and
seismic noises. [We note that GW interferometers can 
already take advantage of input squeezed vacuum even without 
using FD techniques, if the interferometer parameters are carefully optimized. 
For example, an interesting optical configuration 
without FD input-output optics has been explored by Corbitt and Mavalvala~\cite{CM}, 
providing good sensitivities at high frequencies.] 

In the third part of the paper, we apply our FD readout scheme to 
third-generation interferometers, such as EURO/LIGO-III, which are scheduled to
be operative around 2012.  
We assume that on this time scale, due to the implementation of cryogenic techniques 
and the use of kilometer-length KLMTV filters, thermal noise will be 
negligible and loss effects will be rather low. Third-generation interferometers will have to beat the SQL significantly. We compare the performance of SR interferometers with our FD readout
scheme with those of other existing SQL-beating proposals, such as the KLMTV squeezed-variational interferometer~\cite{KLMTV00} and the speed-meter
interferometers~\cite{ligoIII:sm,ligoIII:sag}. We also investigate the
accuracy  of short-arm and short-filter approximations used in
describing GW interferometers and KLMTV filters.  
More dramatic ideas to circumvent the SQL in GW interferometers exist, 
for example, the intra-cavity schemes of Braginsky, Gorodetsky and Khalili~\cite{K03}, 
and the feed-back control scheme of Courty, Heidmann and Pinard~\cite{CHP}. Since 
thorough analyses of these schemes tuned to GW interferometers has not been available yet, 
in this paper we do not compare the performances.

\begin{table*}[t]
\begin{tabular}{cc}
\hline\hline
Quantity & 
Symbol and Value  \\
\hline\hline
Laser frequency  & $\omega_0=1.8\times10^{15}\,{\rm sec}^{-1}$ \\
GW sideband frequency & $\Omega$  \\
Arm-cavity length & $L$ ($4\,{\rm km}$ for LIGO facilities) \\
Mirror mass &  $m$ ($40\,$kg for Advanced LIGO)\\
Input test-mass (ITM) power transmissivity (LIGO only) &  $T$ \\
Arm-cavity circulating power  &   $I_c$ ($840\,$kW for Advanced LIGO)\\
Light power at the beamsplitter  &   $I_0$\\
SR optical resonant (sideband) frequency &   $-\lambda$ \\
SR bandwidth &   $\epsilon$\\
Homodyne detection phase &   $\zeta$ \\
\hline
Input squeeze factor & $r$ \\
Input squeeze angle & $\alpha$\\
\hline\hline
\end{tabular}
\caption{Parameters of the SR interferometer and input squeezed vacuum.\label{tab:I}}
\end{table*}

Readers with particular interests in the astrophysical consequences of 
using input squeezed vacuum and FD schemes could 
go directly to the second part of the paper [Sec.~\ref{sec5}], 
in which an in-depth understanding of the optics is {\it not} required. 
The paper is organized as follows. In Sec.~\ref{sec2}, we write the input-output 
relation of a non-squeezed SR interferometer in terms of the intrinsic FD 
rotation angle and ponderomotive squeeze factor. In Sec.~\ref{sec3} we review the 
KLMTV filters, including the effects of optical losses. In
Sec.~\ref{sec4} we study FD input-output schemes for SR
interferometers. More specifically, in Sec.~\ref{sec4.1}, we write the general input-output
relation of SR interferometers with FD input-output optics; in
Sec.~\ref{sec4.2}, we study all sub-optimal schemes that allow global noise
suppression, proposing the BC scheme; in Sec.~\ref{subsec:smallq} we study 
the regime with low ponderomotive squeezing (high frequency band of Advanced LIGO),
and show the equivalence between the BC and the Harms et al.\
schemes; in Sec.~\ref{subsec:fullopt} we study the fully optimal scheme, showing that 
at low frequencies the BC scheme is a good approximation to it. 
In Sec.~\ref{sec5}, we investigate the improvement in sensitivity to GWs from various 
astrophysical sources. In Sec.~\ref{sec6}, we compare the BC scheme with 
other proposals for third-generation interferometers, and 
study the effect of filter lengths in FD readout schemes. 
Finally, Sec.~\ref{sec7} summarizes our conclusions.

\section{Quadrature rotation and ponderomotive squeezing in signal recycled 
interferometers}
\label{sec2}
A summary of the various parameters of SR interferometers, such as
Advanced LIGO and GEO\,600, is given in
Table~\ref{tab:I}. The input-output relation for the quadrature
fields in signal recycled interferometers  reads
[see, e.g., Eq. (24) in Ref.~\cite{BC5}, with superscript $(1)$ and tilde dropped]
\beq
\left(
\begin{array}{c}
{b}_1 \\
{b}_2
\end{array}
\right)
=\frac{1}{{M}}\,
\left\{
\left(
\begin{array}{cc}
{C}_{11} &{C}_{12} \\
{C}_{21} &{C}_{22} 
\end{array}
\right)
\left(
\begin{array}{c}
{a}_1  \\
{a}_2
\end{array}
\right)
+
\left(
\begin{array}{c}
{D}_{1} \\
{D}_{2}
\end{array}
\right)
\frac{h}{h_{\rm SQL}}
\right\}\,,
\label{inout}
\eeq
where we define 
\beq
{M}={\left[\lambda^2-(\Omega+i\epsilon)^2\right]\,\Omega^2-\lambda\,\iota_c}\,,
\label{denm}
\eeq
and
\bea
\label{c11}
{C}_{11} = {C}_{22} &=& \Omega^2(\Omega^2-\lambda^2+\epsilon^2)+
\lambda\,\iota_c \,, 
\\
{C}_{12} &=& -2 \epsilon\,\lambda\, \Omega^2\,, \\
{C}_{21} &=& 2\epsilon\,\lambda\, \Omega^2 - 2 \epsilon\, \iota_c\,,
\label{coeffc}
\eea
\beq
{D}_{1} = -2\lambda\,\sqrt{\epsilon\, \iota_c}\,\Omega\,, 
\quad \quad {D}_{2} = 2(\epsilon-i\Omega)\,\Omega\,\sqrt{\epsilon\, \iota_c}\,.
\label{coeffd}
\eeq
The parameters $\lambda$ and $\epsilon$ are related to the
 real and imaginary parts of the free\footnote{Here 
``free'' means that the mirrors are all fixed at
 their equilibrium positions.}  optical resonant 
frequency $\omega_{\rm free}$ of the SR interferometer by~\cite{BC5}:
\beq
\label{omega_free}
\omega_{\mbox{\tiny SR}}^{\rm free} = \omega_0 -\lambda -i\epsilon\,,
\eeq
where $\omega_0$ is the laser frequency. The parameter $\iota_c$ is 
defined by
\bea
\label{iotac}
&&\iota_c
= \frac{8\omega_0\,I_c}{m\, L\, c} \nonumber \\
& \simeq& (2\pi\times 100\,{\rm Hz})^3 
\left(\frac{I_c}{840\,{\rm kW}}\right)
\left(\frac{40\,{\rm kg}}{m}\right)
\left(\frac{4\,{\rm km}}{L}\right),
\eea
where $c$ is the speed of light, $m$ is the mirror mass, $L$ is the arm length, and 
$I_c$ is the circulating optical power in the arm cavity, which
in turn depends on the power at beamsplitter, $I_0$, by\footnote{This
  only applies to LIGO; for GEO\,600 we have $I_c=I_0/2$.} 
\beq
\label{ici0}
I_c=\frac{2}{T}I_0\,.
\eeq
The quantity $\iota_c$ must be on the order of $\Omega_{\rm 
GW}^3$ if we want the opto-mechanical coupling to modify the detuned
(assuming $\lambda \sim \Omega_{\rm GW}$) interferometer's sensitivity
in the GW frequency band [see Eqs.~\eqref{c11}--\eqref{coeffc}].  In addition, we have denoted by 
\beq
\label{sql}
h_{\rm SQL}\equiv \sqrt{\frac{8 \hbar}{m \Omega^2 L^2}}\,
\eeq
the free-mass SQL for the gravitational strain
$h(\Omega)$.\footnote{Note that the definitions of $\iota_c$ and
  $h_{\rm SQL}$, written in terms of $m$, $L$ and $I_c$, 
  differs by numerical factors in Advanced LIGO and GEO\,600.}

It is important to note that the input-output relation, as given by
Eqs.~\eqref{inout}--\eqref{coeffd}, has been obtained at the leading order
in $\Omega L/c$ (as well as in $\epsilon L/c$, $\lambda L/c$ and
$\iota_c^{1/3} L/c$). This approximation is called the ``short-arm'' approximation,
since it assumes that the arm length be much smaller than the gravitational 
wavelength. The short-arm approximation simplifies dramatically the
form of the input-output relation, as well as the design of optimal KLMTV
filters (as we shall see later in this paper). 

In the following sections we first write the input-output 
$b_{1,2}$-$a_{1,2}$ relation [the first term inside the
parenthesis on the RHS of Eq.~\eqref{inout}] 
in terms of an intrinsic squeeze factor $q$ and an intrinsic rotation angle $\varphi$. 
We then study how the output quadratures depend on the signal 
[the second term inside the parenthesis on the RHS of Eq.~\eqref{inout}], 
obtaining the quadrature $\zeta_{\rm max}$ 
at which the signal strength is maximal. Finally, we
give the noise spectrum, and express it in terms of $q$, $\varphi$ and
$\zeta_{\rm max}$. 

\begin{figure*}
\begin{center}
\vspace{0.5cm}
\includegraphics[width=0.85\textwidth]{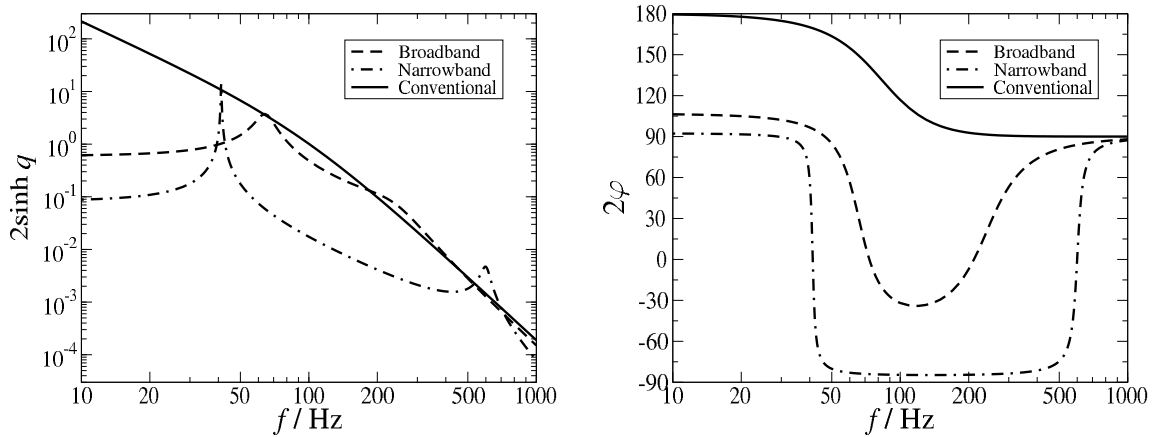}
\caption{We plot $2\sinh q$ (left panel) and $2\varphi$ 
(right panel) for two typical SR optical configurations, the broadband with
$\lambda = 2\pi\times 234.07\,{\rm Hz}$, $\epsilon =2 \pi\times
70.36\,{\rm Hz}$, (dashed line) and narrowband with $\lambda =
2\pi\times 601.43\,{\rm Hz}$, $\epsilon =2 \pi\times 25\,{\rm Hz}$,
(dot-dashed line). We also plot the curves for the conventional interferometer with
$\lambda =0$, $\epsilon = 2 \pi\times 93.75\, {\rm Hz}$ (continuous
line). In all cases we fix $I_c=840\,$kW and $m=40\,$kg. 
\label{Fig1}}
\end{center}
\end{figure*}

\subsection{Rotation of noise quadratures and ponderomotive squeezing}
\label{sec2.1}

As is evident from Eqs.~(\ref{denm})--(\ref{coeffc}), in detuned SR
interferometers (with $\lambda \neq 0$), the input-output relation is
frequency dependent. 
For high-power interferometers like Advanced LIGO, the matrix 
${C}_{ij}$ contains 
both a FD rotation and a FD (ponderomotive) squeezing. 
Let us work these quantities out explicitly. The quantum transfer
matrix (with an overall phase factor removed)   
\beq
{\cal M}_{ij} =\frac{C_{i j}}{|M|}\,, \quad \quad i,j=1,2 \quad (\det {\cal M}_{i j} =1)\,,
\eeq
is a matrix with real elements and with determinant equal to one~\cite{BC2}. 
So, it can always be written 
in the form 
\beq
{\cal M} = \mathbf{R}(\theta)\,\mathbf{S}(\varphi,q)\,,
\label{Mthetaq}
\eeq
as a product of a rotation operator $\mathbf{R}(\theta)$ and a 
squeezing operator $\mathbf{S}(\varphi,q)$, defined by 
\beq
\mathbf{R}(\theta) \equiv
\left(\begin{array}{rr}\cos\theta& -\sin\theta\\ \sin\theta&
  \cos\theta\end{array}\right)
\eeq
and 
\bea
\label{Sq}
\mathbf{S}(\varphi,q) &\equiv& 
\mathbf{R}(\varphi)\,
\mathbf{S}(q)\,
\mathbf{R}(-\varphi)\,, \\
\label{Sqvarphi}
\mathbf{S}(q) &\equiv&
\left(\begin{array}{ll}e^{-q}&\\&e^{q}\end{array}\right)\,,  \quad q
\ge 0\,.
\eea
Here $\theta$ is the rotation angle, $q$ is the squeeze factor, and
$\varphi$ is the squeeze angle. These quantities can in general be
frequency dependent. It can be easily shown that the 
decomposition (\ref{Mthetaq}) is also unique, 
unless $\mathcal{M}$ is a pure rotation (with $q=0$, $\theta$
being the rotation angle, and $\varphi$ being arbitrary). 
{}From Eqs.~(\ref{Mthetaq})--(\ref{Sq}), we have
\beq
\label{trMM}
{\rm Tr} ({\cal M}\,{\cal  M}^\dagger ) 
= 2 \cosh 2q\,,
\eeq
which determines $q$ uniquely once ${\cal M}$ is given 
(recall that $q\ge 0$). If $q$ is zero, 
$\mathbf{S}(\varphi,q)$ reduces to the identity matrix
 regardless of the value of
$\varphi$ [see Eqs.~(\ref{Sq}) and (\ref{Sqvarphi})], and $\theta$
 must be the rotation angle of $\mathcal{M}$;  otherwise, 
in order that
$\mathbf{R}(\theta)\mathbf{S}(q,\varphi)
=\mathbf{R}(\theta')\mathbf{S}(q,\varphi')
$, one must impose  that
\beq
\theta-\theta'=m\pi\,,\qquad  \varphi-\varphi'=n\pi\,,
\eeq
where $m$ and $n$ are integers, and $m-n$ must be an even number. The
uniqueness of the decomposition (\ref{Mthetaq}) assures that the
quantities $\theta$, $q$ and $\varphi$ have unambiguous physical
meaning. 

By comparing Eq.~(\ref{Mthetaq}) with Eqs.~(\ref{inout})--(\ref{coeffc}), 
we can express the angles $\theta,\varphi$ and the factor $q$ 
in terms of SR parameters. Using the identity
(\ref{trMM}), we get the squeeze factor $q$:
\beq
\label{coshq}
\cosh 2q = 1
+ \frac{2 \epsilon^2\,\iota_c^2}{|M|^2}\,.
\eeq
Since for SR interferometers, $C_{11}=C_{22}$, we must impose
$\cos(2\varphi + \theta)=0$, or  
$\theta = \pi/2 - 2\varphi$. This casts $\mathcal{M}_{ij}$ into the following form:
%
\bea
\label{MSR}
&&\mathcal{M} \nonumber \\
&=& \mathbf{R}(\pi/2-\varphi)\,\mathbf{S}(q)\,\mathbf{R}(-\varphi)\nonumber \\
&=&
\left(
\begin{array}{cc}
\sin2\varphi \, \cosh q & -\sinh q - \cos2\varphi\, \cosh q  \\
\sinh q - \cos2 \varphi \,\cosh q & \sin2\varphi \,\cosh q 
\end{array}
\right)\,.\nonumber\\
\eea
%
Thus we have
\bea
\label{tanvarphi}
\tan2\varphi &=& -
\frac{\mathcal{M}_{11}+\mathcal{M}_{22}}{\mathcal{M}_{12}+\mathcal{M}_{21}}
\nonumber \\
&=&
\frac{\iota_c\,\lambda +
\Omega^2\,(\epsilon^2-\lambda^2 + \Omega^2)}{\epsilon\, 
(2 \lambda\,\Omega^2 - \iota_c)}\,, \\
\sinh q &=& \frac{\mathcal{M}_{21}-\mathcal{M}_{12}}{2} = \frac{\epsilon\,\iota_c}{|M|}\,.
\label{sinhq}
\eea
Note that Eq.~(\ref{sinhq}) agrees with Eq.~(\ref{coshq}). 

In absence of SR mirror, or when the SR cavity is either resonant or
antiresonant with the carrier, we have $\lambda =0$, and the above 
equations reduce to the known expressions for conventional interferometers~\cite{KLMTV00}
\beq
\sinh q = \frac{\cal K}{2}\,, \quad
\tan 2 \varphi = -\frac{2}{\cal K}\,,
\label{sinhqconv}
\eeq
with 
\beq
{\cal K} = \frac{2\epsilon\,\iota_c}{\Omega^2\,(\Omega^2 +
  \epsilon^2)}\,,
\eeq
${\cal K}$ being the coupling constant defined by KLMTV in Eq.~(18).
The $\sinh q$ in Eqs.~(\ref{sinhq}) and (\ref{sinhqconv}) is
proportional to $\iota_c$, which  
is in turn proportional to the circulating power $I_c$ and inversely
proportional to the mirror mass $m$. This means the 
squeezing arises from 
the well-known ponderomotive effect~\cite{BM}. 

In Fig.~\ref{Fig1} we plot $2\sinh q$ (left panel) and $2\varphi$
(right panel) as functions of frequency, for two typical SR
configurations and a conventional-interferometer configuration. [Note
that  $2\sinh q \simeq 2q$ as $q\ll 1$ and $2\sinh q \simeq e^{q}$ as $q \gg
1$.]  As we can see from the plots,  both $2\sinh q$ and $2\varphi$ are frequency dependent. 

Let us focus on the detuned configurations (dash and dash-dot curves
in Fig.~\ref{Fig1}). The squeeze factor decreases at high frequencies.  
This can be easily understood from Eq.~\eqref{sinhq}, where we see that $\sinh q$ (hence $q$) 
decreases when $\Omega$ increases, because $M$, Eq.~\eqref{denm}, is a polynomial in $\Omega$, so it  
grows indefinitely as $\Omega$ tends to infinity. 
[The factor $\Omega^2$ in Eq.~(\ref{denm}) can be traced back to the
response of a free mirror to an external force, which decreases
as $1/\Omega^2$; while the factor $[\lambda^2 -(\Omega+i\epsilon)^2]$
increases at high frequencies because the storage time ($1/\epsilon$)
of the interferometer becomes much longer than the GW period.]
Using Eq.~(\ref{denm}) and the fact that $|M| \ge \Im(M)$, we obtain
\beq
\label{Mlb}
|M|\ge 2\,\epsilon\,\Omega^3\,.
\eeq
Combining Eq.~\eqref{Mlb} with Eqs.~(\ref{sinhq}) and~(\ref{iotac}), we have
\bea
\label{qup}
\sinh q =\frac{\epsilon \iota_c}{|M|} &<& \frac{\iota_c}{2\Omega^3}
\nonumber \\
&\simeq& 2\times 10^{-2} \left(\frac{2\pi\times
  300\,{\rm Hz}}{\Omega}\right)^3 \left(\frac{I_c}{840\,{\rm
    kW}}\right) \nonumber \\
&& 
 \times\left(\frac{40\,{\rm kg}}{m}\right)
 \left(\frac{4\,{\rm km}}{L}\right)\,.
\eea
This gives an upperbound for the amount of squeezing achievable with a
given optical power, {\it regardless of resonant features}.  
As we can see from Eq.~(\ref{qup}), even for Advanced LIGO optical power, 
at frequencies larger than $\sim 300$\,Hz, the intrinsic ponderomotive squeezing is
already very small. 

{}From the left panel in Fig.~\ref{Fig1} we observe that the squeeze factor is amplified 
significantly near the ``optical spring'' resonance~\footnote{In detuned SR interferometers there 
are  two resonances in the GW band. One is near the free optical resonant
   frequency of a SR interferometer with fixed mirrors, and we shall
   denote it  ``optical resonant frequency''. The other is 
shifted up from the free pendulum frequency (below
   10\,Hz) into the detection band by the ``optical spring''
   effect~\cite{BC2}. We shall call it the ``optical-spring
   resonant frequency'', or the ``opto-mechanical resonant frequency''. }  
(left peaks), and mildly near the optical 
resonance (right peaks). Those resonant features in $q$ are 
caused by local minima of $M$ around the two resonant frequencies; 
the optical resonance provides less squeezing since squeezing is already 
suppressed at such  high frequencies [see Eq.~\eqref{qup}]. 
The squeeze factor tends to a nonzero constant for $\Omega$ much lower than the
resonant frequencies. By taking the limit of Eq.~(\ref{sinhq}) when 
$\Omega \rightarrow 0$ we obtain that the constant value is: 
\beq
\label{sqz0}
\sinh \left[ q(\Omega=0) \right] = \frac{\epsilon}{\lambda}\,.
\eeq
{}From the right panel of Fig.~\ref{Fig1}, we see that the rotation angle $2\varphi$ 
changes by $180^\circ$ across both the optical-spring resonant
frequency and the optical resonant frequency. 
The above features in $2\varphi$ are typical of resonators and can be explained easily
from Eq.~(\ref{tanvarphi}).

For conventional interferometers ($\lambda =0$; continuous curves 
in Fig.~\ref{Fig1}), the squeeze factor $q$ becomes larger as 
$\Omega$ decreases, providing the strongest squeezing at
almost all frequencies. In particular, $q\rightarrow +\infty$ when
$\Omega \rightarrow 0$, as we can see from Eq.~(\ref{sqz0}).\footnote{In 
reality, $q$ increases only until the test-mass--mirror pendulum frequency is
reached.}  The rotation angle $2\varphi$
changes by $180^{\circ}$  only once over the entire frequency band.

In the low-power limit ($I_c \rightarrow 0$, such that $q \rightarrow
0$), the transfer matrix $\mathcal{M}$ reduces to 
the rotation matrix 
\beq
\mathcal{M}_{\lp} =\mathbf{R}(\pi/2-\varphi_{\lp})\,,
\eeq
with 
\beq
\label{varphilp}
\tan \left (\frac{\pi}{2}-2\,\varphi_{\mbox{\scriptsize low-power}} \right ) =
\frac{2  \lambda\,\epsilon }{\Omega^2-\lambda^2+\epsilon^2}\,.
\eeq
Note that this low-power approximation also applies to high frequencies  
where ponderomotive squeezing is suppressed, even when the power
is the typical high  power in Advanced LIGO  [see Fig.~\ref{Fig1}
  and Eqs.~(\ref{qup})]. 

\subsection{Rotation of signal quadrature}
\label{sec2.2}

Now suppose the output quadrature 
\beq
\label{bzeta}
b_\zeta = b_1\,\sin \zeta + b_2\,\cos \zeta
\eeq
is measured, then the signal part of $b_{\zeta}$ [second term inside
  the parenthesis on the RHS of Eq.~\eqref{inout}] is 
\beq
s_{\zeta} \propto  D_1\, \sin\zeta +  D_2\, \cos\zeta \propto -
\lambda \,\sin\zeta + (\epsilon
-i\Omega)\, \cos\zeta\,.
\eeq
Taking the magnitude squared of the above equation, we obtain 
the signal power in this quadrature,
\bea
&&|s_{\zeta}|^2 \nonumber \\ 
&\propto& (\Omega^2+\lambda^2+\epsilon^2) +
(\Omega^2-\lambda^2+\epsilon^2)\, \cos 2\zeta - 2 \lambda\, \epsilon \,\sin
2\zeta \nonumber \\
&=& S_0+S_1 \cos 2(\zeta - \zeta_{\rm max})\,,
\label{signalquadrature}
\eea
where
\bea
S_0 &\equiv & (\Omega^2+\lambda^2+\epsilon^2)\,,   
\label{eqS0}
\\
S_1 &\equiv & \sqrt{(\Omega^2+\lambda^2+\epsilon^2)^2-4\lambda^2\,\Omega^2}
\,, 
\label{eqS1}
\eea
and
\beq
\label{zetamax}
\zeta_{\rm max} = \frac{1}{2}\arctan\frac{-2\lambda\,\epsilon}{\Omega^2-\lambda^2+\epsilon^2}
\eeq
is the quadrature with maximun signal power. Note that the
relative signal strengths in different quadratures depend 
on $\epsilon$ and $\lambda$ (i.e., on the optical properties of the
interferometer), but not on $\iota_c$ (i.e., on the laser power and
mirror masses).  
Equation~(\ref{zetamax}) suggests a resonant feature of $\zeta_{\rm max}$
near the optical resonant frequency, $\Omega\sim |\lambda|$. 
Equations~(\ref{signalquadrature})--(\ref{eqS1}) also show clearly a known 
result~\cite{BC2}: if $\lambda\,\Omega\neq 0$, 
we have $S_0 > S_1$,  so it is impossible to have an output 
quadrature with no signal.

By comparing Eqs.~(\ref{varphilp}) and (\ref{zetamax}), we can relate 
the frequency dependence of the maximal-signal quadrature of an
interferometer with arbitrary optical power
to the noise-quadrature
rotation of the corresponding low-power interferometer, that is 
\beq
\label{zetavarphi} 
\zeta_{\rm max} = -\frac{1}{2}\left (\frac{\pi}{2}-2\,\varphi_{\mbox{\scriptsize low-power}} \right )\,.
\eeq
As we shall see in Sec.~\ref{sec3}, the factor of $1/2$ in 
front of the RHS of the above equation makes it difficult to design optimal 
FD schemes near the optical resonant frequency.

\subsection{Noise Spectral Density}
\label{sec2.3}

Assuming that ordinary vacuum enters the input port, the SR noise spectral 
density in the $\zeta$ quadrature \eqref{bzeta} 
is given by [see e.g., Eqs. (22)--(36) in Ref.~\cite{BC5}]
\bea
\label{noise_non_squeeze}
&& S_h= \frac{h_{\rm SQL}^2}{\left|D_1\,\sin\zeta+D_2\,\cos\zeta\right|^2}\,\times \nonumber \\
&& [ \left(C_{11}\,\sin\zeta+C_{21}\,\cos\zeta\right)^2
+\left(C_{12}\,\sin\zeta+C_{22}\,\cos\zeta\right)^2]\,. \nonumber \\
\eea
By expressing $S_h$ in terms of the quantities $\varphi$, Eq.~\eqref{tanvarphi}, 
$q$, Eq.~\eqref{sinhq}, $\zeta_{\rm max}$, Eq.~\eqref{zetamax}, and
$S_{0,1}$, Eqs.~\eqref{eqS0} and \eqref{eqS1}, we obtain:
\beq
S_h= |M|^2\,\frac{\left [\cosh 2q - \sinh 2 q\,\cos 2(\zeta-\varphi) \right ]}
{4\epsilon\,\iota_c\,\Omega^2\,\left[S_0 + S_1 \cos 2(\zeta-\zeta_{\rm max})\right]}\,h_{\rm SQL}^2\,.
\eeq

\section{Frequency dependent input-output optics using KLMTV filters}
\label{sec3}

\begin{table}
\begin{tabular}{cc}
\hline\hline
Quantity & 
Symbol   \\
\hline\hline
Filter Length & $L_f$ \\
\begin{tabular}{c}
Input-mirror power transmissivity  \\
and reflectivity
\end{tabular}
& $T_i$, $R_i=1-T_i$ \\
\begin{tabular}{c}
End-mirror power transmissivity  \\
and reflectivity
\end{tabular}
& $T_e$, $R_e=1-T_e$ \\
Resonant (sideband) frequency & $\omega_f$ \\
Bandwidth & $\gamma_f$ \\
\hline\hline
\end{tabular}
\caption{Parameters of an optical filter.\label{tab:II}}
\end{table}

To realize FD homodyne detection and generate squeezed vacuum with 
FD squeeze angle KLMTV~\cite{KLMTV00} proposed to use Fabry-Perot 
cavities, detuned from the laser frequency, with a transmissive input 
mirror and a perfectly reflective end mirror (ideal case).
Later on, PC~\cite{PC02} derived the most general form of 
the FD quadrature rotation achievable by these
filters. We review their work briefly in this section.  

\subsection{Ideal KLMTV filters}
\label{sec3.1}

As shown by PC, the most general quadrature rotation that can be achieved
by a sequence of $n$ ideal KLMTV filters, followed by a frequency independent
rotation, is of the form [see Appendix A of Ref.~\cite{PC02}]:
\beq
\label{KLMTVtanz}
\tan\zeta(\Omega) = \frac{\sum_{k=0}^n B_{k} \Omega^{2k}}{\sum_{j=0}^n
A_{j} \Omega^{2j}}\,,\qquad |A_n+ i B_n|>0\,.
\eeq
The complex resonant frequencies of the filters, $\omega_0 +\Omega_J$, 
$J=1,2,\ldots,n$, are given by the roots (with negative imaginary parts)
of the characteristic equation 
\beq
\label{chareq}
\sum_{k=0}^n(A_k - i B_k)\Omega^{2k}_J =0\,,\qquad \Im(\Omega_J)<0\,.
\eeq 
The constant rotation angle is 
\beq
\label{constrot}
\theta = \arg (A_n + iB_n)\,.
\eeq
[Our Eq.~(\ref{chareq}) is different from Eq.~(A13) of PC, 
because our definition for $\Omega_J$ is different from PC's
definition for $\omega_J$. See their Eq.~(A12).] Like in the
input-output relation of SR interferometers, the filter input-output
relation in this section has also been obtained at the leading order
in $\Omega L/c$ (as well as in $|\Omega_{\rm res}|L/c$), that is, in the 
short-filter approximation. It
is only under this approximation that we can cast the quadrature
rotation of these filters into the elegant form \eqref{KLMTVtanz}.

For low-power interferometers ($\iota_c, q= 0$),
the transfer matrix $\mathcal{M}$ reduces to the pure rotation 
$\mathbf{R}(\pi/2-2\varphi_{\mbox{\scriptsize low-power}})$ with 
\beq
\tan \left (\frac{\pi}{2}-2\varphi_{\mbox{\scriptsize low-power}} \right ) =
\frac{2\lambda\epsilon}{\Omega^2-\lambda^2+\epsilon^2}\,,
\eeq
which is of the form (\ref{KLMTVtanz}) and can be realized by one KLMTV 
filter with complex resonant frequency at $\omega_0
-\lambda -i\epsilon$, which coincides with the free optical resonant
frequency of the SR interferometer [see Eq.~(\ref{omega_free})]. 
Unfortunately, due to the factor of $1/2$ in front of $\arctan$  
in Eq.~(\ref{zetamax}), the frequency-dependent rotation of the 
maximal-signal quadrature cannot be realized  by a sequence
of KLMTV filters.

\begin{figure}
\begin{center}
\begin{tabular}{c}
\includegraphics[width=0.3\textwidth]{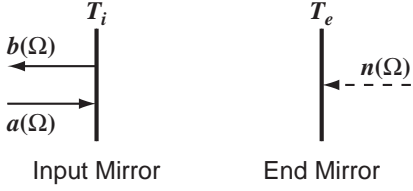} 
\end{tabular}
\caption{Filter cavity with input $a(\Omega)$, output  $b(\Omega)$ and noise 
$n(\Omega)$ field operators. We indicate with $T_i$ and $T_e$ the 
power transmissivity of the input and end mirrors, respectively. \label{lossfilter}}
\end{center}
\end{figure}

\subsection{KLMTV filters with low loss}
\label{sec3.2}

Following KLMTV, we model losses in a filter cavity by assuming
that the end mirror has a non-vanishing power transmissivity $T_e$, and
a power reflectivity of $R_e=1-T_e$. Denoting the front-mirror power 
transmissivity and reflectivity by $T_i$ and $R_i$ ($T_i+R_i=1$), 
the filter input-output relation to {\it first order} in  $T_e/T_i$ reads: 
\beq
\label{KLMTVinout}
\left(
\begin{array}{c}
b_1^{\rm out} \\
b_2^{\rm out}
\end{array}
\right)
=\sqrt{1-\mathcal{E}}\,
\mathbf{R}(\zeta)\,
\left(
\begin{array}{c}
b_1 ^{\rm in}\\
b_2 ^{\rm in}
\end{array}
\right)
+
\sqrt{\mathcal{E}}\, 
\left(
\begin{array}{c}
n_1^{\rm filter} \\
n_2^{\rm filter}
\end{array}
\right)\,,
\eeq
where the rotation $\mathbf{R}(\zeta)$
is the same as in the lossless case, $n_{1,2}^{\rm filter}$ are  vacuum 
quadrature fields leaking in from the end mirror, and the loss factor 
$\mathcal{E}$ is given by  
\beq
\mathcal{E} = \frac{1}{2}\sum_{s=+,-} \mathcal{E}_{s}\,,
\eeq
with
\beq
\label{eqEJ}
\mathcal{E}_{\pm}  = \frac{2 \,T_e}{T_i}\,\frac{2}{1+(\pm \Omega-
    \omega_f)^2/\gamma_f^2} \,,\qquad
\Omega_{\mathrm{res}} = \omega_f - i\gamma_f\,.
\eeq
Here $\omega_f$, $\gamma_f$, $L_f$ are the resonant frequency, bandwidth 
and length of the filter, respectively. The bandwidth $\gamma_f$ is related to $T_i$
and $L_f$ by
\beq
\label{eqTJ}
\gamma_f = \frac{T_i\, c}{4 L_f}\,.
\eeq
[The optical-filter parameters are summarized in Table~\ref{tab:II}.] 
For a sequence of multiple filters, the rotation angles $\zeta$ and loss factors
$\mathcal{E}$ of each filter add up to give the total
rotation angle and loss factor. In this way, the total rotation angle
will be identical to the ideal value, while the total loss factor will
be
\beq
\label{eqEtot}
\mathcal{E} = \frac{1}{2}\sum_{s=+,- \atop J=\rm filters} \mathcal{E}_{s}^J\,.
\eeq
The total loss factor is frequency
dependent, but never exceeds the upper limit 
\beq
\mathcal{E}_{\rm max} = \sum_{J=\rm filters} \frac{4T_e^J}{T_i^J}\,.
\eeq
Moreover, if the filters have eigenfrequencies well-separated from
each other, that is 
\beq
\left||\omega_{f}^I|-|\omega_{f}^J|\right| \gg
\max\{\gamma_{f}^I,\gamma_{f}^J\}\,,
\eeq
and if all filters have high ``quality factors,'' that is
\beq
\omega_f^J \gg \gamma_f^J\,,
\eeq
then, if we evaluate Eq.~(\ref{eqEtot}) around the resonant frequencies 
only one term dominates, yet away from resonances the loss factor is not very 
large. The total loss factor has peaks at the resonant frequencies 
of each filter, with peak value 
\beq
\mathcal{E}_{\rm res}^J = \frac{2T_e^J}{T_i^J}\,, \quad \quad \Omega \simeq 
|\Omega_{\rm res}^J|\,.
\eeq
and width comparable to $\gamma^J_f$. 

Once a filter's bandwidth $\gamma_f$  and the end-mirror
transmissivity $T_e$ are specified,  
we can rewrite the peak value of the total loss factor (near this
filter's resonant frequency) as
\beq
\label{EresL}
\mathcal{E}_{\rm res}^J =\frac{T_e^J c}{2\gamma_f^J L_f}\,, 
\quad \quad \Omega \simeq |\Omega_{\rm res}^J|\,.
\eeq
Thus, the shorter the cavity, the lower the front-mirror
transmissivity and the larger the loss factor.  As an order-of-magnitude
estimate, we show in Table~\ref{tab:loss} the values of $\mathcal{E}_{\rm res}$ 
evaluated for  typical filter lengths
($4000\,$m, $400\,$m, and $30\,$m) and bandwidths $\gamma_f$ ($2\pi \times 100$\,Hz
and $2\pi \times 25\,$Hz), having assumed $T_e = 20\,$ppm~\cite{Whitcomb}.

{\setlength{\tabcolsep}{10pt}
\begin{table}
\begin{tabular}{rrr}
\hline
\multicolumn{1}{c}{$L_f$} & 
\multicolumn{1}{c}{$\gamma_f/(2\pi)$} & 
\multicolumn{1}{c}{$\mathcal{E}_{\rm res}$} \\
\hline
\hline
4000\,m & 100\,Hz & 0.0012 \\ 
 & 25\,Hz & 0.0048 \\
\hline
400\,m & 100\,Hz & 0.012 \\
 & 25\,Hz & 0.048 \\
\hline
30\,m & 100\,Hz & 0.16 \\
 & 25\,Hz & 0.64  \\
\hline
\end{tabular}
\caption{
Peak values of the filter power-loss factor, $\mathcal{E}_{\rm res}$ [Eq.~(\ref{EresL})],
for various filter lengths and bandwidths, assuming an end-mirror transmissivity 
of 20 ppm. \label{tab:loss}}
\end{table}
 }

\subsection{KLMTV filters with significant loss}
\label{sec3.3}

As we can see from Table~\ref{tab:loss}, when the filters are short, 
e.g., on the order of 30\,m, the energy loss factor can become quite 
large, and the leading-order calculation used in Sec.~\ref{sec3.2} 
can no longer be trusted.  Instead, here we give the exact filter
input-output relation.   
By denoting with $a(\Omega)$, $b(\Omega)$ and $n(\Omega)$ the 
(Fourier domain) annihilation operators of the input, output and noise
fields at 
frequency $\omega_0 +\Omega$, we have [see Fig.~\ref{lossfilter}]: 
\bea
b(\Omega) &=&\frac{\sqrt{R_e}\,e^{2i(\Omega-\omega_f) L_f/c} 
-\sqrt{R_i}}{1-\sqrt{R_i
      R_e}\,e^{2i(\Omega-\omega_f)L_f/c}}\, a(\Omega) \nonumber \\
&+&
\frac{\sqrt{T_i T_e}\,e^{i(\Omega-\omega_f)L_f/c}}{1-\sqrt{R_i
      R_e}\,e^{2i(\Omega-\omega_f)L_f/c}}\,n(\Omega)\,.
\label{eqban}
\eea
Here $\omega_0+\omega_f$ is the resonant frequency of the filter cavity
(the one nearest $\omega_0$).
 The quadrature input-output relation can be obtained from Eq.~(\ref{eqban}) 
by using, e.g., Eqs.~(A8) and (A9) of Ref.~\cite{BC5}. Namely, 
the relation
\beq
b(\pm\Omega) = f_{\pm}(\Omega)\, a(\pm \Omega) 
\eeq
valid for annihilation operators is equivalent to the relation 
\beq
\label{anntoquad}
\left(
\begin{array}{c}
b_1 \\
b_2
\end{array}
\right)
=\frac{1}{2}
\left(
\begin{array}{cc}
(f_+ + f_-^*) & i(f_+ - f_-^*) \\
-i(f_+ - f_-^*) & (f_+ + f_-^*)
\end{array}
\right)
\left(
\begin{array}{c}
a_1 \\
a_2
\end{array}
\right)\,,
\eeq
valid for quadrature fields $a_{1,2}$ and $b_{1,2}$. [Note the typo in the (2,1) component of 
Eq.~(A9) of Ref.~\cite{BC5}.] 

Again, we can apply the short-filter approximation, 
$\Omega L/c$, $\omega_{f} L/c$, $\gamma_{f} L/c \ll 1$, $T_e
\stackrel{<}{_\sim} T_i$, and
obtain simpler formulas:
\beq
b=\frac{1-\alpha_f + i\displaystyle\frac{\Omega-\omega_f}{\gamma_f}}{1+\alpha_f -
  i\displaystyle\frac{\Omega-\omega_f}{\gamma_f}}\,a
+\frac{2\sqrt{\alpha_f}}{1+\alpha_f-i\displaystyle\frac{\Omega-\omega_f}{\gamma_f}}\,n\,,
\eeq
where $\alpha_f = T_e/T_i$. 
By converting into the quadrature representation we have:
\beq
\left(
\begin{array}{c}
b_1 \\
b_2
\end{array}
\right)
=
\frac{(\mathcal{R} + i\alpha_f\, \Lambda)
\left(
\begin{array}{c}
a_1 \\
a_2
\end{array}
\right)
+
2\sqrt{\alpha_f} \,\mathcal{N}
\left(
\begin{array}{c}
n_1 \\
n_2
\end{array}
\right)
}{
\left(1+\alpha_f- \displaystyle i\frac{\Omega+\omega_f}{\gamma_f}\right)
\left(1+\alpha_f- \displaystyle i\frac{\Omega-\omega_f}{\gamma_f}\right)
}\,,
\eeq
where $n_{1,2}$ are quadratures of the field $n$, 
\beq
\mathcal{R} = 
\left(
\begin{array}{cc}
1-\alpha_f^2+\frac{\Omega^2-\omega_f^2}{\gamma_f^2} & \frac{2\omega_f}{\gamma_f}
\\
-\frac{2\omega_f}{\gamma_f} & 1-\alpha_f^2
+\frac{\Omega^2-\omega_f^2}{\gamma_f^2}
\end{array}
\right)\,,
\eeq
\beq
\Lambda = \frac{2\Omega}{\gamma }\,\mathbf{I}\,, 
\eeq
and
\beq
\mathcal{N}=
\left(
\begin{array}{cc}
1+\alpha_f -i\frac{\Omega}{\gamma_f} & \frac{\omega_f}{\gamma_f}\\
-\frac{\omega_f}{\gamma_f} & 1+\alpha_f -i\frac{\Omega}{\gamma_f}
\end{array}
\right)\,.
\label{eqcalN}
\eeq

\begin{figure*}
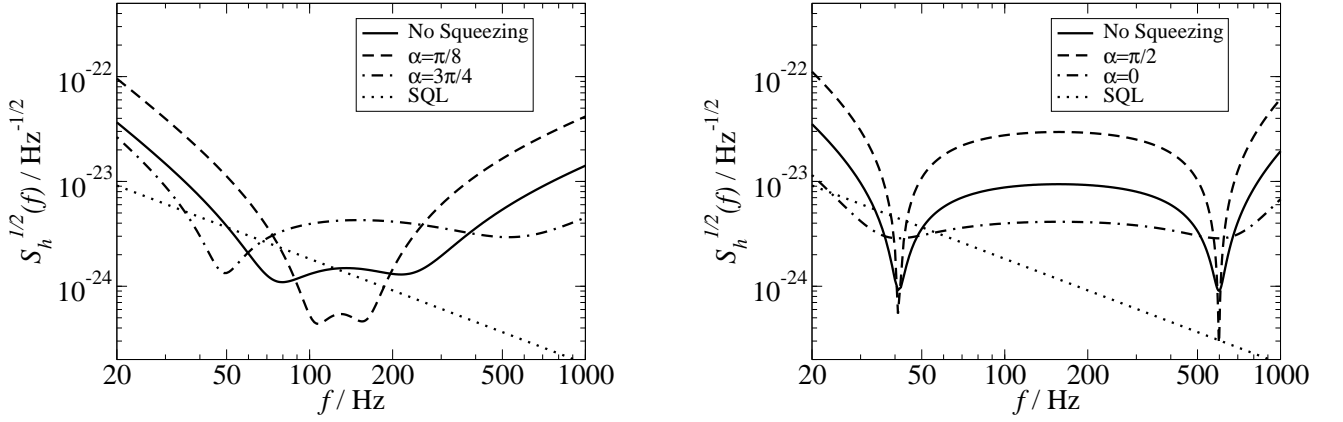

\begin{center}
\vspace{0.5cm}
\begin{tabular}{ccc}
\includegraphics[width=0.45\textwidth]{fig3a.eps} &
\hspace{0.05\textwidth} &
\includegraphics[width=0.45\textwidth]{fig3b.eps} 
\end{tabular}
\vspace{0.2cm}
\caption{Noise curves of SR interferometers with frequency independent input
squeezing and homodyne readout. In the left panel, we show the broadband
configuration: $\lambda = 2\pi\times 234.07\,{\rm Hz}$, $\epsilon =2 \pi\times 70.36\,{\rm Hz}$, 
$\zeta = -0.8037$, with no input squeezing 
($r = 0$, $\alpha=0$) (continuous curve), 
with $e^{2r} = 10$, $\alpha=\pi/8$ (dashed curve) 
and with $e^{2r} = 10$, $\alpha=3\pi/4$ (dash-dot curve). In the right
panel, we plot the narrowband configuration: 
$\lambda = 2\pi\times 600\,{\rm Hz}$, $\epsilon =2 \pi\times 25\,{\rm Hz}$, 
$\zeta = \pi/2$, with no input squeezing ($r=0$,
$\alpha=0$) (continuous curve), 
with $e^{2r} = 10$, $\alpha=\pi/2$ (dashed curve) 
and with $e^{2r} = 10$, $\alpha=0$ (dash-dot curve). In both
 configurations we fix $I_c=840\,$kW and $m=40\,$kg and 
show the SQL curve (dotted line). 
\label{Fig2}}
\end{center}
\vspace{-0.6cm}
\end{figure*}

\section{Squeezed-input and variational-output signal recycled interferometers}
\label{sec4}

\subsection{Input-output relation and noise spectral density}
\label{sec4.1}

As discussed by KLMTV, a GW interferometer with  squeezed vacuum state
$|S(r,\alpha)\rangle$ fed into its input port can be described 
by applying the following unitary transformation, 
\bea
\left(\begin{array}{c}a_1\\a_2\end{array}\right) &\rightarrow&
\mathbf{R}(\alpha)
\mathbf{S}(r)\left(\begin{array}{c}\tilde{a}_1\\\tilde{a}_2\end{array}\right)\,, 
\label{sqztransfop}\\ 
|S(r,\alpha)_a\rangle & \rightarrow&  |0_{\tilde{a}} \rangle\,, 
\eea
in which the quadrature operators undergo a linear transformation,
while the quantum state is transformed back to the vacuum state. 
[Note that $\langle 0_{\tilde{a}}| \tilde{a}_i\,\tilde{a}_j^\dagger
|0_{\tilde{a}} \rangle =2 \pi\,\delta_{ij}\,\delta(\Omega-\Omega')$.]
Equation~(\ref{sqztransfop}) suggests that, in practice, the
squeeze angle $\alpha$ of the squeezed vacuum  injected into the
input port can be obtained by a quadrature-rotating optical 
element, e.g., a KLMTV filter, placed between the 
squeezer and the interferometer. 

Once the unitary transformation is applied, the
input-output relation of the interferometer can be written similarly to
Eq.~(\ref{inout}) as
\begin{widetext}
\beq
\label{inoutgeneric}
\left(\ba{c}b_1\\b_2\ea\right)=
\frac{1}{M}
\left[\left(\ba{cc}
C^{\alpha}_{11}
&C_{12}^{\alpha}
\\C_{21}^{\alpha}
&C_{22}^{\alpha}
\ea\right)
\left(\ba{r} e^{-r} \tilde{a}_1   \\ e^{+r} \tilde{a}_2 \ea\right)+\left(\ba{c}D_1\\D_2\ea\right)\frac{h}{h_{\rm SQL}}\right]\,,
\eeq
where
\beq
\left(\ba{cc}C^{\alpha}_{11}&C_{12}^{\alpha}\\C_{21}^{\alpha}&C_{22}^{\alpha}\ea\right)
\equiv  \left(\ba{cc} C_{11}\,\cos \alpha + C_{12}\,\sin \alpha& 
C_{12}\,\cos \alpha - C_{11}\,\sin \alpha \\
C_{21}\,\cos \alpha + C_{22}\,\sin \alpha& 
C_{22}\,\cos \alpha - C_{21}\,\sin \alpha \ea\right)\,.
\eeq
The quadrature $\tilde{a}_1$ is generally called the ``squeezed
quadrature'' because it enters 
Eq.~\eqref{inoutgeneric} multiplied by $e^{-r}$, while $\tilde{a}_2$ is 
called the ``stretched quadrature'' because it is multiplied by $e^{+r}$. 
If the output quadrature $b_\zeta = b_1\,\sin \zeta + b_2\,\cos \zeta$ is measured, 
the noise spectral density is 
\beq
\label{noise_squeeze}
S_h=\frac{e^{-2r}\left(C^{\alpha}_{11}\,\sin\zeta+C^{\alpha}_{21}\,\cos\zeta\right)^2
+e^{2r}\left(C^{\alpha}_{12}\,\sin\zeta+C^{\alpha}_{22}\,\cos\zeta\right)^2}
{\left|D_1\,\sin\zeta+D_2\,\cos\zeta\right|^2}\,h_{\rm SQL}^2\,,
\eeq
which in terms of the (ponderomotive) squeeze factor $q$, intrinsic
rotation angle $\varphi$, maximal-signal quadrature $\zeta_{\rm max}$ reads:
\beq
\label{noise_squeeze_phiq}
S_h=|M|^2\,\frac{e^{-2r}\,\left [ \cosh q\,\cos(\alpha + \zeta - 2\varphi)-
\sinh q\, \cos (\alpha - \zeta) \right ]^2 + e^{2r}\,
\left [ \cosh q\,\sin(\alpha + \zeta - 2\varphi)-
\sinh q\, \sin(\alpha - \zeta) \right ]^2}
{4\epsilon \iota_c \Omega^2 \left[S_0 + S_1\cos 2(\zeta-\zeta_{\rm max})\right]}\,h_{\rm SQL}^2\,.
\eeq
\end{widetext}
In Eqs.~(\ref{noise_squeeze}) and (\ref{noise_squeeze_phiq}), the
spectral density $S_h$ contains a term proportional to $e^{-2r}$, as
well as one proportional to $e^{2r}$.
We can take advantage of squeezed vacuum only if  $b_{\zeta}$ contains
very little (preferably none) of the stretched quadrature $\tilde{a}_2$. 

In Fig.~\ref{Fig2} we plot some examples of noise spectral densities
with frequency independent input squeezing (constant $\alpha$) and
readout (constant $\zeta$). In this case, squeezing 
can improve the sensitiviy at some frequencies, but at the price of
deteriorating the sensitivity at other frequencies. 
However, as investigated by Corbitt, Mavalvala and
Whitcomb~\cite{CM,CMW}, 
without introducing FD input-output techniques, it
is still possible to take advantage of input squeezing, by choosing carefully 
the SR parameters $(\lambda,\epsilon,\iota_c)$, and/or
by filtering out the squeezed vacuum in the frequency region where the
stretched quadrature increases the noise. On the other hand, if, for a 
substantially detuned
configuration, we would like to obtain a large noise-suppression
factor over the entire frequency band, FD input-output techniques 
should be used. 

\subsection{Cancellation of the stretched quadrature and sub-optimal schemes}
\label{sec4.2}

In order that $S_h$ in Eq.~(\ref{noise_squeeze}) has only 
the term proportional to $e^{-2r}$, we have to impose 
\beq
C_{12}^{\alpha}\sin\zeta+C_{22}^{\alpha}\cos\zeta=0\,,
\eeq
or, more symmetrically in $\alpha$ and $\zeta$,
\beq
\label{subopt}
\left(\begin{array}{cc} \sin\zeta & \cos\zeta \end{array}\right)
\left(
\begin{array}{cc} 
C_{11} & C_{12} \\ C_{21} & C_{22} 
\end{array}
\right) 
\left(\begin{array}{r}
-\sin\alpha \\
\cos\alpha
\end{array}
\right) =0 \,.
\eeq
It is interesting to note that Eq.~(\ref{subopt}) does not depend on $r$.
This happens because the way $e^{-r}\,\tilde{a}_{1}$  and $e^{r}\,\tilde{a}_{2}$ 
are mapped into $b_{\zeta}$ [see Eq.~(\ref{inoutgeneric})] 
depends only on $\alpha$, $C_{ij}$, and $\zeta$, but not on $r$.

Equation (\ref{subopt}) can be satisfied in many ways. 
However, since $C_{ij}$ are frequency dependent, either
$\alpha$ or $\zeta$, or both, will have to be frequency dependent.
Given such a pair of $(\alpha(\Omega),
\zeta(\Omega))$, the noise spectrum can be obtained by inserting them
into Eq.~(\ref{noise_squeeze}), obtaining
\begin{widetext}
\beq
S_h = \frac{
e^{-2r}\left[
C_{11}^{\alpha(\Omega)} \sin\zeta(\Omega) +
C_{21}^{\alpha(\Omega)} \cos\zeta(\Omega)\right]^2
+
e^{2r}\left[
C_{12}^{\alpha(\Omega)} \sin\zeta(\Omega) +
C_{22}^{\alpha(\Omega)} \cos\zeta(\Omega)\right]^2
}{|D_1 \sin\zeta(\Omega) + D_2 \cos\zeta(\Omega) |^2}
h_{\rm SQL}^2\,,
\label{intstep}
\eeq
with the second term in the numerator 
vanishing once Eq.~(\ref{subopt}) is imposed. As a consequence, we
can also write
\bea
S_h &=& e^{-2r} 
\frac{
\left[
C_{11}^{\alpha(\Omega)} \sin\zeta(\Omega) +
C_{21}^{\alpha(\Omega)} \cos\zeta(\Omega)\right]^2
+
\left[
C_{12}^{\alpha(\Omega)} \sin\zeta(\Omega) +
C_{22}^{\alpha(\Omega)} \cos\zeta(\Omega)\right]^2
}{|D_1 \sin\zeta(\Omega) + D_2 \cos\zeta(\Omega) |^2}
h_{\rm SQL}^2 \nonumber \\
&=&
e^{-2r}\frac{
\left[
C_{11} \sin\zeta(\Omega) +
C_{21} \cos\zeta(\Omega)\right]^2
+
\left[
C_{12} \sin\zeta(\Omega) +
C_{22} \cos\zeta(\Omega)\right]^2
}{|D_1 \sin\zeta(\Omega) + D_2 \cos\zeta(\Omega) |^2}
h_{\rm SQL}^2\,.
\label{Shsubopt}
\eea
\end{widetext}
The first equality in Eq.~(\ref{Shsubopt}) says that the noise
spectrum of an input-output scheme [as specified by 
($\alpha(\Omega)$,$\zeta(\Omega)$)] with an input squeeze factor 
$r$ scales as $e^{-2r}$; the second equality in Eq.~(\ref{Shsubopt}) 
must hold since for ordinary vacuum a rotation of the input 
quadratures leaves the system invariant. 
The spectral density, as given by Eq.~(\ref{Shsubopt}), is 
$e^{-2r}$ times that of a (non-squeezed) FD readout scheme with
homodyne phase $\zeta(\Omega)$. Clearly, an additional optimization in $\zeta$ will give the fully 
optimal input-output scheme. However,  
we postpone the discussion of the fully optimal scheme till
Sec.~\ref{subsec:fullopt} and investigate first  the 
{\it sub-optimal} schemes, which have $(\alpha(\Omega),\zeta(\Omega))$ 
satisfying Eq.~(\ref{subopt}) but do 
 not necessarily have the optimal $\zeta(\Omega)$ required by the
 minimization of (\ref{Shsubopt}).  These schemes all provide a global noise
 suppression by the factor $e^{-2r}$.  

The (two) simplest solutions to Eq.~(\ref{subopt}) can be obtained by
 imposing $\zeta$ (or $\alpha$) to be frequency 
independent and solving 
Eq.~(\ref{subopt}) for $\alpha$ (or $\zeta$). 
This means that KLMTV filters are placed either in
the input port or in the output port, but not in both places. 

The first simple solution has been studied by 
Harms et al.~\cite{Harms03}, who proposed to inject squeezed vacuum with
FD squeeze angle into SR interferometers. 
Imposing a frequency independent $\zeta$, they obtained
\beq
\tan \alpha_{\rm subopt}(\Omega)
= \frac{C_{22}\, \cos \zeta + C_{12}\,\sin\zeta}{C_{21}\,\cos\zeta
  +C_{11}\,\sin\zeta}\,. 
\label{alphasubopt}
\eeq
Remarkably, the required $\alpha_{\rm subopt}$ in 
Eq.~(\ref{alphasubopt}) is of the form (\ref{KLMTVtanz}), thus realizable 
by  KLMTV filters. In our notations, the characteristic equation for the
filters is
\beq
\label{charHarms}
\Omega^2(\Omega+\lambda-i\epsilon)
(\Omega-\lambda+i\epsilon) +
\left[\lambda-2ie^{-i\zeta}\epsilon\cos\zeta\right]\,\iota_c=0\,,
\eeq
while the constant rotation following the filters should be
\beq
\theta = \pi/2-\zeta\,.
\eeq
[See Eqs.~\eqref{chareq} and \eqref{constrot}.] Note that, without
making the short-arm and short-filter approximations, both
Eqs.~\eqref{alphasubopt} and \eqref{KLMTVtanz} would have been 
much more complicated, making the identification of filter parameters
  much less straightforward (or even impossible).
 
In this paper, we explore the second simple solution. We assume 
a frequency independent $\alpha$ and requires the FD detection phase
\bea
\tan\zeta_{\rm subopt}(\Omega) 
=- \frac{C_{22}\, \cos \alpha -
C_{21}\,\sin\alpha}{C_{12}\,\cos\alpha -C_{11}\,\sin\alpha}\,.
\label{zetasubopt}
\eea
This detection phase is also of the form 
(\ref{KLMTVtanz}) and realizable by KLMTV filters, with 
characteristic equation 
\beq
\label{charBC}
\Omega^2 (\Omega + \lambda - i\epsilon)
(\Omega - \lambda + i\epsilon)
+\left[\lambda + 2 e^{-i\alpha} \epsilon \sin\alpha\right]\iota_c=0\,,
\eeq
and a subsequent frequency independent rotation 
\beq
\theta = 3\pi/2- \alpha\,.
\eeq
Henceforth, we shall call this scheme the BC scheme. 
The noise spectral density of the BC scheme can be obtained by inserting
Eq.~(\ref{zetasubopt}) into Eq.~(\ref{Shsubopt}); the result is
\beq
\label{Shlossless}
S_h =
\frac{e^{-2r}|M|^2 h_{\rm SQL}^2}{4\epsilon \iota_c \Omega^2
  |\lambda\cos\alpha +(\epsilon-i\Omega)\sin\alpha|^2}\,.
\eeq

\begin{figure*}
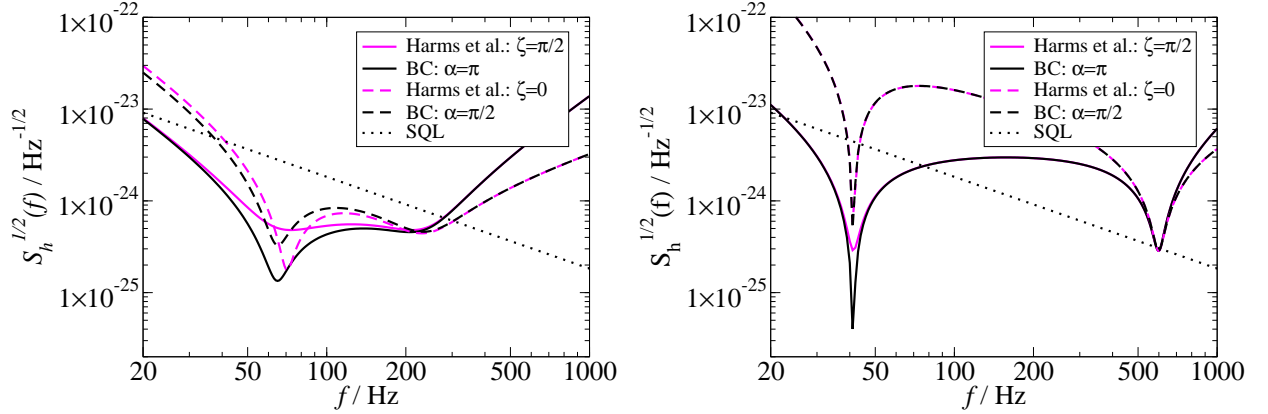

\begin{center}
\vspace{0.5cm}
\begin{tabular}{ccc} 
\includegraphics[width=0.45\textwidth]{fig4a.eps} &
\hspace{0.0\textwidth}
&\includegraphics[width=0.45\textwidth]{fig4b.eps} 
\end{tabular}
\vspace{0.25cm}
\end{center}
\caption{Equivalence of the Harms et al.\ and BC schemes at high
  frequencies.  We assume $I_c=840\,$kW, $m=40\,$kg and  $e^{2r}=10$.  
In the left panel, we plot the broadband configurations with
$\lambda = 2\pi\times 234.07\,{\rm Hz}$, $\epsilon =2 \pi\times
  70.36\,{\rm Hz}$, while in the right panel we show narrowband
  configurations with $\lambda=2\pi\times600\,{\rm Hz}$ and
  $\epsilon=2\pi\times25\,$Hz. In both panels, BC schemes
  with $\alpha=\pi$ and $\pi/2$ are shown in dark continuous and dark
  dashed curves, respectively, while Harms et al.\ schemes with
  $\zeta=\pi/2$ and $\zeta=0$ are shown in light continuous and light
  dashed curves. Noise curves from the two schemes, which are 
  related by $\alpha=\zeta+\pi/2$, do
  agree quite well for frequencies higher than $200\,$Hz in the
  broadband configuration (left
  panel), and at almost all frequencies in the narrowband
  configuration (right panel). Note that at high frequencies $q$ is significantly lower in
  the narrowband configuration, see left panel of Fig.~\ref{Fig2}. 
\label{fig3}
}
\end{figure*}

Additional insight into these sub-optimal schemes can be obtained by
decomposing the input-output $\tilde{a}$-$b$ relation into a product
of rotation and squeezing operators [see Eqs.~(\ref{MSR}), (\ref{sqztransfop})
  and (\ref{inoutgeneric})]:
\begin{widetext}
\beq
\label{qfluc}
\Delta b_{\zeta}=\underbrace{
\Bigg[
\left(
\begin{array}{cc}
0 & 1
\end{array}
\right)
\mathbf{R}(\zeta)\Bigg]}_{\mbox{readout}}
\underbrace{
\Bigg[
 \mathbf{R}(\pi/2-\varphi) \mathbf{S}(q)
 \mathbf{R}(-\varphi)\Bigg]}_{\mbox{interferometer}} 
\underbrace{ 
\Bigg[
 \mathbf{R}(\alpha)\mathbf{S}(r)
\left(
\begin{array}{c}
\tilde{a}_1 \\
\tilde{a}_2
\end{array}
\right)\Bigg]}_{\mbox{input}}\,,
\eeq
Here  
 $\Delta b_{\zeta}$ is the fluctuating (noise) part of $b_{\zeta}$. 
Equation~(\ref{subopt}) can then be put into the following form:
\beq
\underbrace{
\left(
\begin{array}{cc}
0 & 1
\end{array}
\right)
\mathbf{R}(\zeta)}_{\rm readout}
\,
\underbrace{
 \mathbf{R}(\pi/2-\varphi)\, \mathbf{S}(q)
 \mathbf{R}(-\varphi)}_{\rm interferometer}
\,
\underbrace{
 \mathbf{R}(\alpha) 
\left(
\begin{array}{c}
0 \\ 1
\end{array}
\right)}_{\rm input}
=0\,.
\label{suboptimality}
\eeq
\end{widetext}
In the Harms et al.\ scheme, the input quadratures are rotated 
(with FD angle $\alpha$), {\it  before entering the interferometer}, in such a way 
that, after being rotated again and ponderomotively squeezed by 
the interferometer opto-mechanical dynamics, the squeezed quadrature 
 is mapped into a frequency independent output
quadrature, which is detected. 
In the BC scheme a frequency independent squeezed state
enters the interferometer. Due to  rotation and ponderomotive
squeezing inside the interferometer, the squeezed quadrature 
is mapped into a FD output quadrature. We then apply a rotation to the 
field emerging {\it from the interferometer} to counteract this effect 
and bring the (image of the) input squeeze
quadrature back to a frequency independent quadrature and detect it. 

Finally, another interesting sub-optimal scheme can be obtained 
by imposing $\zeta=\alpha=-\varphi$. In this case the noise part 
of the output quadrature field (\ref{qfluc}) is
\beq
\left(
\begin{array}{cc}
0 & 1
\end{array}
\right)
\mathbf{R}(\pi/2)\, \mathbf{S}(q)\, \mathbf{S}(r)
\left(
\begin{array}{c}
\tilde{a}_1 \\
\tilde{a}_2
\end{array}
\right) = e^{-(r+q)}\tilde{a}_1
\eeq
which gives the lowest amount of noise (but does not guarantee a
maximal signal content).  Unfortunately, from Eq.~(\ref{sinhq}) 
we see that $\tan \zeta = -\tan\varphi$ is not 
of the form (\ref{KLMTVtanz}), and thus  not  realizable by KLMTV
filters.

\subsection{Sub-optimal schemes using $q\mbox{--}\varphi$ parametrization: the
low-power limit}
\label{subsec:smallq}

If the ponderomotive squeezing factor $q$ is small, the fully optimal
input-output scheme can be solved easily using the various
quadrature-rotation angles. As seen in Sec.~\ref{sec2}, 
a small $q$ can either arise from a low optical power, or 
from considering high frequencies 
($f \stackrel{>}{_\sim} 300\,{\rm Hz}$ for Advanced LIGO power), see
Eq.~(\ref{sinhq}). However, we shall still refer to this as the
 low-power limit. In this case, the output noise is
proportional to
\beq
\label{qfluclp}
\left(
\begin{array}{cc}
0 & 1
\end{array}
\right)
\mathbf{R}(\zeta + \pi/2 -2\,\varphi_{\mbox{\scriptsize low-power}} +\alpha)\,\mathbf{S}(r) 
\left(
\begin{array}{c}
\tilde{a}_1 \\
\tilde{a}_2
\end{array}
\right)\,,
\eeq
and the minimal noise is obtained whenever
\beq
\zeta+\alpha =2\,\varphi_{\mbox{\scriptsize low-power}}\,.
\eeq
By setting $\zeta$ equal to the maximal-signal quadrature [see Eq.~(\ref{zetavarphi})], 
\beq
\zeta_{\rm max}= -\frac{\pi}{4} +\varphi_{\mbox{\scriptsize low-power}}\,,
\eeq
we find the fully optimal readout scheme:
\beq
\label{fulloptlp}
(\zeta,\alpha)_{\rm opt} = 
\left(-\frac{\pi}{4}+\varphi_{\mbox{\scriptsize low-power}},
\frac{\pi}{4}+\varphi_{\mbox{\scriptsize low-power}}
\right)\,.
\eeq
Simple as it looks, this fully optimal scheme is {\it not} realizable by KLMTV filters because 
$\tan \zeta_{\rm opt}$ and $\tan \alpha_{\rm opt}$ given by 
Eq.~(\ref{fulloptlp}) are not of the form (\ref{KLMTVtanz}). 

We now compare the Harms et al.\ (H)
and BC schemes in the  small-$q$ regime. They can be written in terms of $(\zeta,\alpha)$ as
\bea
(\zeta,\alpha)_{\rm H} &=& (\zeta, 2\,\varphi_{\mbox{\scriptsize low-power}} -\zeta)\,, \nonumber \\
(\zeta,\alpha)_{\rm BC} &=& (2\,\varphi_{\mbox{\scriptsize low-power}}-\alpha , \alpha)\,.
\eea
The two schemes give the same noise output part $e^{-r}\,\tilde{a}_1$ 
[see Eq.~(\ref{qfluclp})], while for the signal power 
they yield [see Eqs.~(\ref{signalquadrature})--(\ref{eqS1})]
\bea
s_{\scriptscriptstyle H} &=& S_0 + S_1 \cos\left[2(\zeta-\zeta_{\rm max})\right] \nonumber
\\
&=& S_0 +S_1 \cos(2\,\zeta+\pi/2-2\,\varphi_{\mbox{\scriptsize low-power}})\,,
\eea
and 
\bea
s_{\scriptscriptstyle BC} &=& S_0 + S_1 \cos\left[2(\zeta-\zeta_{\rm max})\right] \nonumber
\\
&=& S_0 +S_1 \cos(-2\,\alpha+\pi/2 + 2\,\varphi_{\mbox{\scriptsize
    low-power}}) \nonumber \\
&=& S_0 +S_1 \cos(2\,\alpha -\pi/2 - 2\,\varphi_{\mbox{\scriptsize
    low-power}})\,.
\eea
This means, the two sub-optimal schemes have the same ideal performance
in the low-power regime and we can map one into the other by setting 
$\alpha \leftrightarrow \zeta + \pi/2$.

This equivalence can be understood more intuitively 
if we compare the dependence of the various readout quadratures (i.e.,
maximum-signal, Harms et al., and BC) 
on $\varphi_{\lp}$. 

The maximal-signal quadrature $\zeta_{\rm max}$
rotates as ${\rm const}+\varphi_{\lp}$. In the Harms et al.\ scheme,
the detected quadrature is constant, and therefore {\it lags} the
maximal-signal quadrature by ${\rm const} + \varphi_{\lp}$. In the BC 
scheme,  the detected quadrature rotates as $\zeta = {\rm
  const} + 2\varphi_{\lp}$, which {\it advances} the optimal quadrature by
${\rm const}+\varphi_{\lp}$. In this way, if one adjusts the constants
(by adjusting $\zeta$ in the Harms et al.\ scheme and $\alpha$ in the BC
scheme), the detected quadratures in the two schemes can be made to
lie symmetrically  on each side of the maximal-signal
quadrature. Since the detected signal power depends only on 
$\cos\left[2(\zeta-\zeta_{\rm max})\right]$ [see
  Eq.~(\ref{signalquadrature})],  
which is an even function of $(\zeta-\zeta_{\rm max})$, the two schemes must 
detect the same signal power and hence have the same sensitivity.

In Fig.~\ref{fig3}, we give examples of the BC and Harms et al.\ noise
curves for two SR interferometers, a broadband configuration (with
$\lambda = 2\pi\times 234.07\,{\rm Hz}$, $\epsilon =2 \pi\times 
  70.36\,{\rm Hz}$) and a
narrowband one (with $\lambda=2\pi\times600\,{\rm Hz}$ and
  $\epsilon=2\pi\times25\,$Hz). For both interferometers, we use
$I_c=840\,$kW, $m=40\,$kg.  Although the optical power used here is
not low by any practical standards, the two schemes for the broadband
configuration already agree quite well, under the correspondence $\alpha
\leftrightarrow \zeta + \pi/2$, for frequencies above
$\sim$200\,Hz. The two schemes are equivalent for the
narrowband configuration for almost all frequencies. The better
agreement in the narrowband configuration can be understood easily by
realizing that ponderomotive squeezing is weaker in this case, as
shown in the left panel of Fig.~\ref{Fig2}.

\begin{figure*}
\begin{center}
\vspace{1cm}
\includegraphics[width=0.9\textwidth]{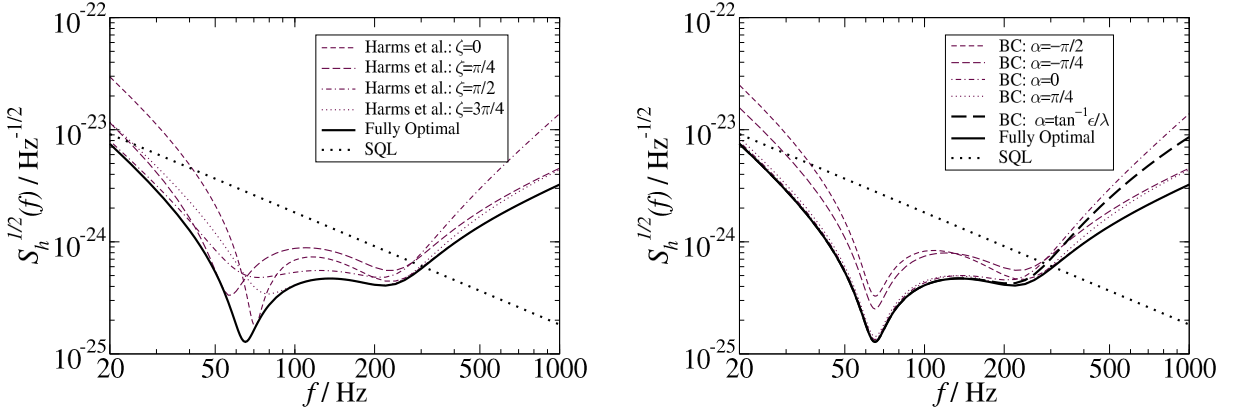}
\vspace{0.2cm}
\end{center}
\caption{Approaching the fully optimal input-output scheme by taking
  the envelope of Harms et al.\ and BC noise curves. We fix 
  $I_c=840\,$kW and $m=40\,$kg, and 
  $\lambda = 2\pi\times 234.07\,{\rm Hz}$, $\epsilon =2 \pi\times
  70.36\,{\rm Hz}$, and assume $e^{2r}=10$. In the left panel, 
  we plot the noise curves of Harms et
  al.\ schemes, with frequency independent readout phase $\zeta =$ 0
  (light short dash), $\pi/4$ (light long dash), $\pi/2$ (light
  dot dash) and $3\pi/4$ (light dot), and FD input squeeze 
  angle given by Eq.~(\ref{alphasubopt}). These curves cross each
  other near both the optical resonance and the optical-spring
  resonance. The fully optimal curve (dark continuous) is obtained by taking the lower
  envelope of the entire family of these curves. In the right
  panel, we plot the BC curves, with frequency independent input
  squeeze angle  $\alpha = -\pi/2$ 
  (light short dash), $-\pi/4$ (light long dash), $0$ (light
  dot dash) and $\pi/4$ (light dot), and FD readout phase given by
  Eq.~(\ref{zetasubopt}).  The lower envelope of these curves also gives the fully optimal
  noise curve (dark continuous, indentical to the one shown in left
  panel). The fact that these curves do not cross each other at low
  frequencies suggests that one member of this family is fully
  optimal in this band. Indeed,
  the BC curve optimized for $\Omega=0$  [dark dashed curve, with
  $\alpha=\arctan\epsilon/\lambda$, see Eq.~(\ref{BCLF})] does agree
  very well with the fully optimal curve for frequencies lower
  than $\sim 200$\,Hz. In both panels we also show the SQL line.
\label{fig4}
}
\end{figure*}

\subsection{The fully optimal scheme and the BC scheme at low frequencies}
\label{subsec:fullopt}
In this section, we consider the fully optimal scheme.
Analytical formulas of the fully optimal detection quadrature 
 has been obtained by Harms et al., but we provide an
alternative approach, yielding results in simpler form and more
related to the BC scheme.

It is straightforward to show that (as also done by Harms et al.\ and
reviewed in Appendix~\ref{appa}), 
fixing $\zeta$ and $r$, the $\alpha$ obtained from
Eq.~(\ref{alphasubopt})  
gives the (constrained) minimum noise. 
On the contrary, fixing $\alpha$ and $r$, 
the readout quadrature $\zeta$ obtained from Eq.~(\ref{zetasubopt})
{\it does not} give the constrained minimum. 
Instead, minimizing $S_h$ [Eq.~(\ref{noise_squeeze})] over $\zeta$
(with $\alpha$ fixed) 
requires a rather complicated readout phase, determined by one of the two roots of
\beq
\label{ultimatezeta}
{\cal F}_2\tan^2\zeta + {\cal F}_1\,\tan\zeta+{\cal F}_0=0\,,
\eeq
where
\begin{widetext}
\begin{subequations}
\bea
\label{F_2opt}
{\cal
  F}_2&=&\left[(C^{\alpha,-r}_{11})^2+(C^{\alpha,r}_{12})^2\right]\,\Re(D_1^*D_2)
  -\left(C^{\alpha,-r}_{11}\,C^{\alpha,-r}_{21}+C^{\alpha,r}_{12}\,C^{\alpha,r}_{22}\right)\,|D_1|^2\,, 
\\ 
\label{F_1opt} 
{\cal F}_1&=&\left((C^{\alpha,-r}_{11})^2+(C^{\alpha,r}_{12})^2\right)\,|D_2|^2
      -\left[(C^{\alpha,-r}_{21})^2+(C^{\alpha,r}_{22})^2\right]\,|D_1|^2\,,\\
{\cal
  F}_0&=&\left(C^{\alpha,-r}_{11}\,C^{\alpha,-r}_{21}+C^{\alpha,r}_{12}\,C^{\alpha,r}_{22}\right)\,|D_2|^2
-\left[(C^{\alpha,-r}_{21})^2+(C^{\alpha,r}_{22})^2\right]\,\Re(D_1^*D_2)\,.
\label{F_0opt}
\eea
\end{subequations}
\end{widetext}
Equations (\ref{ultimatezeta})--(\ref{F_0opt}), 
which we obtained independently~\cite{BCUnpub} from Harms et al., 
are equivalent to Eqs.~(28)--(30) 
of Harms et al.\ once we set $r$ to zero in
Eqs.~(\ref{F_2opt})--(\ref{F_0opt}).  

As said above, the fully optimal scheme, denoted by 
$(\alpha_{\rm opt}(\Omega), \zeta_{\rm opt}(\Omega))$, should satisfy the
sub-optimal condition (\ref{subopt}). As a consequence, 
 the noise spectrum of the fully optimal scheme is also given by
Eq.~(\ref{Shsubopt}), when $\zeta(\Omega)$ is replaced by
 $\zeta_{\rm opt}(\Omega)$. Therefore, $\zeta_{\rm opt}(\Omega)$ can be obtained by minimizing the
$S_h$ in Eq.~(\ref{Shsubopt}), which is 
given by the special case of Eqs.~(\ref{ultimatezeta})--(\ref{F_0opt})
with $r=0$ [or Eqs.~(28)--(30) of Ref.~\cite{Harms03}]; $\alpha_{\rm
  opt}(\Omega)$  can then be obtained from  Eq.~(\ref{alphasubopt}). 
It is evident from Eq.~(\ref{ultimatezeta}) that the fully optimal scheme cannot be realized by KLMTV
filters, except in special cases, e.g., for  conventional interferometers. 
As observed by Harms et al., the optimal noise spectrum can also be obtained graphically, by
plotting all the noise curves with different constant values of $\zeta$,
and then taking the lower envelope of all these curves, as seen in
the left panel of Fig.~\ref{fig4} (and Fig.~4 of
Ref.~\cite{Harms03}). The optimal $\zeta$ at 
each frequency is the one whose noise curve touches the envelope.

We now deduce the optimal scheme in another way. 
Again, since $(\alpha_{\rm opt}(\Omega), \zeta_{\rm opt}(\Omega))$ satisfy
Eq.~(\ref{subopt}), the  fully optimal noise spectral density
can also be obtained by taking the minimum among all BC noise spectral
densities with all possible $\alpha$ --- the minimum 
is achieved automatically in $\alpha_{\rm opt}(\Omega)$, and
for $\zeta_{\rm opt}(\Omega)$ it is given by Eq.~(\ref{zetasubopt}). Similarly, this
can be done graphically by taking the lower envelope
of BC noise curves with all possible $\alpha$, as shown in the right
panel of Fig.~\ref{fig4}. {}From the plot, it is
interesting to observe that, there are no crossings between
different BC noise curves at low frequencies (differently from
 the Harms et al.\
curves in the left panel), suggesting that one BC
curve might be nearly fully optimal at these frequencies!

More quantitatively, since the BC noise spectrum (\ref{Shlossless}) 
has a much simpler dependence on $\alpha$ (than the dependence of the 
Harms et al.\ noise spectrum on $\zeta$), it is much simpler to obtain
the optimal input squeeze angle $\alpha_{\rm opt}$ from this approach
[than to obtain  $\zeta_{\rm opt}$ from the approach starting with the
Harms et al.\ noise spectrum, see Eqs.~\eqref{ultimatezeta}--\eqref{F_0opt}]:
\beq
\label{alphaopt}
\tan 2\alpha_{\rm opt} = \frac{2\lambda\epsilon}{\lambda^2-\epsilon^2-\Omega^2}\,,
\eeq
and 
\begin{widetext}
\beq
S_h^{\rm opt} = \frac{e^{-2r}|M|^2 h_{\rm SQL}^2 }
{\displaystyle 2\,\epsilon\, \iota_c\, \Omega^2
  (\lambda^2+\epsilon^2+\Omega^2)\left[1+\sqrt{1-\left(\frac{2\lambda\Omega}{\lambda^2+\epsilon^2+\Omega^2}\right)^2}\right]} \,.
\label{Shopt}
\eeq
\end{widetext}
These simple explicit expressions of $S_h^{\rm opt}$ and $\alpha_{\rm 
opt}(\Omega)$ have not been previously obtained. 
The optimal readout phase $\zeta_{\rm opt}$ can be obtained from
Eq.~(\ref{zetasubopt}). {}From Eq.~(\ref{alphaopt}), we can see that
the fully optimal scheme cannot be achieved by
KLMTV filters. The only exception is when $\lambda=0$ 
(i.e., for a conventional interferometer). 
In this case we have $\alpha_{\rm opt} = 0$, and
the $\zeta_{\rm opt}(\Omega)$ is given by Eq.~(\ref{zetasubopt}) and it is reliazable 
by KLMTV filters. This is 
exactly the KLMTV squeezed-variational
scheme. 

Although
the form of $\alpha_{\rm opt}$ is not achievable by KLMTV
filters, we note that, at low frequencies (lower than the optical 
resonant frequency), the variation in $\alpha_{\rm opt}$ is mild. In
fact, by setting in the BC scheme
\beq
\label{BCLF}
\alpha 
= \alpha_{\rm opt}(\Omega=0) = \arctan \left(\frac{\epsilon}{\lambda}\right)\,,
\eeq
we obtain\,,
\bea
&&S_h^{\rm BC\,\,low\mbox{-}freq}  \nonumber \\
&=&  \frac{e^{-2r}|M|^2 h_{\rm SQL}^2}
{\displaystyle 2\,\epsilon\,\iota_c\,\Omega^2\left[2(\lambda^2+\epsilon^2+\Omega^2)-\frac{2\lambda^2\Omega^2}{\lambda^2+\epsilon^2}\right]}\,.
\eea
Taking the ratio between $S_h^{\rm BC\,\,low\mbox{-}freq}$ and
$S_h^{\rm opt}$, and expanding in $\Omega$, we have
\beq
\label{correction}
\frac{S_{h}^{\rm BC\,\,low\mbox{-}freq}}{S_h^{\rm opt}} = 1+
\left(\frac{\lambda\epsilon}{\lambda^2+\epsilon^2}\right)^2
\left(\frac{\Omega^2}{\lambda^2+\epsilon^2}\right)^2\,.
\eeq
The correction factor in Eq.~(\ref{correction}) is usually small at 
low frequencies.  For example, by maximizing over
either $\epsilon$ or $\lambda$, it is easy to show that 
\beq
\left(\frac{\lambda\epsilon}{\lambda^2+\epsilon^2}\right)^2
\left(\frac{\Omega^2}{\lambda^2+\epsilon^2}\right)^2
< \frac{27}{256}\left(\frac{\Omega}{\max \{ \lambda ,\epsilon\}}\right)^4
\eeq
at worst. The correction in the noise spectral density cannot exceed
$\sim 10\%$ (in power) for $\Omega \sim \max \{ \lambda ,\epsilon \} $. 
For substantially detuned configurations ($\lambda$ exceeding $\sim 200\,$Hz), 
this makes the BC scheme essentially fully optimal up to 
$\sim 200\,$Hz. This result is confirmed by  the right panel of 
Fig.~\ref{fig4}, in which $S_h^{\rm BC\;low\mbox{-}freq}$ is plotted
(dark dashed curve) in comparison with $S_h^{\rm opt}$ (dark
continuous curve).

\begin{table*}
\begin{scriptsize}
\centerline{
\begin{tabular}{cccrrrr|cc|cc|cc}
\multicolumn{7}{c|}{Interferometer Configuration} &
\multicolumn{2}{c|}{Filter I} &
\multicolumn{2}{c|}{Filter II} & 
\multicolumn{2}{c}{Performance} \\
\hline
\begin{tabular}{c} Input-Output \\ Scheme \end{tabular} &
\begin{tabular}{c} Mirror \\Type \end{tabular} &
$e^{-2r}$ &
\multicolumn{1}{c}{
$\displaystyle \frac{\epsilon}{2\pi\,{\rm Hz}}$}
& 
\multicolumn{1}{c}{
$\displaystyle \frac{\lambda}{2\pi\,{\rm Hz}}$}
&
\multicolumn{1}{c}{$\alpha$} & \multicolumn{1}{c|}{$\zeta$} & 
\multicolumn{1}{c}{$\displaystyle \frac{\Omega_{\rm res}^{\rm I}
  }{2\pi\,{\rm Hz}}$} &
\begin{tabular}{c} $T_i^{\rm I}$ \\ (ppm) \end{tabular} & 
\multicolumn{1}{c}{$\displaystyle \frac{\Omega_{\rm res}^{\rm II} }{2\pi\,{\rm Hz}}$} &
\begin{tabular}{c} $T_i^{\rm II}$ \\ (ppm) \end{tabular} & 
\begin{tabular}{c} SNR \\ 300\,Mpc \end{tabular} &
\begin{tabular}{c} Event Rate \\ Improvement \end{tabular}
\\
\hline
No Squeezing & Spherical & 1 & 70.4
& 234.1 &  & $-0.804$ &  &  &  &  & 5.44 &
$1.00$ \\  
No Filters &  &  0.1 & 561.8 & 55.2 & $1.522$ & $-0.040$ &
& & & &  6.73 & 1.89 \\
Harms et al.\ & &  0.1 & $280.4$ & $296.3$ & FD & $-0.381$ &
$296.0-285.4\,i$ & 
$717$ &
$-59.8-23.7\,i$ &
$59.6$
& $7.15$ ($7.81$) & $2.27$ ($2.96$)\\
BC & & $0.1$ & $157.4$ & $355.2$ & $-2.090$ &
FD & $352.0-161.0\,i$ & $404$ &
$-59.5-12.3\,i$ & $30.9$ & $6.77$ ($7.84$) & $1.93$ ($2.99$)\\
\hline
No Squeezing & 
\begin{tabular}{c} Mexican \\ Hat \end{tabular}& 1 & $14.4$ & $179.1$ &  & $-1.010$ & &
& & & $9.29$ & $1.00$ \\
No Filters & & $0.1$ & $275.3$ & $101.3$ & $1.395$ & $-0.131$ & 
& & & & $10.44$  & $1.42$
\\
Harms et al.\ & & $0.1$ & $106.5$ & $233.9$ & FD & $-0.518$ &
$227.2-115.4\,i$ & $290$ & $-73.2-18.0\,i$ & $45.1$ & $11.73$ ($15.08$) &
$2.01$ ($4.27$)\\
BC & & $0.1$ &$55.7$ & $240.5$ & $-2.179$ & FD &
$230.8-61.2\,i$ & $154$ & $-72.7-9.8\,i$ & $24.6$ & $10.45$ ($15.63$) &
$1.42$ ($4.76$)\\
\hline
\end{tabular}}
\end{scriptsize}
\caption{Optimizations of SR interferometers with (i) no squeezing, 
(ii) frequency independent squeezing and homodyne detection (``no filters'')
(iii) FD 
squeezing but frequency independent readout (the Harms et al.\ scheme), and (iv) frequency
  independent squeezing but FD readout (the BC scheme)
  for neutron-star--binary 
  inspirals, with quantum noise, seismic noise and thermoelastic
  noise (with spherical and Mexican-Hat mirrors) included. 
The only optical losses included are those from the 30-m
  optical  filters. 
The round-trip loss of each filter is set to be
  $20\,$ppm~\cite{Whitcomb}. Noise curves of configurations listed
  here are plotted in Fig.~\ref{BBopt}. For the Harms et al.\ and BC
  schemes, we also optimize the SNR when there is no optical
  losses, those SNRs and the corresponding event-rate improvements
  are quoted inside brackets.  
\label{NSNS}}
\end{table*}

\section{Applications to Advanced LIGO} 
\label{sec5}

In this section, we discuss the possibility of applying the above FD
techniques to Advanced LIGO interferometers.  
As shown by KLMTV~\cite{KLMTV00}, a major difficulty 
in making those techniques practical for advanced interferometers 
is the issue of  optical losses. Given a certain bandwidth and mirror quality 
(i.e., round-trip loss in the filter cavities), 
the shorter the filters, the higher their optical losses 
(see Table~\ref{tab:loss}). In fact, in order to achieve 
third-generation performance, optical filters
in the squeezed-variational scheme will have to be $\sim$ kilometer in
lengths. 
In Advanced LIGO, kilometer-scale filter cavities are not
practical and only short filters can fit into the corner-station building. 
A plausible length scale is $\sim 30\,$meters; and the realistic
round-trip loss is around 20\,ppm~\cite{Whitcomb}. 
With such short (and lossy)  filters, we shall assume most of the time
that filter losses will dominate and ignore 
internal interferometer losses [see Sec.~V in
  Ref.~\cite{BC2} for treatment of lossy SR interferometers]. We shall
only comment briefly on the effect of internal losses when discussing narrowband sources.
The noise spectrum with filter losses are
obtained by using the exact input-output relation of KLMTV filters
(Sec.~\ref{sec3.3}). 

In Secs.~\ref{subsec:bbopt}, ~\ref{subsec:nbopt} and
~\ref{subsec:wbopt}, respectively, 
we shall discuss the broadband configuration 
optimized for the detection of NS-NS binary inspiral waveforms, 
the narrowband configuration targeting GWs from specific accreting NS's and 
the wideband configuration that can be used to observe several kind of  sources.
[For an exhaustive discussion and summary of GW sources for advanced interferometers 
see, e.g., Ref.~\cite{CT}.]  

\begin{figure*}
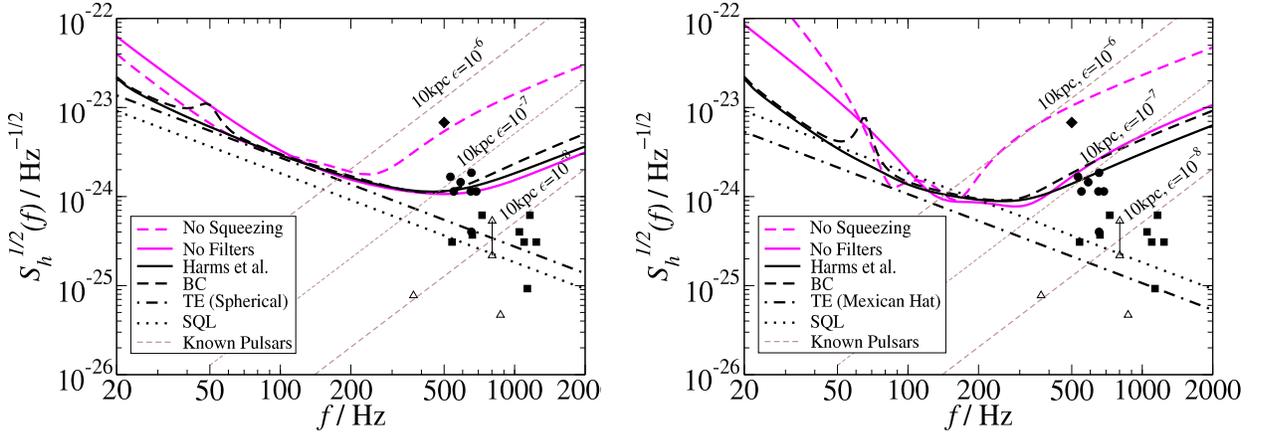

\vspace{1cm}
\begin{tabular}{ccc}
\includegraphics[width=0.45\textwidth]{fig6a.eps} &
\hspace{0.0\textwidth} &
\includegraphics[width=0.45\textwidth]{fig6b.eps}
\end{tabular}
\vspace{0.5cm}
\caption{Noise curves optimized for NS-NS 
binaries, for no squeezing (``no squeezing'', light dashed curves), 
frequency-independent squeezing and homodyne detection  (``no
filters'', light continuous  curves),  
the Harms et al.\ scheme (dark continuous curves) and  the BC scheme 
(dark dashed curves). Parameters of each configuration
are listed in Table~\ref{NSNS}.
Quantum noise, seismic noise and thermoelastic noise of
sapphire (also shown, in dash-dot curves) are included to give the
total noise curves. We have used predictions for the thermoelastic
noise of spherical mirrors (left panel) and that of 
Mexican-Hat mirrors~\cite{TE} 
(right panel). In addition, we have shown the characteristic strengths
of possible GWs from LMXBs (diamond, solid circles, solid squares and
open triangles) and known radio pulsars (thin dashed lines).
\label{BBopt}}
\end{figure*}

\subsection{Broadband configuration: NS-NS binary inspiral}
\label{subsec:bbopt}

Inspiral waves from compact binaries (NS-NS, NS-BH or BH-BH) are among
the most promising sources for Advanced LIGO. In this section, we
discuss the so-called broadband configuration obtained by maximizing the
signal-to-noise ratio for NS-NS inspiral waveforms, proportional to
\beq
\sqrt{\int_{f_c}^{+\infty} \frac{|\tilde{h}(f)|^2}{S_h(f)} df}\,,
\eeq
where
\beq
|\tilde{h}(f)| = A f^{-7/6} \Theta(f_{\rm ISCO} - f)
\eeq
is the frequency-domain amplitude of the leading (Newtonian) order 
inspiral signal 
in the stationary-phase approximation. The cutoff frequency is chosen to be 
$f_{\rm ISCO}$, the GW frequency corresponding to the Innermost Stable 
Circular Orbit (ISCO) of a Schwarzchild black hole with mass
$2M_{\rm NS} = 2\times 1.4\,M_{\odot}=2.8\,M_{\odot}$,  
which is equal to $1570\,$Hz. In the optimization, we have also
included the seismic noise 
\beq
\sqrt{S_{h}^{\rm seis}} = 5.3 \times 10^{-21} 
\left(\frac{10\,{\rm Hz}}{f}\right)^{9.2}\,\frac{1}{\sqrt{\rm Hz}}\,,
\eeq
and the thermoelastic noise of sapphire mirrors with spherical
surfaces, as in the baseline design, 
\beq
\sqrt{S_{h}^{\rm TE}} = 2.7 \times  10^{-24} \left(\frac{100\,{\rm
    Hz}}{f}\right)\,\frac{1}{\sqrt{\rm Hz}}
\eeq 
as well as when the so-called Mexican-Hat mirrors are used, which are designed
to reduce this noise~\cite{TE},
\beq
\sqrt{S_{h}^{\rm TE\,MH}} = 1.1\times 10^{-24} \left(\frac{100\,{\rm
    Hz}}{f}\right)\,\frac{1}{\sqrt{\rm Hz}}\,.
\eeq 
In Table~\ref{NSNS} we list the values of $\lambda$, $\epsilon$,
$\alpha$ (frequency independent squeezing angle for BC scheme), 
$\zeta$ (frequency independent detection quadrature for Harms et al.'s scheme), 
obtained by optimizing the SNR of  NS-NS binary inspirals at 300\,Mpc, 
and the corresponding optimal SNR. 
We assume $I_c=840\,$kW, $m=40\,$kg, and $e^{-2r}=10$ and 
did the optimization for (i) non-squeezed SR interferometers, (ii) SR
interferometers with frequency independent squeezing and homodyne detection
(``no filters''),\footnote{Corbitt, Mavalvala and Whitcomb~\cite{CMW} 
are currently investigating this scheme.
}  (iii) squeezed SR interferometers with the Harms et al.\ scheme 
(FD input squeezing + ordinary homodyne detection) and (iv) squeezed 
SR interferometers with the BC scheme (ordinary squeezing + FD homodyne detection).  
In the Table we also give the improvements in the predicted 
event rate with respect to non-squeezed configurations, as the {\it cube} of the improvements in SNR at a fixed distance.

As we can read from Table~\ref{NSNS}, with frequency independent
squeezing (i.e., no filters), 
it is already possible to improve the NS-NS event rate by a
significant amount, $89\%$ (spherical mirror) or $42\%$ (MH
mirror). The Harms et al.\ scheme provides further improvement in the event 
rate with respect to the no-filter case, by $20\%$ (spherical mirror) or  
$42\%$ (MH mirror). The BC scheme, however, being more susceptible to
filter optical losses, does not yield as good a performance. 
In order to appreciate how much the filter optical losses affect the 
sensitivity, we have also optimized the SNR for the FD schemes {\it without}
including  
filter optical losses (but with thermal and seismic noises included), 
the results are quoted in brackets in Table~\ref{NSNS}. FD schemes
without losses can outperform frequency independent squeezing
significantly. For example, the ideal BC scheme can have $58\%$
(spherical mirror) or $235\%$ (MH mirror) more event 
rates than the no-filter case. [In this case the 
the BC scheme can also provide slightly 
higher event rates than the Harms et al.\ scheme, 
due to better sensitivity at low frequencies (but mostly still masked
by the thermal noise), by 
$1\%$ (spherical mirror) or $11\%$ (MH mirror).] 

Noise curves corresponding to the optical 
configurations listed in Table~\ref{NSNS} are plotted in
Fig.~\ref{BBopt}. We notice  that due to optical losses 
the BC noise spectral densities have a
peak around the optical-spring resonant frequency.
The noise spectrum of the ``no filters'' scheme 
(squeezing with frequency-independent input-output optics) is
comparable to the Harms et al.\ and BC schemes at high frequencies,
but becomes worse at low frequencies. These ``no-filter'' curves are quite
similar to the wideband noise curves proposed by Corbitt and 
Mavalvala~\cite{CM}, especially in the case of spherical mirrors. 
 
The squeezing noise curves optimized for NS-NS binaries also have better
high-frequency sensitivity than non-squeezed configurations, 
although they were not optimized specifically for high frequencies.  
{}From Fig.~\ref{BBopt}, we see that for frequencies higher than 
$\sim$500\,Hz (spherical mirrors) or $\sim$300\,Hz (Mexican-Hat mirrors), 
the squeezed configurations are $\sim 5$ -- $8$ (spherical mirrors) or $\sim 3$ -- $5$
(Mexican-Hat mirrors) as sensitive  (in amplitude) as the non-squeezed
configurations. [The Mexican-Hat mirrors produce lower thermoelastic
  noise, so the noise spectral densities are better optimized at low
  frequencies, reducing the bandwidth. This is why in this case the noise curves
  optimized for NS-NS binaries yield worse high-frequency
  sensitivity than those with spherical mirrors.]

\begin{table*}
\begin{footnotesize}
\begin{tabular}{|c|c|c|c|c|c|c|c|c|}
\hline
\multicolumn{1}{|c|}{} &
\multicolumn{3}{c|}{GW parameters} &
\multicolumn{5}{c|}{
$\sqrt{S_{h_c}^{\mbox{\tiny 20-day}}/S_h(f_{\mbox{\tiny GW}})}$} \\
\hline
\multicolumn{1}{|c|}{}
& 
\multicolumn{1}{c|}{$f_{\rm GW}$}
& 
\multicolumn{1}{c|}{ $h_c$ } &
\multicolumn{1}{c|}{ $\sqrt{S_{h_c}^{\mbox{\tiny 20-day} }}$}
& 
\multicolumn{1}{c|}{No} &
\multicolumn{2}{c|}{5\,dB} & 
\multicolumn{2}{c|}{10\,dB}
\\
& (Hz) & 
$(10^{-27})$ & $(10^{-24}/\sqrt{\rm Hz})$ &
Squeezing  & Harms et al.\ & BC  & Harms
et al.\ & 
 BC \\
\hline
\input LMXBSNR.tex
\hline
\end{tabular}
\end{footnotesize}
\caption{Frequency, characteristic amplitude $h_c$, 
and characteristic strengths ($\sqrt{S_{h_c}^{20\mbox{-}\rm day}}$, 
20-day coherent integration
with $1\%$ false alarm) of possible GWs from several known
LMXBs [including Z sources (the first 8 sources), Type-I bursters (the
  next 8 sources) and accreting millisecond
pulsars (the last 3 sources)]~\cite{LB02}, and sensitivities
achievable by non-squeezed, Harms et al.\ and BC schemes. 
Both 5\,dB ($e^{-2r}=0.316$) and 10\,dB ($e^{-2r}=0.1$) squeezing are
considered. 
Sensitivity is measured by taking the ratio between the characteristic
strength and square root of the noise spectral density at the
predicted GW frequency. Bold face is used for ratios larger than
unity, in which case the GW is detectable. In this Table we use 
  the baseline assumption (a1) (i.e., $f_s=f_d$ for Z sources, and
  mass-quadrupole emission). To obtain the predictions for other mechanisms 
of GW emission and determination of the spin frequency, see 
  Table~\ref{tab:conversion}. Note that $\sqrt{S_{h_c}}$ should
  convert by the same factor as $h_c$. 
For the source SAX\,J1808.4$-$3658 in particular, we have also shown
in parenthesis values that correspond to a  4-month coherent
integration. 
 \label{tab:LMXBsnr}}
\end{table*}

In Fig.~\ref{BBopt}, we also plot (in light thin dashed lines) 
the {\it characteristic GW strengths} from  known radio pulsars. 
Following the notation of Cutler and
Thorne~\cite{CT}, the characteristic strength $S_{h_{c}}$ is defined as  
the maximum allowed noise spectral density $S_h(f_{\rm source})$ (at and near the
source frequency $f_{\rm source}$) such that the source is  detectable.  Note that $S_{h_{c}}$ will in general depend on the data analysis technique and statistical criteria used, e.g., integration time, confidence level, etc.; sometimes it is also obtained by averaging over unknown source parameters, such as the spin orientation of pulsars [see App.~\ref{app:LMXB} for more details]. 
Here for known radio pulsars at 10\,kpc distance, with ellipticity $\epsilon=10^{-6}$, $10^{-7}$ and
$10^{-8}$,  we have been assuming 1$\%$ of false-alarm probability in a 
coherent search of $10^7$\,s of data (coherent search for such a long time 
can only be done for pulsars whose sky positions and phase evolutions are known~\cite{BCCS,BC}).

{}From Fig.~\ref{BBopt} we see that the NS-NS optimized noise spectra  for spherical mirrors can
detect known pulsars at 10\,kpc with $\epsilon \stackrel{>}{_\sim} 10^{-7}$ if
the GW frequency is higher than 500\,Hz, while those for MH mirrors
can detect $\epsilon \stackrel{>}{_\sim} 2\times 10^{-7}$ if GW frequency is
higher than 300\,Hz. 

We have also shown in Fig.~\ref{BBopt} the frequencies and the
estimated characteristic GW strengths from LMXBs (Sco X-1 in
diamond, the Z sources in solid dots, Type-I bursters in solid
squares, and accreting millisecond pulsars in open triangles). 
We shall explain those sources in more detail in the next section and in App.~\ref{app:LMXB}. 
All the squeezed-input configurations are able to detect Sco X-1 
with large margins, while configurations with spherical mirrors might also 
be able to detect the group of six Z sources near 600\,Hz.

\subsection{Narrowband configuration: LMXB}
\label{subsec:nbopt}

Low-Mass X-ray Binaries (LMXBs) are systems formed by a neutron star
and a low-mass stellar companion, from which the neutron star accrets
material. Observations of LMXBs 
have provided evidence of a NS spin-frequency ``locking'' in 
the range $260\,{\rm Hz} < f_s < 600\, {\rm Hz}$ 
(much lower than the breaking frequency of $\sim 1.5$\,kHz~\cite{UBC}). 
These systems are rather old and believed to have been spun up
by accretion torque. Thus, to explain the locking it has been 
conjectured that accretion torque could be balanced by angular-momentum 
loss due to GW emission~\cite{old,B98,AKS}. In Table~\ref{tab:LMXBsnr}, we
list a number of LMXBs that are promising GW sources: the first group
contains the so-called Z sources, the second group the Type-I
bursters, and the third group accreting millisecond pulsars (all
data are taken from Refs.~\cite{B98,UBC,LB02}). 

The spin frequency of the NS in these LMXBs is not
unambiguously determined, except for accreting millisecond pulsars,
whose X-ray fluxes pulsate at their spin frequencies, i.e. $f_P=f_s$. 
For Type-I
bursters, the spin frequency can be inferred from the millisecond
oscillations in their X-ray fluxes observed after bursts ($f_B$) and from the
kHz QPO difference frequency ($f_d$). However, for different 
sources, it has been observed that either $f_d=f_B$ or $f_d= f_B/2$, and it is not firm yet
whether $f_s$ should be equal to $f_B$ or $f_d$. Recently, X-ray
bursts have been observed~\cite{SAX} from the source  SAX\,J1808.4--3658 (an
accreting millisecond pulsar, with spin frequency known from 
$f_s=f_P$~\cite{CM98}), and X-ray flux after the bursts is observed to oscillate
at the spin frequency (i.e.,  $f_B=f_P=f_s$). Moreover, for this source
the kHz QPO difference frequency is observed to be 
half this value: $f_d=f_P/2$. This might favor the argument that
$f_s=f_B$ for all Type-I bursters, as assumed by Refs.~\cite{B98,UBC,LB02} and
used in Table~\ref{tab:LMXBsnr} [henceforth we shall always adopt this
  assumption].  For Z sources, only kHz QPOs have been
observed; this makes it difficult to determine the NS spin frequency:
it could be either (a) $f_s=f_d$ or (b) $f_s=2 f_d$ [note
that for different Type-I bursters either (a) or (b) could
be true].

\begin{table}
\vspace{0.5cm}
\begin{tabular}{l|cc}
\hline
 & a ($f_s=f_d$) & b ($f_s=2 f_d$) \\
\hline
1 (MQ)
& $\left(f^{(a1)}_{\rm GW},h_c^{(a1)}\right)$ &
$\left(2f^{(a1)}_{\rm GW},\sqrt{\frac{1}{2}}h_c^{(a1)}\right)$ \\
2 (CQ)
& $\left(\frac{2}{3}f^{(a1)}_{\rm GW}, \sqrt{\frac{3}{2}}h_c^{(a1)}\right)$ &
$\left(\frac{4}{3}f^{(a1)}_{\rm GW},\sqrt{\frac{3}{4}}h_c^{(a1)}\right)$
 \\
\hline
\end{tabular}
\caption{Conversion of predicted GW frequencies and characteristic GW
  amplitudes $(f_{\rm GW},h_c)$ from LMXBs between different
  assumptions on spin frequency and GW emission mechanism. In
  particular, our baseline assumption, (a1) for Z sources and (1) for
  Type-I bursters and accreting millisecond pulsars, has been used by
  Refs.~\cite{LB02,UBC,CT} to give numerical estimates for GW
  frequency and characteristic amplitudes/strengths. [For Type-I
  bursters and accreting millisecond pulsars, the conversion from (1)
  to (2) follows the rule from (a1) to (a2) in the table.]
\label{tab:conversion}} 
\end{table}

\begin{table*}
\begin{scriptsize}
\centerline{
\begin{tabular}{crcccc|rc|rc|ccc}
\multicolumn{6}{c|}{Interferometer Configuration} &
\multicolumn{2}{c|}{Filter I} &
\multicolumn{2}{c|}{Filter II (unapplied)} & 
\multicolumn{3}{c}{Performance} \\
\hline
Scheme &
$e^{-2r}$ &
\multicolumn{1}{c}{
$\displaystyle \frac{\epsilon}{2\pi\,{\rm Hz}}$}
& 
\multicolumn{1}{c}{
$\displaystyle \frac{\lambda}{2\pi\,{\rm Hz}}$}
&
\multicolumn{1}{c}{$\alpha$} & \multicolumn{1}{c|}{$\zeta$} & 
\multicolumn{1}{c}{$\displaystyle \frac{\Omega_{\rm res}^{\rm I}
  }{2\pi\,{\rm Hz}}$} &
\begin{tabular}{c} $T_i^{\rm I}$ \\ (ppm) \end{tabular} & 
\multicolumn{1}{c}{$\displaystyle \frac{\Omega_{\rm res}^{\rm II} }{2\pi\,{\rm Hz}}$} &
\begin{tabular}{c} $T_i^{\rm II}$ \\ (ppm) \end{tabular} & 
\begin{tabular}{c} $\sqrt{S_h}/(10^{-24}/\sqrt{{\rm Hz}})$ \\
at 600\,Hz \end{tabular} &
\begin{tabular}{c} BW \\ (Hz) \end{tabular} &
\begin{tabular}{c} NS-NS \@ 300\,Mpc \\ Spherical/MH \end{tabular}
\\
\hline
No Squeezing & $1$ (0dB) & $25$ & $601.4$ &   & $-0.748$ & & & & & $0.89$
& $41$ & $2.93/4.43$ \\
\hline
Harms et al.\ &  $0.316$ (5dB)  
& $60$ & $601.2$ & FD & $-0.749$ & 
$599.7-60.3\,i$ & 152 & $-43.0-2.2\,i$ & 5.6 & $0.94$ 
& $152$ & 4.59/6.70 \\
BC & & $60$ & $601.3$ & $0.806$ & FD & 
$599.8-60.3\,i$ & 152 & $-43.0-2.3\,i$ & 5.8 & $0.95$
& $153$ & 4.59/6.64\\
\hline
Harms et al.\ & $0.1$ (10dB)
& $100$ & $597.9$ & FD & $-0.722$ &
$596.4-100.5\,i$ & 253 & $-44.1-3.8\,i$ & 9.7 & $0.88$ &
$356$ & 6.03/8.15\\
BC & 
& $100$ & $598.3$ & $0.843$ & FD &
$596.7-100.5\,i$ & 253 & $-44.1-3.9\,i$ & 9.7 & $0.89$ &
$361$ & 5.98/7.96\\
\hline
\end{tabular}}
\end{scriptsize}
\caption{Optimization of SR interferometers with (i) no squeezing, (ii) FD 
squeezing but frequency-independent readout (the Harms et al.\ scheme), and (iii) 
frequency-independent squeezing but FD readout (the BC scheme)
  for narrowband sources around 600\,Hz. We have considered both
  5\,dB ($e^{-2r}=0.316$) and 10\,dB ($e^{-2r}=0.1$) squeezing.  
 In both of the FD schemes, filter II, 
which has impractically high finesse, does not affect high frequency performance, and is not applied.
  The 600\,Hz sensitivity,
  bandwidth, and SNR for NS-NS binaries at 300\,Mpc are given as
  performance indices. Here bandwidth is defined as the difference in
  the two frequencies at which $\sqrt{S_h(f)} = \sqrt{2 S_h(600\,{\rm Hz})}$. 
Noise curves of configurations listed
  here are plotted in Fig.~\ref{NBopt}.
\label{NBtab}}
\end{table*}

\begin{figure}
\vspace{0.5cm}
\centerline{\includegraphics[width=0.45\textwidth]{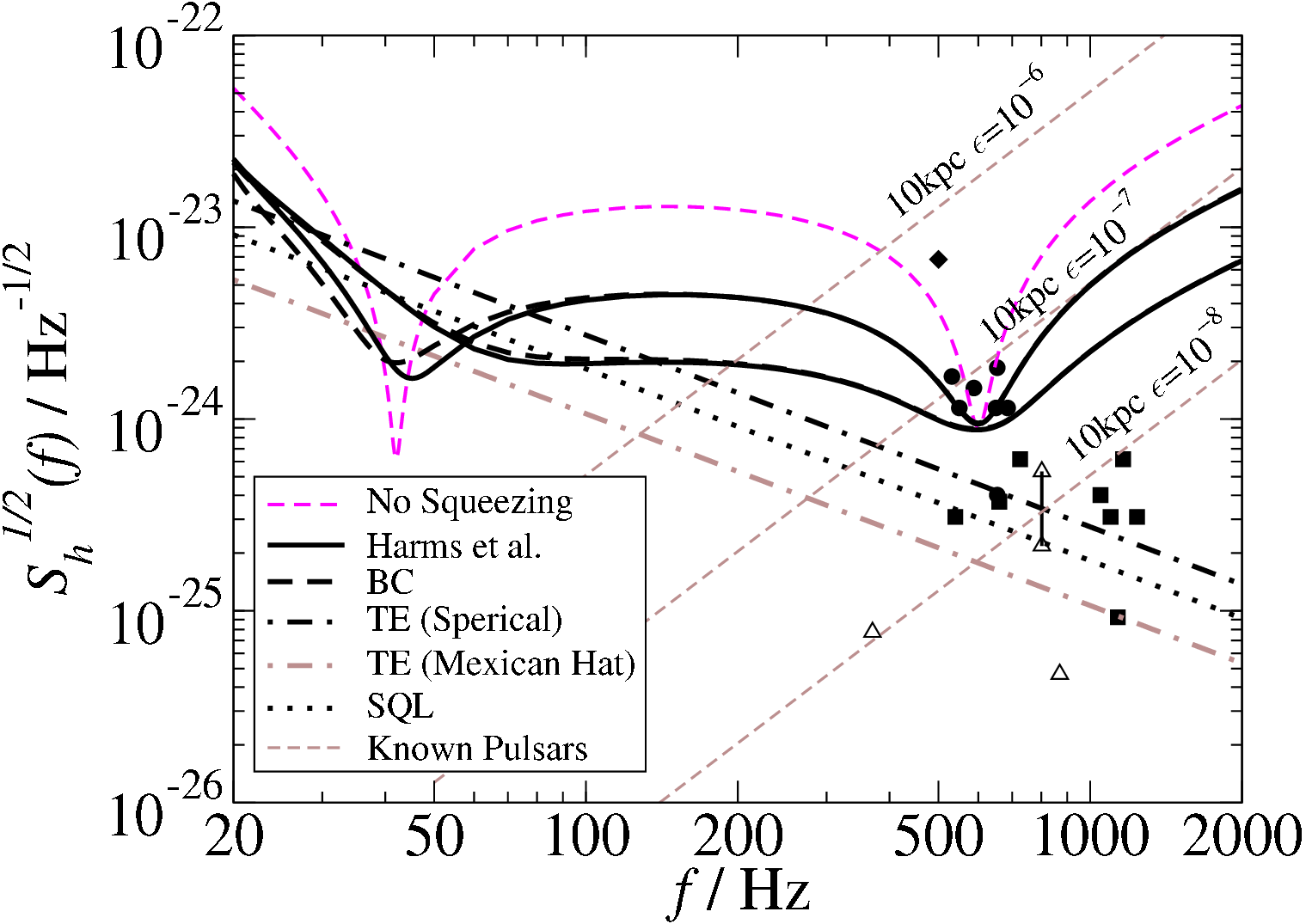}} 
\caption{Noise curves of non-squeezed (light dashed curve), 
Harms et al.\ (dark continuous curves) and BC (dark dashed curves)
configurations optimized for narrowband sources, for 5\,dB and
10\,dB squeezing. We apply only one filter, that is the one with resonant frequency 
near the free optical resonant frequency of the SR interferometer, or filter I 
(see Table~\ref{tab:LMXBsnr}).  The interferometer noise curves contain only
quantum noise but include filter losses. The thermoelastic noise of
spherical and MH mirrors, and the SQL are plotted for comparison.
We also show the frequencies and characteristic strengths 
(20-day coherent integration, $1\%$ false alarm~\cite{BCCS,BC})
of possible GWs from LMXBs~\cite{B98} [using the baseline assumption (a1),
  namely $f_s=f_d$ and mass-quadrupole emission]:  Z sources in solid
circles (Sco X-1 in diamond), Type-I X-ray bursters in solid squares 
and accreting millisecond pulsars in open triangles. 
For the accreting millisecond pulsar SAX\,J1808.4--3658, 
for which the orbital parameters and GW phase evolutions are known~\cite{SAX},  
we show in another open triangle (linked to the 20-day one with a
vertical segment of solid line) the characteristic strength assuming a 
4-month integration. \label{NBopt}}

\vspace{1cm}
\centerline{\includegraphics[width=0.45\textwidth]{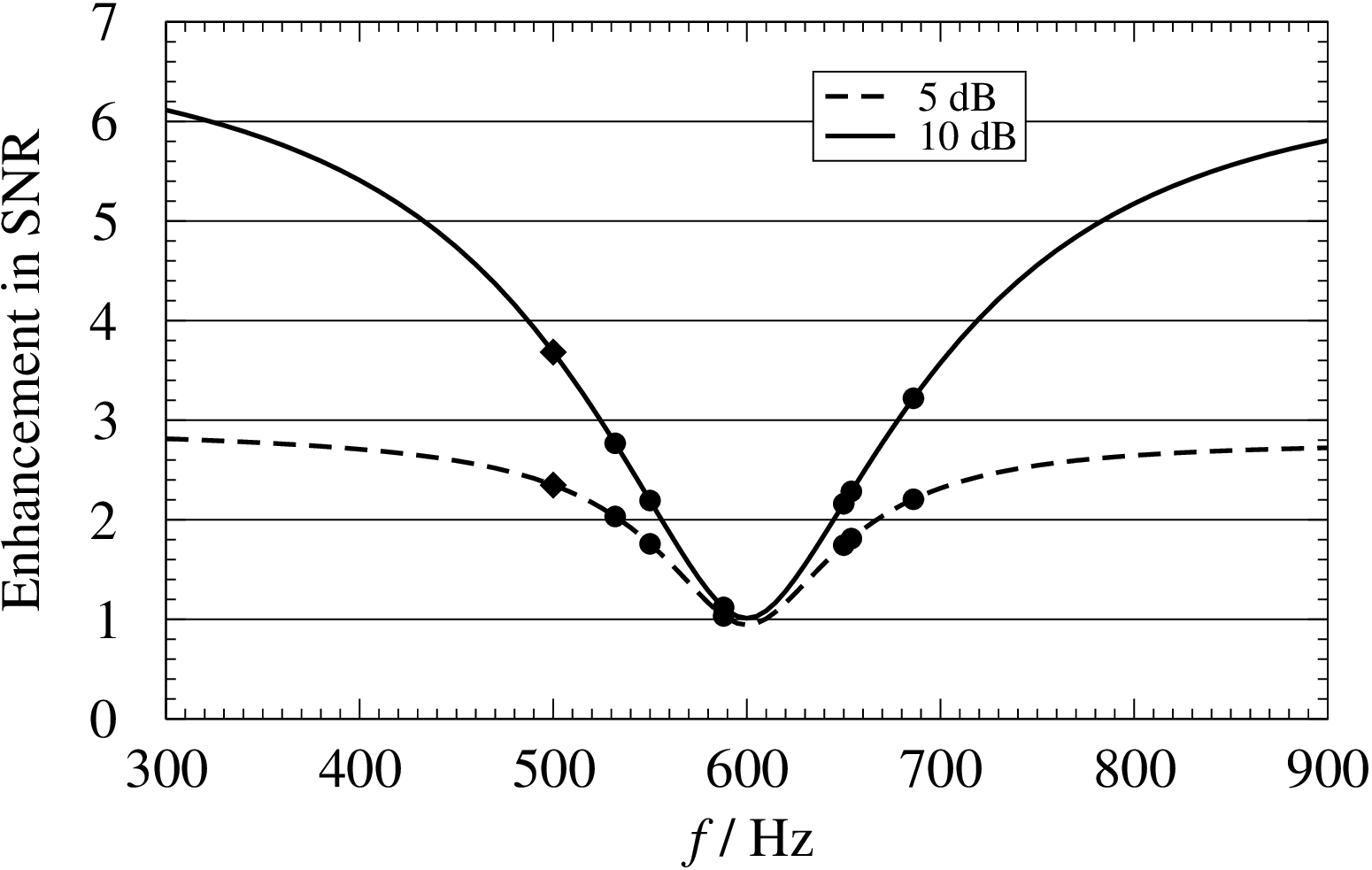} 
} \caption{Relative increase in SNR for LMXB sources around 600\,Hz, 
with 5\,dB ($e^{-2r}=0.316$, dashed curve) and 10\,dB ($e^{-2r}=0.1$, 
continuous curve) squeezing. Since the Harms et al.\ and BC schemes
are extremely close to each other only one curve is shown for each
squeeze factor. The various detectable LMXBs [under the baseline
  assumption] listed in
Table~\ref{tab:LMXBsnr} are also shown (Sco\,X$-$1  in solid diamond,
the rest in solid  
circles) .  \label{NBinc}}
\end{figure}

Moreover, two plausible physical mechanisms for GW emission 
from accreting NS's have been proposed: (1) mass quadrupole radiation from 
deformed NS crusts ($f_{\rm GW} = 2 f_s$)~\cite{old,B98}; and (2) 
current quadrupole radiation from unstable (with respect to
gravitational radiation) pulsation  
modes (r-modes) in NS cores ($f_{\rm GW} = 4 f_{s}/3$)~\cite{rmode,OLCSVA,AKS}. 
Suppose one of the two emission mechanisms to dominate, then along with  uncertainties 
in spin frequencies, we have four possibilities for Z sources, 
(a1), (a2), (b1) and (b2); [and two possibilities for accreting 
millisecond pulsars and Type-I bursters, (1) and (2)]. 
In the following, we consider (a1)  for Z sources and (1) for 
accreting millisecond pulsars and Type-I bursters our {\it baseline assumption}, as done in 
Refs.~\cite{B98,UBC,LB02,CT}, and comment on what happens if the
other options turn out to be true. 

In the second column of Table~\ref{tab:LMXBsnr}, we list GW
frequencies obtained from the baseline assumption; GW frequencies
based on other assumptions can be obtained from the (a1) or (1) value
by using Table~\ref{tab:conversion}.  The {\it characteristic GW
amplitude} $h_c$ from LMXBs has been estimated~\cite{UCB,OLCSVA} by
assuming a balance between GW angular momentum loss and accretion
torque, with the latter estimated from X-ray flux, and by subsequent
averaging over the (unknown) spin orientation [see
Appendix~\ref{app:LMXB_a} for a detailed explanation on the averaging
process and the associated uncertainties].  However, the value of
$h_c$ can also be different due to the various assumptions on spin
frequency and GW emission mechanism we can make.  Values listed in the third
column of Table~\ref{tab:LMXBsnr} has been obtained in
Refs.~\cite{B98,UBC,LB02} using the baseline assumption; conversions
from (a1) or (1) to the other assumptions can be made easily using
Eq.~(8) of Ref.~\cite{UCB} and Eqs.~(4.4)--(4.6) of
Ref.~\cite{OLCSVA}, and are given in Table~\ref{tab:conversion}. 
By assuming $1\%$ false-alarm probability and 20-day coherent
integration time [due to unknown orbital motion and frequency drifts
caused by fluctuations in the mass accretion rate] $S_{h_c}$ can be
obtained from $h_c$ (listed on the fourth column of
Table~\ref{tab:LMXBsnr}, see App.~\ref{app:LMXB_b} for details), note
that for the different assumptions $\sqrt{S_{h_c}}$ changes by the
same factor as $h_c$.  For the accreting millisecond pulsar
SAX\,J1808.4--3658, for which the orbital motion is known~\cite{CM98},
assuming that GW frequency evolution can be obtained, we also show (in
parenthesis) the characteristic strength obtained with a 4-month
integration.

It is important to realize that there are still uncertainties as to
whether a particular source will be detectable, even if the noise
curve is below $S_{h_c}$ --- as explained in
Appendix~\ref{app:LMXB}. However, the main aim of this paper is to discuss
interferometer configurations, rather than the data analysis of
narrowband sources, so we shall use $S_{h_c}$, as done by
Cutler and Thorne~\cite{CT} despite the subtleties, as a playground to
compare sensitivities of different interferometer/filter
configurations. Conclusions drawn in our discussions on whether these
sources will be detectable should definitely be refined by more
rigorous investigations. 

In Fig.~\ref{NBopt} we plot the noise curves
obtained for a non-squeezed 
SR interferometer and  for squeezed 
SR interferometers with the Harms et al.\ and BC schemes by optimizing
their sensitivities in a narrow band around 600\,Hz. 
Peak sensitivities and bandwidths are adjusted to incorporate the 
signal strengths of a group of 7 Z sources (including Sco\,X-1). The
baseline assumption is used in obtaining $f_{\rm GW}$ and $S_{h_c}$
for these sources.

\begin{figure*}
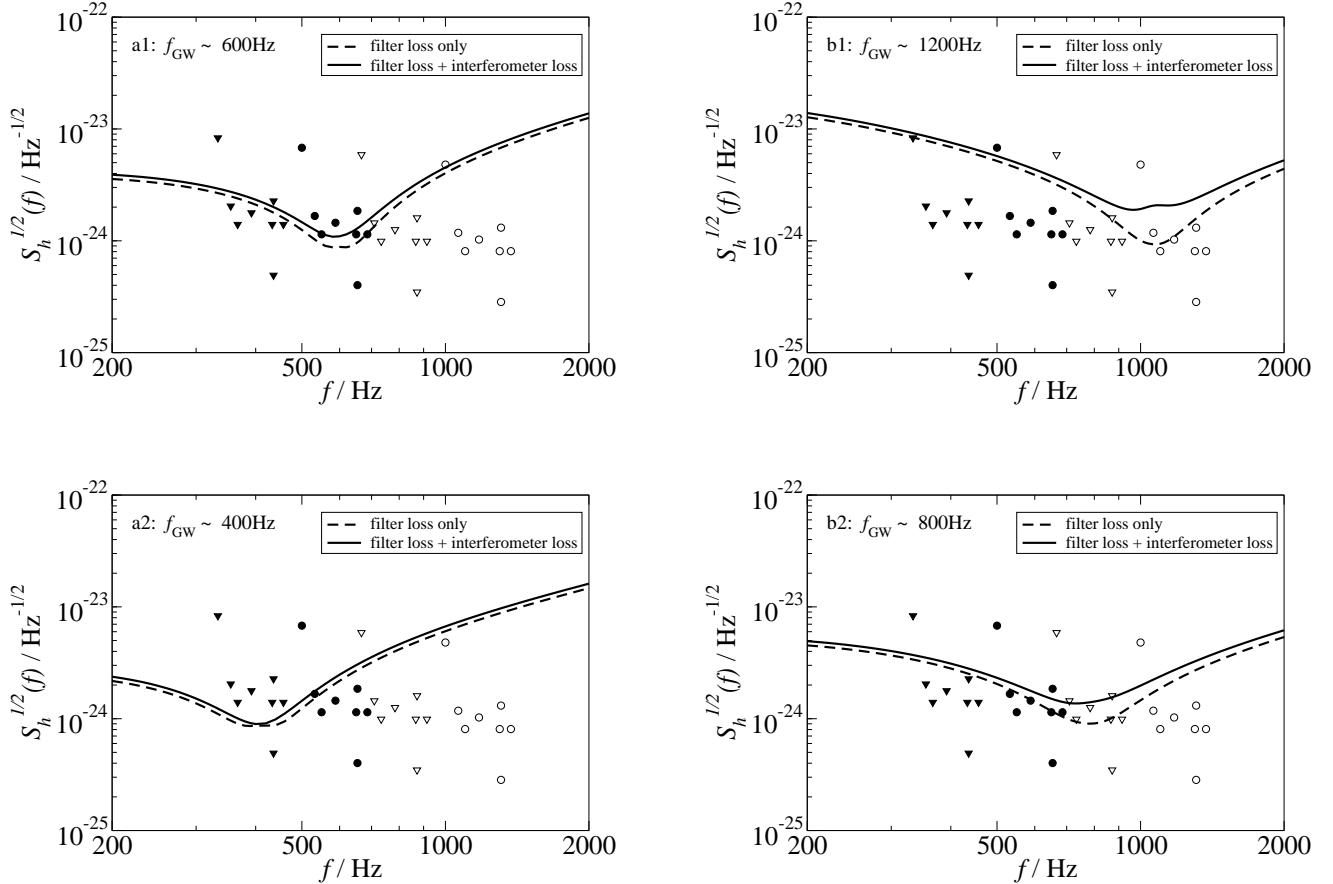

\vspace{0.5cm}
\begin{tabular}{ccc}
\includegraphics[width=0.45\textwidth]{fig9_a1.eps} &
\hspace{0.05\textwidth} &
\includegraphics[width=0.45\textwidth]{fig9_b1.eps} \\
\vspace{0.5cm}
\\
\includegraphics[width=0.45\textwidth]{fig9_a2.eps} &
\hspace{0.05\textwidth}  &
\includegraphics[width=0.45\textwidth]{fig9_b2.eps} \\
\end{tabular}
\vspace{0.5cm}
\caption{
Consequences of spin-frequency and emission-mechanism
  uncertainties on the detection of LMXBs (Z sources) using narrowband
  configurations. We plot the GW
  frequency and characteristic GW strength (for 20-day coherent
  integration), under approximations (a1) (solid circles with center
  frequency around 600\,Hz), (b1) (open
  circles, with center frequency around 1200\,Hz), (a2) (solid
  triangles, with center frequency around 400\,Hz) and (b2) (open
  triangles, with center frequency around 800\,Hz), along with Harms
  et al.\ (equivalent to BC) noise curves (with 10\,dB squeezing) 
tuned to those
  frequencies, with (solid curves) and without (dashed curves) internal losses of the
  interferometer. We assume an ITM power transmissivity of $0.033$,
  SR-cavity round-trip loss of $1\%$  and photodetection loss of $2\%$; $I_c=840$\,kW and $m=40$\,kg. The rest of the parameters are listed in
  Table~\ref{tab:morenb}. 
\label{fig:assumptions}}
\end{figure*}

{
\begin{table}
\begin{tabular}{crrrc}
\multicolumn{1}{c}{assumption} & 
\multicolumn{1}{c}{$\displaystyle\frac{f_{\rm central}}{\rm Hz}$} & 
\multicolumn{1}{c}{$\displaystyle\frac{\epsilon}{2\pi{\rm Hz}}$}  & 
\multicolumn{1}{c}{$\displaystyle\frac{\lambda}{2\pi{\rm Hz}}$}
  & 
\multicolumn{1}{c}{$\zeta$}  \vspace{0.1cm}
\\ \hline
(a1) & 600 & 30  & 600.4 & $-0.756$ \\
(b1) & 1200 & 90 & 1057.3 & $-0.065$ \\
(a2) & 400 & 25 & 412.6 & $-0.749$ \\
(b2) & 800 & 90 & 769.0 & $-0.306$\\
\hline
\end{tabular}
\caption{Parameters of narrowband configurations tuned to LMXB sources
 when different assumptions on spin frequency and GW emission
 mechanism are adopted. Noise curves of these configurations, 
with and without interferometer
 losses included, are shown in Fig.~\ref{fig:assumptions}, and compared
 to the corresponding GW characteristic strengths. \label{tab:morenb}}
\end{table}
}

For the non-squeezed interferometer, we obtain a
noise curve similar to the ``narrowband'' curve in 
Fig.~1 of Cutler and Thorne~\cite{CT}, provided originally by
Ken Strain. For squeezed interferometers, 
we have considered both 5\,dB ($e^{-2r}=0.316$) and 10\,dB
($e^{-2r}=0.1$) squeezing. 
Since in narrowband configurations, the seismic and thermal noises
do not affect significantly the choice of the SR parameters, 
the noise curves in Fig.~\ref{NBopt} have been optimized using only 
the quantum-optical noise (but we include filter optical losses). 
[For comparison we plot in Fig.~\ref{NBopt} the thermoelastic noises.] 
We obtain the parameters $\epsilon$, $\lambda$ and $\zeta$ for the
squeezed configurations following a heuristic procedure. Since 
the filters are very lossy, it is desirable to increase $\epsilon$ 
from the non-squeezed value, $2\pi\times25\,$Hz,  so that the noise due to
filter losses decreases and although the ideal
minimum of $S_h^{\rm lossless}$ increases, it is still 
buried by the noise due to filter losses. As we increase $\epsilon$
from $2\pi\times 25\,$Hz, we search for the $\lambda$ and
$\zeta$ that minimize $S_h$ at 600\,Hz; we find that the sensitivity
at 600\,Hz remains roughly the same, while the bandwidth increases.  
Trying to include as many sources as possible, 
we set  $\epsilon=2\pi\times 100\,$Hz for 5\,dB squeezing and 
$2\pi\times60\,$Hz for 10\,dB squeezing. 
The interferometer and filter parameters used in these configurations 
are listed in Table.~\ref{NBtab}.

As we see from Fig.~\ref{NBopt}, the Harms et al.\ 
[two dark continuous lines, one for 5\,dB squeezing 
the other for 10\,dB squeezing] and the BC [two
  dark dashed lines] schemes are extremely close to each other. The peak
sensitivities in the 5\,dB and 10\,dB cases are chosen to be
comparable to each other, while 10\,dB squeezing gives a broader
band. Although the FD techniques cannot increase the peak sensitivity
much due to filter losses, they do increase the bandwidth of observation. 
This will allow the observation of
multiple possible sources with a fixed  configuration. For example, with
the frequency and GW strengths estimates we used in Fig.~\ref{NBopt},
with 10\,dB squeezing, we can detect simultaneously 7 sources near
600\,Hz (including Sco~X-1), while with 5\,dB squeezing we can detect 6 of them
simultaneously (including Sco~X-1). In
Fig.~\ref{NBinc}, we plot the increase in SNR by the squeezed schemes,
as compared to the non-squeezed schemes, for LMXBs around the
resonant frequency; both 5\,dB and 10\,dB squeezing are shown. 
In Table~\ref{tab:LMXBsnr}, columns 5--9 we list the sensitivies of these
configurations.

As in the case of NS binary inspirals, SR interferometers with
frequency-independent squeezing and readout phase can also be
optimized for the detection of LMXBs. However, squeezing combined
with frequency-independent input-output optics cannot easily improve
peak sensitivity and bandwidth at the same time for narrowband
configurations. As a consequence, as we optimize the
frequency-independent scheme with
5\,dB squeezing, we obtain narrowband configurations that can detect at most 4 sources out of the group of 7 (including
Sco~X-1). [With 10\,dB squeezing, when a similar optimization
  is done for frequency-independent schemes, one finds that a wideband
  interferometer with 
  frequency-independent scheme\footnote{
Corbitt, Mavalvala and Whitcomb~\cite{CMW} are currently investigating 
this optical configuration.} 
can detect all 7 sources --- no narrowbanding is necessary, as we shall
  see in the next section.]

Now we look at the interferometer performances if assumptions other
than (a1) turn out to be true. In Fig.~\ref{fig:assumptions}, we show
the predicted GW strengths from the Z sources under the four
assumptions (obtained using Tables~\ref{tab:LMXBsnr} and
~\ref{tab:conversion}): (a1) (solid circles, with center frequency
around 600\,Hz), (b1) (open circles, with center frequency around
1200\,Hz), (a2) (solid triangles, with center frequency around
400\,Hz) and (b2) (open triangles, with center frequency around
800\,Hz).  Given these hypothetical groups of sources, we tune
squeezed-input SR interferometers (with Harms et al.\ or BC schemes,
which are equivalent at high frequencies) to each of them: around 600
Hz, 1200 Hz, 400 Hz, and 800 Hz, with interferometer parameters listed
in Table~\ref{tab:morenb} and noise curves shown in
Fig.~\ref{fig:assumptions}.  We remark that assumptions that yield
lower $f_{\rm GW}$'s tend to make the sources more detectable. 
In this study, we have also taken into account interferometer
losses, which has been neglected up till now.  We assume the ITM power
transmissivity to be $T=0.033$, SR-cavity round-trip loss to be $1\%$
(denoted by $\lambda_{\rm SR}$ in Ref.~\cite{BC2}) and photodetection
loss to be $2\%$ (denoted by $\lambda_{\rm PD}$ in
Ref.~\cite{BC2}).\footnote{The $1\%$ SR-cavity loss is the major
interferometer loss, according to Ref.~\cite{BC2}. In addition, we did
not use the value $T=0.005$ in the baseline design of Advanced LIGO:
assuming the same amount of loss per round trip inside the SR cavity,
a much smaller $T$ will make the effect of this loss much larger.}
These numbers are crude estimates; given the effects of interferometer
losses suggested by Fig.~\ref{fig:assumptions}, especially in higher
frequencies [i.e., if assumptions (b1) or (b2) turns out to be true],
more refined understanding of realistic interferometer losses, as well
as a more systematic study of interferometer parameters will be
crucial in fully understanding whether and how Advanced LIGO can
detect these narrowband sources.

\begin{table*}
\begin{scriptsize}
\centerline{
\begin{tabular}{ccccc|rc|rc|ccc}
\multicolumn{5}{c|}{Interferometer Configuration} &
\multicolumn{2}{c|}{Filter I} &
\multicolumn{2}{c|}{Filter II } & 
\multicolumn{3}{c}{Performance 10\,dB [5\,dB]} \\
\hline
Scheme &
\multicolumn{1}{c}{
$\displaystyle \frac{\epsilon}{2\pi\,{\rm Hz}}$}
& 
\multicolumn{1}{c}{
$\displaystyle \frac{\lambda}{2\pi\,{\rm Hz}}$}
&
\multicolumn{1}{c}{$\alpha$} & \multicolumn{1}{c|}{$\zeta$} & 
\multicolumn{1}{c}{$\displaystyle \frac{\Omega_{\rm res}^{\rm I}
  }{2\pi\,{\rm Hz}}$} &
\begin{tabular}{c} $T_i^{\rm I}$ \\ (ppm) \end{tabular} & 
\multicolumn{1}{c}{$\displaystyle \frac{\Omega_{\rm res}^{\rm II} }{2\pi\,{\rm Hz}}$} &
\begin{tabular}{c} $T_i^{\rm II}$ \\ (ppm) \end{tabular} & 
\multicolumn{1}{c}{
\begin{tabular}{c} $\sqrt{S_h}/(10^{-24}/\sqrt{{\rm Hz}})$ \\
at 600\,Hz \end{tabular}
} &
\multicolumn{2}{c}{
\begin{tabular}{c} NS-NS at 300\,Mpc \\ Spherical/MH \end{tabular}}
\\
\hline
No Filters  & $600$ & $0$ & $\pi/2$ & $0$ & 
& & & & $0.99$  [$1.77$] & 6.47/8.70 & [5.69/7.60] \\
Harms et al.\   & $600$ & $0$ & FD & $0$ & 
$2.8-600.0\,i$ & 1508 & 
$-41.3 -41.0\,i$ & 103 & $1.12$  [$1.82$] & 7.00/10.65 & [6.01/8.76] \\
BC &  $600$  & $0$ & $\pi/2$ & FD & 
$2.8-600.0\,i$ & 1508 &
$-41.3 -41.0\,i$ & 103 & $1.13$  [$1.85$] & 6.68/9.68 & [5.79/8.09]\\
\hline
\end{tabular}}
\end{scriptsize}
\caption{Parameters of wideband configurations, including: squeezing
  with frequency independent input-output optics, the Harms et al.\ scheme
  and the BC scheme. Since there is no detuning in these SR
  interferometers, they are equivalent to conventional
  interferometers, which KLMTV studied. We also give the noise
  spectral densities at 600\,Hz, and the SNR for NS-NS binaries at
  300\,Mpc distance.  Both 5\,dB and 10\,dB squeezing are considered,
  with 5\,dB numbers quoted in parentheses, ``[...]''.  In cases with
  10\,dB squeezing, the 
  sensitivties at 600\,Hz is only slightly worse than the narrowband
  configurations (see Table~\ref{NBtab}).  
\label{wbconfig}}
\end{table*}

\begin{figure}
\vspace{1cm}
\centerline{
\includegraphics[width=0.475\textwidth]{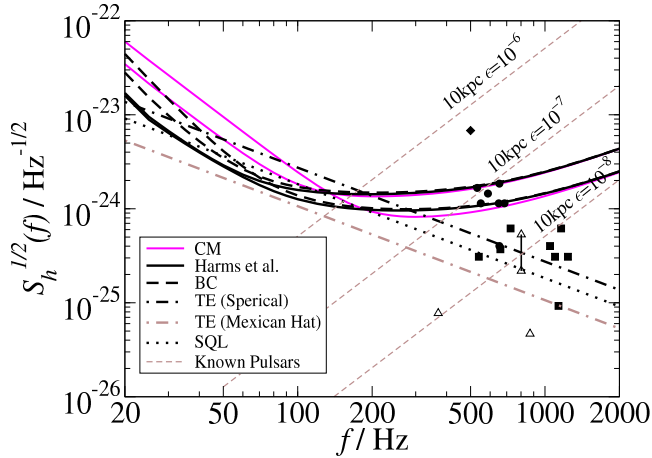}
}
\caption{Noise spectral densities of wideband configurations, with $\epsilon=2\pi\times 600\,$Hz
  and $\lambda=0$, frequency-independent squeezing ($\alpha=\pi/2$, $\zeta=0$, light continuous curve),
  the Harms et al.\ scheme ($\zeta=0$, dark continuous curve) and the BC scheme
  ($\alpha=\pi/2$, dark dashed curve) are used. Only the quantum-optical noise (taking filter
  losses into account) is included. Thermoelastic noises of
  spherical and MH mirrors are also shown, in dark and light dash-dot
  lines, respectively. The SQL is shown in dashed line. Possible GW signals 
  from LMXBs [under the baseline assumption] and known radio pulsars are also shown. \label{fig:WBopt}}
\end{figure}

\begin{figure}
\vspace{1cm}
\includegraphics[width=0.475\textwidth]{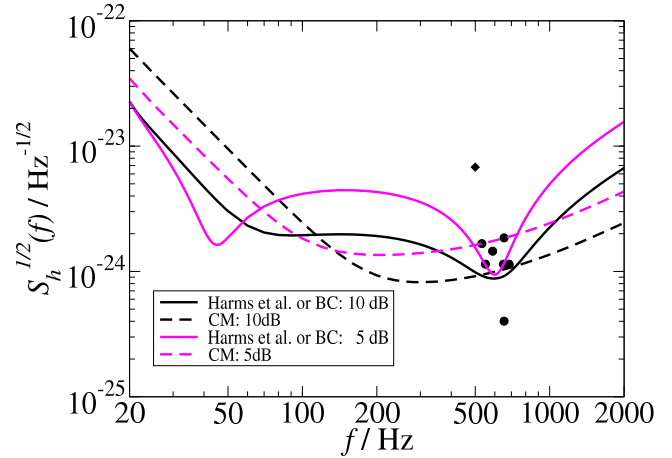}
\caption{Comparison of sensitivities to LMXB sources. We plot the noise
  curves of Harms et al.\ (or equivalently BC) scheme (continuous curves)
  and with frequency independent squeezing and readout scheme (the CM
  configuration,  dashed curves), with 5\,dB (light curves) and 10\,dB (dark curves)
  squeezing. We also plot the estimated frequencies and
  characteristic GW strengths of Sco X-1 and several other Z sources
  [under the baseline assumption (a1)].
\label{LMXBcompare}}
\end{figure}

\begin{figure}
\vspace{0.5cm}
\begin{center}
\includegraphics[width=0.45\textwidth]{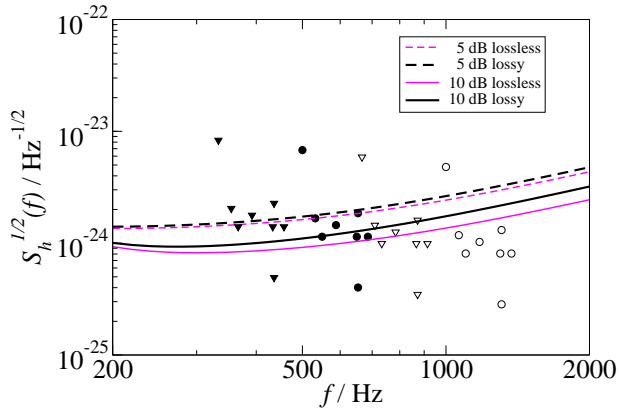} 
\end{center}
\vspace{0.25cm}
\caption{
Consequences of spin-frequency and emission-mechanism
  uncertainties on the detection of LMXBs (Z sources) using the wideband
  configuration. We plot the GW
  frequency and characteristic GW strength (for 20-day coherent
  integration), under approximations (a1) (solid circles), (b1) (open
  circles), (a2) (solid
  triangles) and (b2) (open
  triangles), along with CM
 noise curves with 5\,dB (dashed curves) and 10\,dB (continuous curves)
  squeezing, with (dark curves) and without (light curves)
  interferometer losses.
\label{fig:wbmore}}
\end{figure}

\subsection{Wideband configuration}
\label{subsec:wbopt}

The so-called wideband configuration of SR interferometers can be obtained 
setting $\lambda$ small and $\epsilon$ rather high. These 
configurations can be used to detect a broad range of generic sources, 
including: coalescence of  NS-NS binary, tidal disruption of NS 
by the BH companion, accreting NS's and radio pulsars. 
There are no specific criteria for the noise spectrum of the wideband
configuration. For simplicity we set $\epsilon=2\pi\times600\,$Hz 
and $\lambda=0$ (since this configuration is similar to the one 
by Corbitt and Mavalvala~\cite{CM}, we denote it by CM). 
The various parameters used are summarized in Table~\ref{wbconfig}, along
with SNR achievable for NS binaries at 300\,Mpc and sensitivities at
600\,Hz. Both 5\,dB and 10\,dB squeezing are considered, with 5\,dB
numbers quoted in parentheses, ``[...]''. We plot 
the corresponding noise curves in Fig.~\ref{fig:WBopt}. 

For frequencies higher than 200\,Hz, the CM noise curves are always
better than those with FD techniques. This is because, as observed by
Corbitt and Mavalvala, at high frequencies, the optimal squeeze 
angle and detection phase depend very mildly on the frequency. 
Therefore, the FD schemes, having additional filter losses, 
give worse performances. At high frequencies, the wideband schemes 
give a sensitivity of $3.2$ (10\,dB squeezing) or 
$1.8$ (5\,dB squeezing) times better (in amplitude) 
than the wideband configuration without
squeezing. With 10\,dB squeezing, the wideband
configurations can detect known pulsars at 10\,kpc with $\epsilon
\stackrel{>}{_\sim} 10^{-7}$ if $f_{\rm GW} \stackrel{>}{_\sim}
420\,$Hz, with $\epsilon
\stackrel{>}{_\sim} 3\times 10^{-8}$ if $f_{\rm GW} \stackrel{>}{_\sim}
1\,$kHz. [With 5\,dB squeezing, the minimum detectable $\epsilon$ will be 1.8
  times larger than the 10\,dB value.] However, if we also require good 
sensitivities below 200\,Hz, then the FD wideband schemes are preferable 
to the CM configuration . 

In addition, in the 10\,dB squeezing case, when spherical mirrors are used, 
the SNR for binaries are all above $96\%$ the optimal values obtained 
in the broadband case (see Table~\ref{NSNS}). However, for Mexican-Hat mirrors, the SNR is
less optimal, equal to $83\%$ (no filters), $91\%$ (Harms et al.) and
$93\%$ (BC) the optimal values (of the same scheme). [See Table~\ref{NSNS}.] 
These can be understood by going back to Sec.~\ref{subsec:nbopt} 
and observing in (the left panel of) Fig.~\ref{BBopt} 
that for spherical mirrors, the optimal noise curves are very wideband. 

It is also interesting to note that, with 10\,dB squeezing,
the sensitivities of  wideband configurations around 600 Hz, 
are only slightly worse, $\sim 10\%$ in amplitude, than the narrowband 
configurations. As a consequence, with 10\,dB squeezing, the wideband
configurations, even without FD techniques, can detect the same groups
of LMXBs discussed in the last section (see Fig.~\ref{fig:WBopt}). 
However, it should be noted that, if 10\,dB
squeezing is not achievable, then one cannot detect these sources with
the wideband configuration. For example, 5\,dB squeezing will barely allow
one or two more LMXBs than Sco X-1 to be detected.  
The narrowband configuration (with FD input-output schemes), by
contrast, will only miss one source 
in the group of 7. In Fig.~\ref{LMXBcompare}, we compare the sensitivities
of narrowband FD schemes and wideband frequency independent schemes to  LMXB
sources, with 5\,dB and 10\,dB squeezings.

Finally, by taking into account all other assumptions on spin
frequency and GW emission mechanism, we plot in Fig.~\ref{fig:wbmore}, 
the predictions of (a1), (b1), (a2) and (b2), along with
CM noise curves with 5\,dB (dashed curve) and 10\,dB squeezing
(continuous curve), with (dark curves) and without (light curves) 
interferometer losses included. 

\section{Third-generation interferometers}
\label{sec6}

We now assume that on time scales of third-generation  GW 
interferometers (around 2012), thermal noise of mirrors will 
be reduced by a large factor, for example by using cryogenic techniques, 
and we can take full advantage of the improvements in quantum noise 
obtained by FD input-output techniques. In addition, we assume that 
long filters 
can be fit into the existing vacuum tubes (which house the arm cavities) 
of the LIGO facility and made 4\,km long, so that optical losses will 
be significantly lowered (see Table~\ref{tab:loss}). 
As discussed in Sec.~\ref{subsec:fullopt}, the BC scheme is
nearly optimal for frequencies lower than the optical resonance [see
Fig.~\ref{fig4} and Eq.~\eqref{correction}], thus 
in the following we shall restrict our analysis to the BC scheme. 
However, before showing the performances, we want to discuss 
the limitations of the so-called short-cavity approximation, so far used 
in the literature to describe kilometer-scale filter cavities~~\cite{KLMTV00,ligoIII:sm}.

\subsection{Breakdown of short-cavity approximation}
\label{sec6.1} 

Up till now in this paper, we have been using the short-cavity
approximation, which imposes that $\Omega L/c \ll 1$. 
[Note that when refered to the interferometer, $L$ is 
 the arm length,  $\Omega$ is the GW
sideband frequency
or the optical resonant frequency $-\lambda-i\epsilon$; 
when refered to  filter cavities, $L$ is the 
 filter length, 
$\Omega$ is the GW sideband frequency or the filter 
resonant frequency $\Omega_{\rm res}$.] 
As we saw
in Secs.~\ref{sec2} and \ref{sec3}, the short-cavity approximations,
applied to SR interferometers and KLMTV filters, simplify
significantly their input-output relations [see
  Eqs.~\eqref{c11}--\eqref{coeffc}, \eqref{KLMTVtanz} and
  \eqref{KLMTVinout}], allowing a 
straightforward determination of filter parameters in the Harms et
al.\ and BC schemes via characteristic equations [Eqs.~\eqref{charHarms} and
  \eqref{charBC}]. 

On the contrary, without this approximation (i.e., when cavity lengths
are too long for this approximation to work), the filter parameters
cannot be determined easily --- it is not even clear whether the
optimal/suboptimal frequency dependence required by (the exact
input-output relation of) SR interferometers can at all be realized by
(those of) KLMTV filters. 

Since we have derived the exact input-output relation of the
filters [Eqs.~(\ref{eqban})--(\ref{anntoquad})],
as well as (partially\footnote{In Ref.~\cite{BC5} we treated exactly
  the propagation of light 
  inside the interferometers,  approximated the
  radition-pressure--induced motion of the ITM as being equal to that
  ot the ETM.})
that of the interferometer~[Eqs.~(99)--(104) of Ref.~\cite{BC5}], we can 
investigate the range of validity of the short-cavity approximations.

\begin{figure}
\vspace{1cm}
\includegraphics[width=0.475\textwidth]{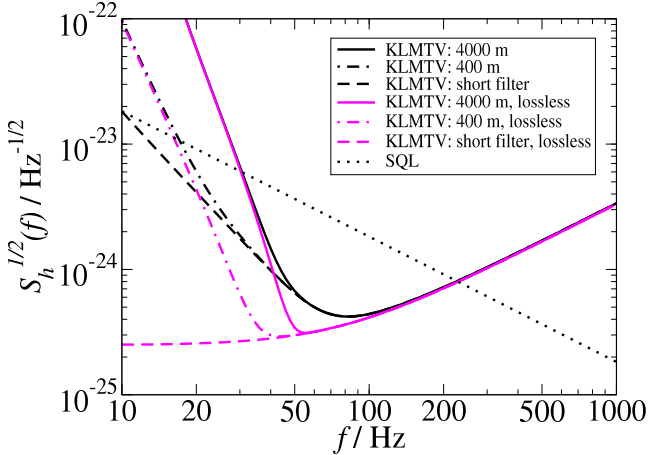}
\caption{ 
To show the break-down of the short-cavity approximation 
we plot several noise curves for the KLMTV (conventional) squeezed-variational 
interferometer~\cite{KLMTV00} with $I_c=840$\,kW, $m=40\,$kg and
$e^{-2r}=0.1$, fixing a bandwidth of $\epsilon=2\pi\times 75$\,Hz.  
The dark continuous curve refers to the nominal filter-cavity length 
$L_f =4$\,km, and the round-trip filter loss $T_e=20\,$ppm; while the
short-filter approximation ($L_f=0.1\,{\rm m}$) predicts the dark dashed curve. For
comparison, we also plot noise curves  of configuration with $L_f=400\,$m,
$T_e=2\,$ppm (dark dash-dot curve), along with
those of the lossless optical configurations with $L_f=4\,$km (light continuous
curve), $L_f=400\,$m (light dash-dot curve) and with short-filter
approximation ($L_f=0.1\,{\rm m}$) (light dashed curve).  
\label{KLMTV_long}
}
\end{figure}

\begin{figure}
\vspace{1cm}
\includegraphics[width=0.475\textwidth]{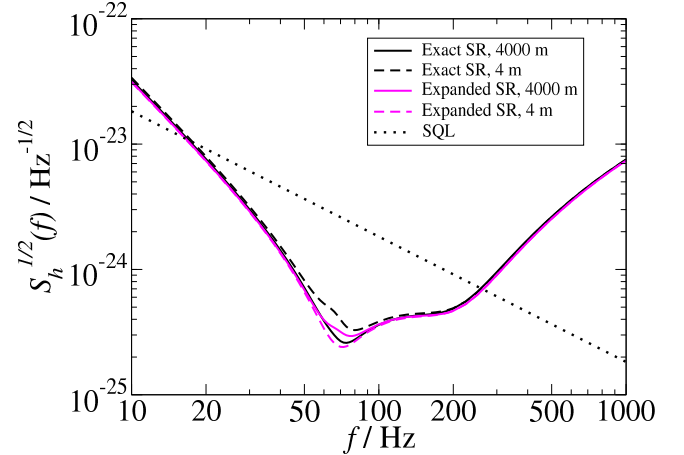}
\caption{To investigate the short-filter and short-arm approximation in 
SR squeezed-variational interferometers we plot several 
noise curves fixing $I_c=840$\,kW, $m=40\,$kg, $e^{-2r}=0.1$, 
$\epsilon=2\pi\times 80\,$Hz and $\lambda=2\pi\times 200\,$Hz. 
In particular we show the noise curve obtained with 
exact input-output relation and 4\,km filter cavities (dark continuous curve), 
and 4\,m filter (dark dashed curve); the noise curves obtained with first-order 
expanded input-output relation and 4\,km filter (light continuous curve) and 4\,m filter 
(light dashed curve). We use $T_e =20$ ppm for the 4\,km configurations, and 
$T_e =0.02$ ppm  for 4\,m configurations, such that the overall loss factor remains 
the same. The disagreements between curves with the same pattern 
(continuous or dashed) but different color (dark or light) is due to 
the inaccuracy of short-filter approximation; while the disagreements
between curves with the same color but different pattern is due to 
the inaccuracy of short-arm approximation.
\label{SR_long}}
\end{figure}

Let us start with conventional interferometers. 
As we have checked in this case, 
the short-arm approximation is still quite accurate, in the
sense that, for a given readout scheme (i.e., a given set of input or
output filters), using exact and
short-arm--approximated   
input-output relation do not give very different results. Yet, the
short-filter approximation seems to lose accuracy at low
frequencies. We study this effect in Fig.~\ref{KLMTV_long}, by
plotting several noise curves for   
squeezed-variational conventional interferometers~\cite{KLMTV00} with
$I_c=840\,$kW, $m=40\,$kg, $e^{-2r}=0.1$, and $\epsilon=2\pi\times
75\,$Hz, using the {\it exact} interferometer input-output
relation. In doing so, we use filters with bandwidths and resonant frequencies
obtained from the short-filter approximation, but with different
actual lengths and losses. In the figure, we show the noise curve for
filters with $L_f=4\,$km and $T_e=20\,$ppm in dark continuous curve,
and also lossy filters with decreasing length but the same $T_e/L_f$
ratio: $L_f=400\,$m in dark dotted curve and $L_f=0.1\,$m (to simulate
short-filter limit) in dark dashed
curve. The noise spectrum improves as the filter length  decreases.  
[In fact, since in this case the short-arm
approximation is accurate,  short filters must give the
optimal performance.]  In Fig.~\ref{KLMTV_long}, we also show
noise curves for lossless configurations with $L_f=4\,$km in light
continuous curve, $L_f=400\,$m in light dotted curve and $L_f=0.1\,$m
(to simulate short-filter limit) in light dashed curve. The reason for
such dramatic noise increase at low frequencies can be attributed
to the strong ponderomotive squeezing generated by conventional
interferometers at these frequencies (note that $q\rightarrow+\infty$ as
$\Omega\rightarrow 0$, see left panel of
Fig.~\ref{Fig1}). The stronger the squeezing, the higher the accuracy
requirement on the FD readout phase; yet the accuracy of short-filter
approximation does not increase indefinitely when $\Omega \rightarrow
0$.  

By contrast, as we have checked, the short-cavity approximations still
apply very well to squeezed-input conventional interferometer which at
low frequencies does not have as good an ideal sensitivity as the
squeezed-variational conventional interferometer.

In Fig.~\ref{SR_long} we investigate the short-arm and short-filter
approximations  
for SR squeezed-variational interferometers (the BC scheme)
with $I_c=840\,$kW, $m=40\,$kg, $e^{-2r}=0.1$, 
$\epsilon=2\pi\times 80$\,Hz and $\lambda=2\pi\times 200\,$Hz. 
In this case, both the short-filter and short-arm approximations introduce 
some inaccuracies, but they are by far not as significant as in the 
 squeezed-variational conventional interferometers. 
In particular, in Fig.~\ref{SR_long}, we plot noise curves obtained 
using exact interferometer input-output relation, with 4\,km (dark continuous curve) 
and 4\,m filters (dark dashed curve), and
 noise curves obtained using short-arm--approximated interferometer
 input-output relation,  with 4\,km (light continuous
curve) and 4\,m (light dashed curve) filters. We fix $T_e=20\,$ppm for
4\,km configurations, and $0.02\,$ppm for 4\,m configurations, keeping  
the same overall loss factor. [Filter resonant frequencies and
  bandwidths are still  obtained
from the characteristic equation \eqref{charBC}, which in turn has been
derived based on both short-arm
and short-filter approximations].
Noise curves with the same color (light or dark) 
use the same interferometer input-output relation, so the difference between them reflects the inaccuracy of the short-filter
approximation; those with the same pattern (continuous or dash)
share the same filter input-output relation, so their difference 
reflects the inaccuracy of the short-arm approximation. 
We conclude that the errors arising from the short-arm and short-filter 
approximations somewhat cancel each other, making the 
exact noise curve differ only slighly from the curve with both
short-arm and short-filter approximations applied. 
 The mild noise increase around the optical-spring
resonance in this case can
also be understood from the ponderomotive squeezing factor. As we see from
the left panel of Fig.~\ref{Fig1} (the dashed curve represents a
similar configuration), ponderomotive squeezing is the
strongest near this resonance, yet even here the squeeze factor is
still small compared to that of the conventional interferometer at lower
frequencies.

\begin{figure}
\vspace{1cm}
\centerline{\includegraphics[width=0.475\textwidth]{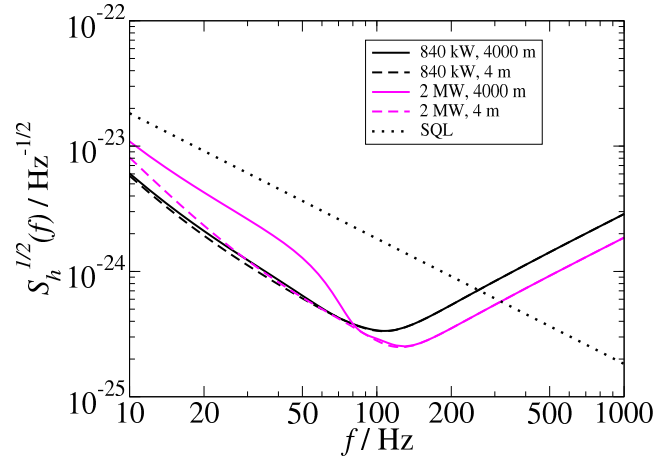}}
\caption{We plot the noise curves of squeezed-variational speed meters
with   $\Omega_{\rm s}=2\pi\times 95.3\,$Hz, $\delta=2\pi\times 100\,$Hz,
and $e^{-2r}=0.1$, assuming $I_c=840\,$kW (dark curves) and
$I_c=2\,$MW (light curves), and $L_f=4000\,$m (continuous curves)
and $L_f=4\,$m (dashed curves). The round-trip losses are $T_e=20$\,ppm for 
4000\,m filters and $T_e=0.02\,$ppm for 4\,m filters. 
The optical losses are included following Ref.~\cite{PC02}.
\label{SM_long}}
\end{figure}

We now discuss the short-cavity approximation in squeezed-variational 
speed meters~\cite{ligoIII:sm,ligoIII:sag}. 
We consider the configuration with $\Omega_{\rm s}=2\pi\times 95.3\,$Hz (the
``sloshing frequency'', as denoted by $\Omega$ in Ref.~\cite{PC02})
and $\delta=2\pi\times 100\,$Hz (bandwidth), assuming $e^{-2r}=0.1$. We 
include optical losses as done in Ref.~\cite{PC02}. 
As for conventional squeezed-variational interferometers,  the
short-arm approximation is rather accurate here. 
[This is true if the enhanced formula (i.e., expanded to
  next-to-leading order in $\Omega L/c$)  for the quantity $\kappa$ is used, see footnote~5 of 
Ref.~\cite{PC02}.] However, the short-filter approximation is not accurate
enough, if we increase the optical power further from Advanced LIGO
value. 
In Fig.~\ref{SM_long}, we plot four noise curves with $I_c=840\,$kW 
(dark curves) and $2\,$MW (light curves), and
filter lengths 4000\,m (and $T_e=20\,$ppm, continuous curves) and 4\,m
(and $T_e=0.02\,$ppm, dashed curves). [Again, resonant frequencies and
  bandwidths of the filters are obtained in the same way as in
  Ref.~\cite{PC02}, based on short-arm and short-filter
  approximations.] As we see, a filter length of 4000\,m 
increases the noise significantly  as $I_c$ becomes on the order of 
2\,MW. The increase is rather constant (and now as dramatic as in
KLMTV squeezed-variational conventional interferometers) at low frequencies, 
because speed meters have a constant ponderomotive squeezing 
factor at low frequencies~\cite{PC02}.  

We notice that in all the above cases where the short-cavity approximations 
break down, using filter parameters obtained from the characteristic equations 
(as we have done above), which are derived assuming those approximations, can no
longer be optimal.  Instead, one must optimize filter parameters
numerically using exact filter and interferometer input-output
relations.  We do not have quantitative results yet on how much
sensitivity can be gained by this re-optimization, but it does
not seem likely that the sensitivity can reach the optimal level 
(i.e., having the FD rotation from the filters matching exactly the 
interferometer's requirement).

\subsection{Performances of SR squeezed-variational interferometers}

\begin{figure}
\vspace{0.5cm}
\centerline{
\includegraphics[width=0.475\textwidth]{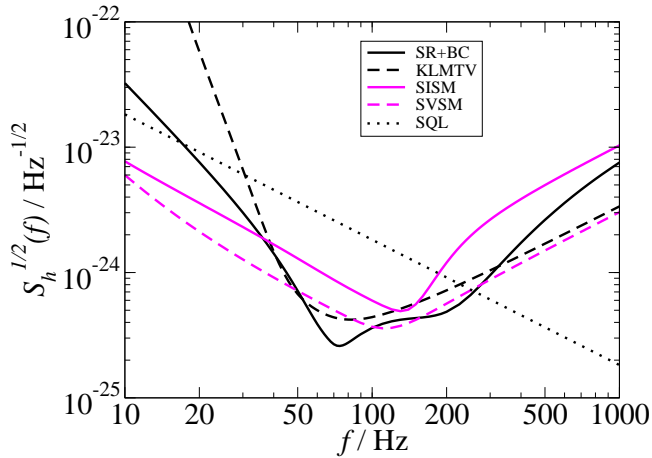}
}
\caption{Noise spectral densities of SR interferometers with 
BC scheme (SR+BC, dark continuous line), conventional 
squeezed-variational interferometer (KLMTV, dark dashed line), 
the Purdue-Chen speed-meter interferometer with ordinary homodyne 
detection (light continuous line) and FD homodyne detection 
(light dashed curve). We assume $I_{c}=840\,$kW, $m=40\,$kg and $e^{-2r}=0.1$; 
the SR interferometer with BC scheme has $\lambda=
2\pi\times 200\,$Hz, $\epsilon=2\pi\times 80\,$Hz and
$\alpha=\arctan(\epsilon/\lambda)$; the speed meter with ordinary
homodyne detection has $\Omega_{s}=173.2\,$Hz and 
$\delta=2\pi\times 200$\,Hz, while the speed meter with FD homodyne
detection has  $\Omega_{s}=2\pi
\times 95.3\,$Hz and $\delta=2\pi\times 100\,$Hz. Optical filters are assumed to be 4\,km long, with
20\,ppm round-trip loss. 
\label{ligo3}}
\end{figure}

Using exact filter and interferometer input-output relations (i.e., without applying
short-cavity approximations), and assuming that 4\,km filters will be 
used in third-generation interferometers, we compare in Fig.~\ref{ligo3}
the noise spectral densities of conventional squeezed-variational 
interferometers, SR squeezed-variational interferometers (BC scheme), and 
the squeezed-variational and -input speed meters. The BC scheme (which 
requires two additional km-scale cavities) has better sensitivity 
than the conventional squeezed-variational scheme (which also 
requires two additional km-scale cavities) for all frequencies 
below $\sim$350\,Hz. It has also better performances than the squeezed-input speed
meter (which requires one additional km-scale cavity) 
for all frequencies above $\sim$40\,Hz. The BC scheme 
has comparable (or slightly better) sensitivities with respect to 
the squeezed-variational speed meter (which requires three additional 
km-scale cavities)\footnote{We do not 
discuss the Sagnac interferometer which is also a speed meter {\it without} 
adding any km-scale cavities~\cite{ligoIII:sag}. A Sagnac interferometer can 
achieve sensitivities equivalent to the Michelson Purdue-Chen speed meters, and its 
squeezed-variational version requires only {\it two} additional
km cavities.} for frequencies between $\sim 50$\,Hz and $\sim 300\,$Hz.  

\section{Conclusions}
\label{sec7}

In this paper, we generalized the study of KLMTV~\cite{KLMTV00} on FD input-output
optics to SR interferometers, and discussed possible applications to
second- and third-generation GW interferometers. 

In the first part of the paper (Secs.~\ref{sec2}~--~\ref{sec4}), we
studied the quantum optical properties of SR interferometers and FD input-output schemes. 
We wrote the input-output relations of SR interferometers as a product of ponderomotive squeezing and
quadrature rotations, deriving explicit formulas for the intrinsic rotation angle and squeeze factor 
[see Eqs.~\eqref{tanvarphi} and \eqref{sinhq}], and investigating their features for several 
optical configurations. We found that ponderomotive squeezing becomes very weak 
in SR interferometers for frequencies higher than $\sim 300\,$Hz, 
regardless of the optical configuration [see Eq.~\eqref{qup}].  
Then, we built and analyzed the performances of the input-output scheme 
which combine FD homodyne detection (via KLMTV filters) with 
ordinary input squeezed vacuum (BC scheme), and compared it to the recent 
FD scheme proposed by Harms et al.~\cite{Harms03}. In the low-power 
limit (which also describes the high-frequency band of Advanced LIGO) 
we worked out the fully optimal input-output scheme [see Eq.~(\ref{fulloptlp})].  
In the general case, we derived simple analytical formulas for the 
fully optimal noise spectrum [Eq.~(\ref{Shopt})] and the optimal input 
squeeze angle [Eq.~(\ref{alphaopt})], and found that at low frequencies, 
the BC scheme can approximate the fully optimal noise curve 
very well [see Eq.~(\ref{correction})], providing better perfomances than the Harms et
al.\ scheme. These results for SR interferometers are quite similar
to the conventional interferometer case, in which as shown by KLMTV, a frequency
independent squeezed vacuum is already fully optimal (with FD readout), yet a frequency
independent readout cannot give as good a sensitivity (even with FD squeezing).  
[The BC and Harms et al.\ schemes generalize to SR interferometers 
the squeezed-variational and squeezed-input schemes introduced by KLMTV for 
conventional interferometers.]

In the second part of the paper (Sec.~\ref{sec5}), assuming that squeezed vacuum 
in the GW band would become available during the operation of Advanced LIGO, 
we evaluated the improvement in astrophysical sensitivity to specific
sources achievable by these FD schemes,
 under the facility limitation that the filters cannot be 
longer than $30\,$meters. It is important to note that, 
as  has been realized by Corbitt and Mavalvala~\cite{CM},
for nearly tuned SR interferometers with a large bandwidth (wideband configuration), 
the optimal input-output scheme is nearly frequency independent at high
frequencies. So, in this case it is possible to use squeezing 
optimally without introducing FD techniques. The Corbitt-Mavalvala (CM) wideband configuration can be used to
detect simultaneously various types of sources in the high-frequency band,
e.g., NS-NS merger, tidal disruption in NS-BH systems, or GWs from known radio pulsars. 
In addition, if 10\,dB squeezing can be realized, this wideband configuration can already detect a
group of 7 LMXBs (including Sco~X-1) around 600\,Hz. 

However, for specific sources with known spectral features, it is more
convenient to use optical configurations which are not wideband.  In
this case, the FD techniques can provide more flexibility and somewhat
better sensitivity, despite significant optical losses due to short
filters.  The Harms et al.\ scheme is shown to provide a better
sensitivity than the BC scheme in general in the Advanced LIGO era,
due to the BC scheme's higher susceptibility to losses at low
frequencies and the filter-length limitation to 30 meters. 
For NS-NS inspirals, assuming 10\,dB squeezing, {\it without}
using FD filters, one can improve the event rate by $89\%$ (spherical
mirror) or $42\%$ (MH mirror) with respect to the non-squeezed case;
further improvements of $20\%$ (spherical mirror) and $42\%$ (MH
mirror) can be obtained by the Harms et al.\ scheme. For
LMXBs, using 5\,dB squeezing, without using FD techniques, the
broadband configuration can only detect 3 sources simultaneously.  By
adjusting the SR parameters, we find that a frequency independent
input-output scheme cannot detect more than 4 sources, since in this
case we cannot gain sensitivity and bandwidth at the same time.  The
Harms et al.\ and BC schemes, instead can allow the detection of 6
sources simultaneously, by opening up the bandwidth with FD filters
(although the peak sensitivity cannot be improved much due to
significant filter losses).  With 10\,dB squeezing, the Harms et al.\
and BC schemes can open up the bandwidth further, including a group of
7 sources near 600\,Hz. [However, with this level of squeezing, the
broadband configuration can also detect the same sources, though with
slightly less sensitivity.] It is important to mention that, the above
detailed results about LMXBs are obtained by assuming (baseline
assumption) that the Z sources, which are the most promising GW
sources among all the LMXBs, are spinning at the QPO difference
frequency ($f_s=f_d\approx 300\,$Hz) and that the GW is dominantly
mass-quadrupole emission ($f_{\rm GW}=2 f_s \approx 600\,$Hz).  Recent
results~\cite{SAX}, however, may suggest that the spin frequencies
could be twice the QPO difference frequency and therefore lie around
$600\,$Hz instead. In addition, it is not clear yet whether the
dominant GW emission is mass quadrupole (e.g., due to deformation in
the crust) or current quadrupole (e.g., due to r-mode).  In
Figs.~\ref{fig:assumptions} and \ref{fig:wbmore}, we have briefly
explored the sensitivities of narrowband and wideband configurations
to these alternative scenarios.  Moreover, we also realized that
optical losses inside the SR interferometer become rather crucial in
deciding whether these sources will be detectable, especially if the
predicted waves are at higher frequencies. A more careful study of
optical losses will be reported in a forthcoming paper~\cite{BC7}.  It
should also be remembered that we have been relying on the
characteristic GW strength~\cite{CT} of LMXBs to provide a very rough
criterion of detectability --- in order to make a straightforward yet
specific comparison between different noise curves. As a consequence,
the true detectability of these sources by configurations studied in
this paper should be refined by a more rigorous study. 

In the third part of the paper (Sec.~\ref{sec6}), we 
investigated the performances of squeezed SR interferometer with FD 
output using two 4-km KLMTV filter cavities.  
We found [see Fig.~\ref{ligo3}] that SR interferometers 
with input squeezing and sub-optimal FD readout scheme (the BC scheme) can have
competing  sensitivities to existing proposals for third-generation
interferometers~\cite{ligoIII:sm,ligoIII:sag}, especially in the
middle frequency band of $50$ -- $350\,$Hz (see Fig.~\ref{ligo3}) .  We also discuss 
the limitation of the short-arm and short-filter approximations 
adopted by most of the past works~\cite{KLMTV00,ligoIII:sm}. 
Should at least one of these approximations break down, the optimal (or suboptimal) filter parameters provided by the characteristic equation \eqref{chareq} would not give the required FD rotation --- which, in fact, may not even be realizable by any sequence of detuned FP cavities. 
Indeed,  we found that for  squeezed-variational conventional interferometers
and speed meters (with high power), the approximation 
breaks down at low frequencies if applied to km-scale filters (see Fig.~\ref{KLMTV_long} and 
Fig.~\ref{SM_long}), which we attribute to the high level of ponderomotive squeezing and therefore more stringent requirement rotation accuracy in these configurations;  while for SR squeezed-variational interferometers (BC scheme),  the error introduced by the approximations is rather mild (see Fig.~\ref{SR_long}).

\acknowledgments
We wish to thank Nergis Mavalvala for reading and commenting on our
manuscript, and Lars Bildsten, Thomas Corbitt, Teviet Creighton, 
Lee Lindblom,  Kip Thorne and Stan Withcomb for useful discussions. 
We thank Curt Cutler for clarifying many points in the data analysis of narrowband sources.
AB thanks LIGO Caltech Laboratory under NSF cooperative agreement PHY92-10038 for support 
and the Theoretical Astrophysics and Relativity group for hospitality during her 
visit at Caltech during the Summer 2003. Research of YC is supported by NSF 
grants PHY-0099568, and by the David and Barbara Groce Fund at the San Diego
Foundation. 

\appendix
\section{Proof that the fully optimal scheme satisfies the sub-optimal condition}
\label{appa}

Suppose $\zeta$ is the readout phase. As seen in Sec. IV, with the sub-optimal $\alpha$ given 
by Eq.~(\ref{subopt}), only the squeezed quadrature 
enters the detected quadrature. We now show that given $\zeta$, this
$\alpha$ is also the {\it optimal} squeeze quadrature in the sense that it 
 minimizes the output noise. Since when we vary $\alpha$ alone, 
the signal strength in the output quadrature 
$b_{\zeta}$ remains constant, we only need to minimize the noise in 
$b_{\zeta}$. We write Eq.~(\ref{qfluc}) schematically as 
\beq
\Delta b_{\zeta} = \left(\begin{array}{cc} A_1 & A_2 \end{array}\right)
\mathbf{R}(\alpha)
\left(
\ba{r}
e^{-r}\tilde{a}_1 \\
e^{r} \tilde{a}_2
\ea
\right)\,,
\eeq
with $(A_1\,\,A_2)$ the product of the readout and the interferometer  parts, $A_{1,2} \in \Re$. 
The noise spectrum depends on $\alpha$ as 
\beq
S_h \propto
\left(\begin{array}{cc} A_1 & A_2 \end{array}\right) 
\mathbf{R}(\alpha) 
\left(
\begin{array}{cc}
e^{-2r} &  \\
 & e^{2r}
\end{array}
\right
)
\mathbf{R}(-\alpha) 
\left(\begin{array}{c} A_1 \\ A_2 \end{array}\right) \,.
\eeq
Minimizing $S_h$ then requires 
\beq
 \left(\begin{array}{cc} A_1 & A_2 \end{array}\right) 
\mathbf{R}(\alpha)
\left(\ba{c}
0 \\ 1
\ea
\right)=0
\,,
\eeq
which is equivalent to Eq.~(\ref{suboptimality}) and hence to Eq.~(\ref{subopt}).

\section{On the detectability of narrowband sources}
\label{app:LMXB}

In this appendix, we briefly review some subtleties that are not taken into 
account in our discussion of narrowband sources. 
We restrict the analysis to the mass-quadrupole radiation mechanism.  

\subsection{Characteristic amplitude of monochromatic gravitational waves: $h_c$}
\label{app:LMXB_a}

Let us consider a monochromatic source emitting GWs at (angular) 
frequency $\Omega_0 \neq 0$: 
\bea
\label{ht}
 h(t) &=& F_+ \,h_+\, \cos \Omega_{0} t + F_\times\, h_\times\, \sin\Omega_{0} t\,, \\
 h_+ &=& h_0\,(1 + \cos^2 i)\,, \\
 h_\times &=& 2 h_0\,\cos i\,,
\eea
where we denote with $F_+, F_\times$ the antenna patterns~\cite{KT}
and with $i$ the angle that the line-of-sight forms with the spin
direction of the neutron star. The quantity $h_0$ is an intrinsic GW
amplitude depending on the ellipticity of the isolated pulsar, or on 
the X-ray flux emitted by the LMXB (through the balance between accretion
torque and GW radiation-reaction torque), as well as on the distance of
the source.  Suppose the signal is observed in the time interval
$-T_0/2<t<T_0/2$, then the signal-to-noise ratio (SNR) using
optimal matched filtering is:
\bea
\label{SNR}
{\rm SNR}^2 &=&\int_{-\infty}^{+\infty}
\frac{d\Omega}{2\pi}\frac{2|\tilde{h}(\Omega)|^2}{S_h(\Omega)} \nonumber \\
&\simeq & \left[\frac{2T_0}{S_h(\Omega_0)}\right]\left[\frac{F_+^2\,
h_+^2+F_\times^2\, h_\times^2}{2}\right]\,.  
\eea 
Note that Eq.~\eqref{SNR} differs from Eq.~(29) of Ref.~\cite{KT} ---
the latter is wrong by a factor of 2~\cite{KT_private}. As we shall
see in the next section, when statistical issues are considered,
signals above a certain {\it threshold} SNR will be detectable.

In reality $F_{+,\times}$ vary due to earth's motion, or are 
unknown for some sources; the inclination angle $i$, although may stay
constant, could also be unknown, even for known sources, e.g., LMXBs.
As a consequence, with a fixed $h_0$, the SNR achievable can be
different. If we are interested in expected event rates, 
we should {\it average} the SNR over different source and detector orientations.
If we want to understand the detectability of a particular source 
and extract an {\it upper limit}, we should consider unfavorable geometries.
In this paper the predictions for LMXBs and isolated pulsars 
have been obtained averaging the SNR. As said, this is not 
appropriate for evaluating the detectability of individual sources.
In what follows we shall briefly review the average procedure, and comment
on what might be done in order to extract upper limits from a specific 
source.

Let us first consider the variation or uncertainty in
$F_{+,\times}$. The most straightforward {\it ansatz} for taking this
into account is to use the r.m.s.\ average of SNR over the entire sky
--- as viewed by the detector. The ansatz gives:
\bea \sqrt{\langle {\rm SNR}^2
\rangle_{\rm det}} &=& \sqrt{2}\,\langle F_+^2 \rangle^{\frac{1}{2}}
\sqrt{\frac{T_0}{S_h(\Omega_0)}} \sqrt{\frac{h_+^2+ h_\times^2}{2}}
\nonumber \\ && \equiv \frac{1}{h_n(\Omega_0)}\sqrt{\frac{h_+^2+
h_\times^2}{2}}\,.  
\eea
To obtain the RHS in the above equation we use 
$\langle F_+^2 \rangle=\langle F_\times^2 \rangle=1/5$, and 
define [as done in Eq.~(51) of Ref.~\cite{KT}]: 
\beq 
\label{eqhn}
h_n(\Omega_0) \equiv
\sqrt{\frac{S_h(\Omega_0)/T_0}{2 \langle F_+^2\rangle}}\,.  
\eeq 
Now let us consider the dependence of $\sqrt{\langle{\rm SNR}^2\rangle_{\rm det}}$  on $i$. 
There are two plausible averaging prescriptions. The first, which is the easiest, and most appropriate 
for a known source at a fixed distance, averages ${\rm SNR}^2$ uniformly over source angles, as
\bea
\label{snrrms}
\sqrt{\langle {\rm SNR}^2 \rangle_{\rm det\, \& \,src}} &=&
\frac{1}{h_n}\sqrt{\int \frac{d\Omega_{\rm src}}{4\pi}
\left(\frac{h_+^2+h_\times^2}{2}\right) } \nonumber \\
&=& \frac{1}{h_n}\,\left(\sqrt{\frac{8}{5}}\,h_0\right)\,.
\eea
It is then natural to define as {\it characteristic amplitude}:
\beq
h_a = \sqrt{\frac{8}{5}}\,h_0 \approx 1.26\,h_0\,.
\label{eqha}
\eeq
Another way of averaging was proposed by Thorne in Ref.~\cite{KT},
which has the property that $\mbox{event rate} \propto \langle {\rm
SNR} \rangle^3$, if we assume uniform distribution of sources in the universe,
\bea
\label{cubic}
\langle {\rm SNR}\rangle_{\rm det\, \& \, src} &=&\left[ \int \frac{d\Omega_{\rm src}}{4\pi} 
\left[\langle {\rm SNR}^2 \rangle_{\rm det}\right]^{\frac{3}{2}}\right]^{\frac{1}{3}} \nonumber \\
&=&
\frac{1}{h_n}\left[
\int \frac{d\Omega_{\rm src}}{4\pi}\left( \frac{h_+^2+h_\times^2}{2}\right)^{\frac{3}{2}}
\right]^{\frac{1}{3}}\,.\quad\quad
\eea
The integral~\eqref{cubic} cannot be performed analytically, so Thorne
introduced a {\it kludge} factor
\beq 
\left[ \int \frac{d\Omega_{\rm src}}{4\pi}\left(
\frac{h_+^2+h_\times^2}{2}\right)^{\frac{3}{2}} \right]^{\frac{1}{3}} \approx
\sqrt{\frac{4}{3}} \left[ \int \frac{d\Omega_{\rm src}}{4\pi}
\frac{h_+^2+h_\times^2}{2} \right]^{\frac{1}{2}} \,, 
\eeq 
yielding
\beq
\label{hc}
\left[\langle {\rm SNR}\rangle_{\rm det\, \& \, src}\right]_{\rm kludge} = 
\frac{1}{h_n}\left( \sqrt{\frac{32}{15}}\,h_0\right)\,.
\eeq
The above expression originated the following definition 
for the characteristic strength:
\beq
\label{eqhckludge}
[h_c]_{\rm kludge} = \sqrt{\frac{32}{15}}\,h_0 \approx 1.46\,h_0\,.
\eeq
The kludged characteristic strength $[h_c]_{\rm kludge}$ has been used
by many authors, including us in this paper. In particular,
$[h_c]_{\rm kludge}$ of a pulsar at distance $r$ with ellipticity
$\epsilon$ and frequency $f$ can be obtained from Eq.~(3.6) of
Ref.~\cite{BCCS}, and $[h_c]_{\rm kludge}$ from LMXBs with mass-quadrupole
emission mechanisms that balances the accretion torque can be obtained from
Eq.~(4) of Ref.~\cite{B98}.

However, the kludge factor $\sqrt{4/3}$ is not accurate. 
A simple numerical calculation gives:
\bea
&&\frac{ \displaystyle \left[ \int \frac{d\Omega_{\rm
src}}{4\pi}\left( \frac{h_+^2+h_\times^2}{2}\right)^{\frac{3}{2}}
\right]^{\frac{1}{3}} } {\displaystyle \left[ \int \frac{d\Omega_{\rm
src}}{4\pi} \frac{h_+^2+h_\times^2}{2} \right]^{\frac{1}{2}} } \nonumber \\
&=&
\frac{\displaystyle \left[\frac{1}{2}\int_0^\pi \sin i \left[(1+\cos^2
i)^2+(2\cos i)^2\right]^{3/2} di\right]^{\frac{1}{3}}} {\displaystyle
\left[\frac{1}{2}\int_0^\pi \sin i \left[(1+\cos^2 i)^2+(2\cos
i)^2\right] di\right]^{\frac{1}{2}} }\nonumber \\
&\approx& 1.047\,.  
\eea
or
\beq
\label{eqhc}
h_c = 1.32\, h_0\,.
\eeq
On the other hand, if we are interested in setting upper limits, 
we should use an $i$ that has the lowest possible $\langle {\rm SNR}
\rangle_{\rm det}$; this implies $i=\pi/2$ and 
\beq
\label{eqhUL}
\min_i \big\{\langle {\rm SNR} \rangle_{\rm det}\big\} = \frac{h_{\rm
UL}}{h_n}\,,\qquad h_{\rm UL} = \sqrt{\frac{1}{2}}\,h_0 =0.707\,h_0\,.
\eeq 
For a known source with constant $h_0$ but uncertain orientation 
(uniformly distributed $\cos i$), we can also
ask for the probability that $\langle {\rm SNR} \rangle_{\rm det}$
exceed $h_a/h_n$, $[h_c]_{\rm kludge}/h_n$ and $h_c/h_n$. The answers
are $41\%$, $29\%$ and $37\%$, respectively.

To summarize, we have managed to write the SNR for a given source
or a given set of sources (with fixed intrinsic amplitude $h_0$,
unknown $i$) in the form of 
\beq
\label{SNRhchn}
\mathrm{SNR} = \frac{h_c}{h_{n}}\,,
\eeq
where $h_c$ is the characteristic amplitude --- with four different
relations to $h_0$, \eqref{eqha}, \eqref{eqhckludge}, \eqref{eqhc} and
\eqref{eqhUL}, yielding SNRs that are either averaged in different
ways over different $i$'s, or taken as the minimum. [The quantitiy
$h_n$ is defined in Eq.~\eqref{eqhn} in terms of $S_h$ and integration
time.] Given the characteristic amplitude based on a particular prescription, values based on other prescriptions can be obtained from Eqs.~\eqref{eqha}, \eqref{eqhckludge}, \eqref{eqhc} and
\eqref{eqhUL}
using the fact that $h_0$ is the same in all of them. For example, given $[h_c]_{\rm
kludge}$ (which is used in this paper), we have
\bea h_a & = & 0.866\, [h_c]_{\rm kludge}\,,\\ 
h_c & = & 0.907\,[h_c]_{\rm kludge}\,,\\ 
h_{\rm UL} & = & 0.484\,[h_c]_{\rm kludge}\,. 
\eea 
Note that the more conservative
$h_{\rm UL}$ is a factor of $\sim 2$ smaller than $[h_c]_{\rm
kludge}$. 

\subsection{Factors that determine the detection threshold: from $h_c$ to $S_{h_c}$}
\label{app:LMXB_b}

With data analysis methods and desired statistical confidence, 
a threshold (minimum) SNR can be obtained; hence from
Eq.~\eqref{SNRhchn}, for a certain $h_c$, a maximum $h_n$, and thus a
maximum $S_h$, or $S_{h_c}$ can be obtained. [Here $h_c$
should be specified from an intrinsic GW amplitude $h_0$, through a
$h_c\mbox{--}h_0$ relation, like one of Eqs.~\eqref{eqha},
\eqref{eqhckludge}, \eqref{eqhc} and \eqref{eqhUL}.]

Refs.~\cite{BCCS,BC} introduces the canonical sensitivity $h_{3/\rm
yr}=4.2\,\sqrt{S_n(f) \times 10^{-7}\,{\rm Hz}}$, which is the
characteristic amplitude of the weakest source detectable with $99\%$
confidence level (i.e., 1$\%$ false alarm) in a coherent search of
$10^7$ s of data, if the frequency and phase evolution of the source
is known [see Eq.~(1.4) of Ref.~\cite{BCCS}]. This readily gives the
$S_{h_c}$ for known pulsars. However, it should be noted that
$S_{h_c}$ obtained using this $h_{3/{\rm yr}}$ only guarantees that
the {\it expectation value} (or {\it average}) of the detection statistic be
higher than the detection threshold~\cite{BCCS}, and gives a high
false-dismissal rate of about $50\%$~\cite{CC_private,S1continuous}. In other
words, even if the noise curve touches $S_{h_c}$ for a particular
source, there is still around $50\%$ chance this source will not make
the detection threshold.

For LMXBs, frequency and phase evolution of the GW due to orbital
motion is unknown and one must build an appropriate bank of templates
to search for these parameters, resulting in a threshold higher than
$h_{3/\rm yr}$; in addition, variations in accretion rate, which
induces ``random walks'' in the spin, and hence in the GW frequency,
further complicates the data analysis procedure, increasing the
threshold further.  Brady and Creighton studied these issues, and
devised a two-step hierarchical scheme for detecting such
signals~\cite{BC}. They use the relative sensitivity $\Theta_{\rm
rel}$ to measure the increase in the threshold for the characteristic
amplitude: $h_{\rm th} = h_{3/\rm yr}/\Theta_{\rm rel}$. As a
consequence, in our notation, we have $\sqrt{S_{h_c}(f)} =
h_c\,\Theta_{\rm rel}/(4.2\,\sqrt{10^{-7}\,{\rm Hz}})$.  Brady and
Creighton have shown that, for Sco X-1, with realistic computational
power, $\Theta_{\rm rel}=0.41$ [Sec.~VIIC of Ref.~\cite{BC}]. This
yields a value of $S_{h_c}$ comparable to that of a coherent
integration of 20 days. In the paper we use this prescription for all LMXBs
(except for SAX\,J1808.4--3658), and denote this characteristic
strength by $S_h^{20\mbox{-}\rm day}$.

\end{document}

%% file: LMXBSNR.tex
GX\,349$+$2 & 532 &   5.40 &   1.67 &   0.65 & {\bf   1.33} & {\bf   1.32} & {\bf   1.80} & {\bf   1.79} \\ 
4U\,1820$-$30 & 550 &   3.70 &   1.14 &   0.58 & {\bf   1.02} & {\bf   1.01} & {\bf   1.27} & {\bf   1.25} \\ 
GX\,17$+$2 & 588 &   4.70 &   1.45 & {\bf   1.48} & {\bf   1.52} & {\bf   1.51} & {\bf   1.65} & {\bf   1.63} \\ 
4U\,0614$+$06 & 654 &   1.30 &   0.40 &   0.19 &   0.35 &   0.34 &   0.44 &   0.43 \\ 
GX\,5$-$1 & 654 &   6.00 &   1.85 &   0.88 & {\bf   1.60} & {\bf   1.58} & {\bf   2.01} & {\bf   2.00} \\ 
Cyg\,X$-$2 & 686 &   3.70 &   1.14 &   0.36 &   0.80 &   0.79 & {\bf   1.17} & {\bf   1.16} \\ 
GX\,340$+$0 & 650 &   3.70 &   1.14 &   0.58 & {\bf   1.01} & {\bf   1.00} & {\bf   1.25} & {\bf   1.24} \\ 
Sco\,X$-$1 & 500 &  22.00 &   6.79 & {\bf   1.87} & {\bf   4.40} & {\bf   4.38} & {\bf   6.89} & {\bf   6.83} \\ 
\hline 
4U\,1702$-$429 & 660 &   1.20 &   0.37 &   0.16 &   0.31 &   0.31 &   0.40 &   0.40 \\ 
4U\,1728$-$34 & 726 &   2.00 &   0.62 &   0.14 &   0.34 &   0.34 &   0.57 &   0.56 \\ 
4U\,1916$-$053 & 540 &   1.00 &   0.31 &   0.13 &   0.26 &   0.26 &   0.34 &   0.33 \\ 
KS\,1731$-$260 & 1048 &   1.30 &   0.40 &   0.03 &   0.07 &   0.07 &   0.16 &   0.16 \\ 
Aql\,X$-$1 & 1098 &   1.00 &   0.31 &   0.02 &   0.05 &   0.05 &   0.11 &   0.11 \\ 
MXB\,1658$-$298 & 1134 &   0.30 &   0.09 &   0.01 &   0.01 &   0.01 &   0.03 &   0.03 \\ 
4U\,1636$-$53 & 1162 &   2.00 &   0.62 &   0.03 &   0.09 &   0.09 &   0.21 &   0.21 \\ 
4U\,1608$-$52 & 1238 &   1.00 &   0.31 &   0.01 &   0.04 &   0.04 &   0.09 &   0.09 \\ 
\hline 
SAX\,J1808.4$-$3658 & 802 &   0.71 &   0.22 (0.53)&   0.03 (0.08) &
0.08 (0.20)&   0.08  (0.20) &   0.16 (0.39) &   0.16 (0.39) \\ 
XTE\,J1751$-$305 & 870 &   0.15 &   0.05 &   0.00 &   0.01 &   0.01 &   0.03 &   0.03 \\ 
XTE\,J0929$-$314 & 370 &   0.25 &   0.08 &   0.01 &   0.03 &   0.03 &   0.06 &   0.05 \\ 

%% file: paper.bbl
\begin{thebibliography}{99}
%
\bibitem{LIGO} A. Abramovici et al., {\it Science} {\bf 256}, 325 (1992);
  \url{http://www.ligo.caltech.edu}.
%
\bibitem{S1inst} The LIGO Scientific Collaboration, {\it Detector
  Description and Performance for the First Coincidence Observations
  between LIGO and GEO}, \url{gr-qc/0308043}
%

\bibitem{VIRGO} B. Caron et al., {\it Class. Quantum Grav.} {\bf 14},
  1461 (1997); \url{http://www.virgo.infn.it}.
%
\bibitem{GEO} B.~Willke et al., {\it Class.~Quantum Grav.} {\bf 19}, 1377 (2002); 
\url{http://www.geo600.uni-hannover.de}.
%
\bibitem{TAMA} M. Ando et al., {\it Phys. Rev. Lett.} {\bf 86}, 3950
  (2001); \url{http://tamago.mtk.nao.ac.jp}.
%
\bibitem{S1continuous} The LIGO Scientific Collaboration, {\it Setting
   upper limits on the strength of periodic gravitational waves using
   the first science data from the GEO600  and LIGO detectors,}
   \url{gr-qc/0308050}. 
\bibitem{S1inspiral} The LIGO Scientific Collaboration, {\it Analysis of
  LIGO data for gravitational waves from binary neutron stars,}
  \url{gr-qc/0308069}. 
\bibitem{S1burst} The LIGO Scientific Collaboration, {\it  First upper limits from LIGO on gravitational wave bursts,} \url{gr-qc/0312056}.
\bibitem{S1stochastic} The LIGO Scientific Collaboration, {\it Analysis of First LIGO Science Data for Stochastic Gravitational Waves,} \url{gr-qc/0312088}. 
%
\bibitem{ALIGO} P. Fritschel, {\it 
Second generation instruments for the Laser Interferometer
Gravitational-wave Observatory (LIGO), in Gravitational Wave
Detection,} {\it Proc.~SPIE} {\bf 4856-39}, p. 282 (2002); \url{gr-qc/0308090}.
%
\bibitem{SR} B.J. Meers, Phys. Rev. D {\bf 38}, 2317 (1988); 
J.Y. Vinet, B. Meers, C.N. Man and A. Brillet, {\it Phys. Rev. D} {\bf 38}, 433 (1988);
B.J.~Meers and K.A.~Strain, {\it Phys. Rev. A} {\bf 44}, 4693 (1991); 
R. W. P. Drever, in ``The detection of gravitational waves,'' ed. by D. G. Blair, 
(Cambridge University  Press, Cambridge, England, 1991); 
%
\bibitem{M95} J.~Mizuno, ``Comparison of optical configurations for
  laser-interferometer gravitational-wave detectors,'', Ph.D.\ thesis,
  Max-Planck-Institut f\"ur Quantenoptik, Garching, Germany (1995). 
%

\bibitem{Caves} C.\ M.\ Caves, {\it Phys.\ Rev.\ D} {\bf 23}, 1693 (1981). 
%
\bibitem{Unruh} W.\ G.\ Unruh, in {\it Quantum Optics, Experimental 
Gravitation, and Measurement Theory},
eds.\ P.\  Meystre and M.\ O.\ Scully, (Plenum, 1982), p.\ 647.
%
\bibitem{SQL} V.B.~Braginsky and F.Ya.~Khalili, {\it Rev.~Mod.~Phys.}{\bf 68}, 1 (1996).
%
\bibitem{Others} M.\ T.\ Jaekel and S.\ Reynaud,
{\it Europhys. Lett.} {\bf 13}, 301 (1990); 
A.\ F.\ Pace, M.\ J.\ Collett and D.\ F.\ Walls, {\it Phys.\ Rev.\ A}
{\bf 47}, 3173 (1993).
%
\bibitem{SQZ} See, e.g., the following special issues of journals: 
{\it J.~Opt.~Soc.~Am.~B} 
{\bf 4}, 1453 (1987) 
and {\it Quantum Noise Reduction in Optical Systems,} edited by C.~Fabre and E.~Giacobino [Appl.~Phys.~B {\bf 55}, 189ff (1992)].
%
\bibitem{SQZexp} M.~Xiao, L.-A.~Wu, and H.~J.~Kimble, 
{\it Phys.~Rev.~Lett.}~{\bf 59}, 278 (1987); P.~Grangier, R.E.~Slusher, B.~Yurke, and A.~LaPorta, 
{\it Phys.~Rev.~Lett.} {\bf 59}, 2153 (1987). 
%
\bibitem{FDH} S.P.~Vyatchanin and Matsko, {\it JETP} {\bf 77}, 218 (1993); 
S.P.~Vyatchanin and E.A.~Zubova, {\it Phys.~Lett.~A} {\bf 203}, 269 (1995); 
S.P.~Vyatchanin, {\it ibid.}\ {\bf 239}, 201 (1998).
%
\bibitem{KLMTV00} H.J.~Kimble, Yu.~Levin, A.B.~Matsko, K.S.~Thorne
and S.P.~Vyatchanin, {\it Phys. Rev. D} {\bf 65}, 022002 (2002).
%
\bibitem{PC02} see Purdue and Chen in Ref.~\cite{ligoIII:sm}.
%
\bibitem{ANU} K.~McKenzie, D.A.~Shaddock, D.E.~McClelland, B.C.~Buchler, P.K.~Lam, \prl {\bf 88}, 231102 (2002).
%
\bibitem{CM} T.~Corbitt and N.~Mavalvala, ``Quantum noise in
  gravitational-wave interferometers: overview and recent
  developments,'' in {\it Noise an Fluctuations in Photonics and
  Quantum Optics}, Proceedings of SPIE 5111-23 (2003),
  \url{gr-qc/0306055}; T.~Corbitt and N.~Mavalvala, {\it Optimization of the
  Advanced LIGO detector to include squeezing,} (in preparation). 
%
\bibitem{AEI} R.~Schnabel, S.~Chelkowski, J.~Harms, A.~Franzen,
  H.~Vahlbruch and K.~Danzmann, ``Squeezed light enhanced Michelson
  interferometer,'' talke given at the {\it 2003 Aspen Winter Conference on
  Gravitational Waves and their Detection,} LIGO Document Number
  LIGO-G030220-00-Z.   
%
\bibitem{Harms03} J.~Harms, Y. Chen, S. Chelkowski, A. Franzen, 
H. Vahlbruch, K. Danzmann and R. Schnabel, \prd {\bf 68}, 042001
(2003). 
%
\bibitem{ligoIII:sm} V.B. Braginsky, F.Ya. Khalili, {\it Phys. Lett. A} 
{\bf 147}, 251 (1990); V.B. Braginsky, M.L. Gorodetsky,
F.Ya. Khalili, and K.S. Thorne, {\it Phys. Rev. D} {\bf 61} 044002 (2000); 
P. Purdue, {\it Phys. Rev. D} {\bf 66}, 022001 (2002);
P. Purdue and Y. Chen,  {\it Phys. Rev. D} {\bf 66}, 122004 (2002).
%
\bibitem{ligoIII:sag}
Y. Chen, \prd {\bf 67}, 122004 (2003); F.Ya. Khalili, ``Quantum
speedmeter and laser interferometric gravitational-wave antennae,'' 
\url{gr-qc/0211088}; S.L. Danilishin, ``Sensitivity limitations in optical speed meter 
topology of gravitational-wave antennae,'' \url{gr-qc/0312016}.
%
\bibitem{K03} V.B.~Braginsky, M.L.~Gorodetsky, F.Ya.~Khalili, {\it Phys.~Lett.~A} {\bf 232}, 340 (1997); F.Ya.~Khalili, {\it Phys.~Lett.~A}, 308 (2002); F.Ya.~Khalili, {\it Phys.~Lett.~A} {\bf 317}, 169 (2003).
%
\bibitem{CHP} J.M. Courty, A. Heidmann and M. Pinard, {\it Europhys. Lett.} {\bf 63}, 226 (2003). 
%
\bibitem{BC5} A.~Buonanno and Y.~Chen, \prd  {\bf 67},  062002 (2003). 
%
\bibitem{BC2} A.~Buonanno and Y.~Chen, {\it Phys. Rev. D} {\bf 64}, 042006 (2001); 
%
\bibitem{BM} V.B. Braginsky and A.B. Manukin, {\it Zh. \'Eksp. Teor. Fiz.} 
{\bf 52}, 987 (1967) [{\it Sov. Phys. JETP} {\bf 25}, 653 (1967)].
%

\bibitem{Whitcomb} S.~Whitcomb, (private communication).
%

\bibitem{CMW} T.~Corbitt, N.~Mavalvala and S.~Whitcomb, {\it Optical
  cavities as amplitude filters for squeezed fields,} (in preparation). 
%
\bibitem{BCUnpub} A. Buonanno and Y. Chen, (unpublished).
%
\bibitem{CT} C. Cutler and K. Thorne, ``An overview of gravitational-wave 
sources'', \url{gr-qc/0204090}.
%
\bibitem{TE} E. D'Ambrosio, R. O'Shaughnessy, V. Strigin, K.S. Thorne 
and S.P. Vyatchanin, (in preparation).
%
\bibitem{BCCS} P.R.~Brady, T. Creighton, C. Cutler and B.F. Schutz, {\it Phys. Rev. D} 
{\bf 57}, 2101 (1998).
%
\bibitem{BC} P.R.~Brady and T. Creighton, {\it Phys. Rev. D} {\bf 61}, 082001 (2000). 
%
\bibitem{old} R. V. Wagoner, {\it Atrophys.~J.} {\bf 278}, 345 (1984). 
\bibitem{B98} L.~Bildsten, {\it Astrophys.~J.} {\bf 501}, L89 (1998). 
\bibitem{AKS} N. Andersson, K. Kokkotas and N. Stergioulas, {\it Astrophys.~J.} {\bf 510}, 854 (1999).
\bibitem{UBC} G. Ushomirsky, L. Bildsten and C. Cutler, ``Gravitational waves from 
low-mass X-ray binaries: a status report,'' \url{gr-qc/0001129}.
\bibitem{LB02} L. Bildsten, ``Arresting accretion torques 
with gravitational radiation,'' \url{gr-qc/0212004}.
%
\bibitem{SAX} D. Chakrabarty, E.H. Morgan, M.P. Muno, 
D.K. Galloway, R. Wijnands, M. van der Klis and C.B. Markwardt, {\it Nature} 
{\bf 424}, 42 (2003); R. Wijnands, M. van der Klis, J. Homan, 
D. Chabrabarty, C.B. Markwardt and E.H. Morgan {\it Nature} {\bf 424},
43 (2003).  
%
\bibitem{CM98} D.~Chakrabarty and E.H.~Morgan, {\it Nature} {\bf 394}, 346 (1998).
%
\bibitem{rmode} N.~Andersson, {\it Astrophys.~J.}, {\bf 502} 708
  (1998); J.L.~Friedman and S.M.~Morsink, {\it Astrophys.~J.}, {\bf
  502}  714 (1998); L.~Lindblom, B.J.~Owen and S.M.~Morsink, {\it
  Phys.~Rev.~Lett.} {\bf 80}, 4843 (1998). 
%
\bibitem{UCB} G. Ushomirsky, C.~Cutler and L.~Bildsten, {\it
  Mon.~Not.~R.Astron.~Soc.} {\bf 319}, 902 (2000). 
%
\bibitem{OLCSVA} B.J.~Owen, L.~Lindblom, C.~Cutler, B.F.~Schutz,
  A.~Vecchio and N.~Andersson, {\it Phys.~Rev.~D}, {\bf 58}, 084020
  (1998). 
%
\bibitem{BC7} A.~Buonanno and Y.~Chen, (in preparation).
%
\bibitem{KT} K.S. Thorne, {\it Gravitational radiation}, in 
{\it 300 Years of Gravitation} ed. by S.W. Hawking and 
W. Israel (Cambridge University Press, Cambridge, 1987).
\bibitem{KT_private} K.S.~Thorne, (private communication). 
\bibitem{CC_private} C.~Cutler, (private communication).
\end{thebibliography}
